\documentclass[12pt]{article}
\pdfoutput=1

\usepackage{comment}
\usepackage{amssymb}
\usepackage{amsmath}
\usepackage{amsthm}
\usepackage{mathtools}
\usepackage[usenames,dvipsnames]{xcolor}
\usepackage{epsfig}
\usepackage{dcolumn}
\usepackage{upgreek}
\usepackage{setspace}
\usepackage{enumitem}
\usepackage{array, multirow,bigdelim,arydshln}
\usepackage{appendix}
\usepackage[export]{adjustbox}
\usepackage{xparse}
\usepackage[utf8]{inputenc}
\usepackage{microtype}
\usepackage{bm}
\usepackage{braket}
\usepackage{dsfont}
\usepackage{nccmath}
\usepackage{datetime}
\usepackage{bm}
\usepackage{multirow}
\usepackage{mathrsfs,amsfonts,xfrac,pifont,bbold,physics}
\usepackage{tensor}
\usepackage{float}
\usepackage{multicol}
\usepackage{subfig}
\usepackage{graphics}
\usepackage{amstext} 
\usepackage{tikz}
\usepackage{jheppub}
\usepackage{booktabs}
\usepackage[usenames,dvipsnames]{xcolor}
\usepackage{hyperref}
\hypersetup{
    pdfencoding=unicode,
	colorlinks=true,
	urlcolor=Maroon,
	linkcolor=RoyalBlue,
	citecolor=Maroon,
	pdftitle={Emergent Supersymmetry at Large N},
	pdfauthor={Shiroman Prakash},
	pdfdisplaydoctitle=true,
	pdfstartview=FitH,
	linktocpage=true
}
\DeclareMathAlphabet{\mathbbold}{U}{bbold}{m}{n}

\usepackage{physics}
\usepackage{lastpage}

\allowdisplaybreaks
\graphicspath{{figures/}}
\numberwithin{figure}{section}
\def \be {\begin{equation}}
\def \ee {\end{equation}}
\def \nn {\nonumber}

\def \leq {\leqslant}
\def \geq {\geqslant}

\newcolumntype{L}{>{$}l<{$}}

\newcommand{\RN}[1]{\textup{\uppercase\expandafter{\romannumeral#1}}}

\title{Emergent Supersymmetry at Large \texorpdfstring{$N$}{N}}
\author[1]{Shiroman Prakash}
\author[2]{Shubham Kumar Sinha}
\affiliation[1]{Dayalbagh Educational Institute,  Agra - 282005, India}
\affiliation[2]{Indian Institute of Science Education and Research Mohali,
Knowledge City, Sector 81, SAS Nagar, Punjab 140306, India}
\emailAdd{sprakash@dei.ac.in}
\emailAdd{shbkrsi@gmail.com}

\abstract{%
We search for infrared fixed points of Gross-Neveu Yukawa models with matrix degrees of freedom in $d=4-\varepsilon$. We consider three models -- a model with $SU(N)$ symmetry in which the scalar and fermionic fields both transform in the adjoint representation, a model with $SO(N)$ symmetry in which the scalar and fermion fields both transform as real symmetric-traceless matrices, and a model with $SO(N)$ symmetry in which the scalar field transforms as a real symmetric-traceless matrix, while the fermion transforms in the adjoint representation. These models differ at finite $N$, but their large-$N$ limits are perturbatively equivalent. The first two models contain a supersymmetric fixed point for all $N$, which is attractive to all  classically-marginal deformations for $N$ sufficiently large. The third model possesses a stable fixed point that, although non-supersymmetric, gives rise to many correlation functions that are identical to those of a supersymmetric fixed point when $N$ is sufficiently large. We also find several non-supersymmetric fixed points at finite and large-$N$. Planar diagrams dominate the large-$N$ limit of these fixed points, which suggests the possibility of a stringy holographic dual description. 
}

\begin{document}
	
\setcounter{tocdepth}{3}
\maketitle
\setcounter{page}{2}
	
\section{\label{sec:introduction}Introduction}

't Hooft \cite{'tHooft:1973jz} showed that the large-$N$ limits of quantum field theories whose fields are $N\times N$ matrices interacting via single-trace interactions are dominated by planar Feynman diagrams, with subleading corrections in $1/N^2$ corresponding to higher-genus Feynman diagrams. Such large-$N$ quantum field theories are therefore expected to be dual to string theories. For the case of conformal field theories, this correspondence has been made precise via the AdS/CFT correspondence. The main examples of the AdS/CFT correspondence involve highly supersymmetric field theories, such as $\mathcal N=4$ super Yang-Mills theory, which is dual to type IIB string theory on $AdS_5 \times S_5$ and the $\mathcal N=6$ ABJM theory, which is dual to type IIA string theory on $AdS_4 \times CP_3$ \cite{Maldacena:1997re, ABJM}. Do there exist strongly interacting conformal field theories with little or no supersymmetry whose large-$N$ limit is dominated by planar diagrams? 

The infrared fixed point of $\phi^4$ theory \cite{Wilson:1971dc, Wilson:1973jj} is one of the simplest examples of an interacting fixed point in $d=3$. It can be generalized to the critical $O(N)$ vector model \cite{Moshe:2003xn} by promoting $\phi$ to a vector $\phi_i$, but, in the large-$N$ limit, most planar diagrams are suppressed, and the resulting fixed point (like all vector models), contains an infinite tower of nearly conserved currents, dual to higher spin gauge fields in the bulk \cite{KlebanovPolyakov}. To obtain a fixed point dominated by all planar diagrams, it is natural to consider generalizations of the $\phi^4$ theory to contain scalar fields that transform as matrices under a symmetry group with rank $N$. Natural examples of such theories would be based on scalar fields in the adjoint representation of $SU(N)$, in the traceless symmetric rank-2 matrix representation of $SO(N)$, or in the bifundamental representation of $O(N)\times O(N)/\mathbb Z_2$. Such theories admit a single trace coupling of the schematic form $\tr \phi^4$. However, perturbative results in $d=4-\varepsilon$ appear to rule out the existence of real fixed points for these theories with non-vanishing single-trace coupling in $d=3$ at large-$N$.\footnote{Results for the bifundamental model are summarized in \cite{Osborn:2017ucf, Kapoor:2021lrr}. The nonexistence of fixed points within the epsilon expansion for the adjoint/symmetric traceless $\phi^4$ model follows from the calculations in this paper. See also \cite{Manenti:2021elk}, \cite{Henriksson:2020fqi}, and \cite{Reehorst:2020phk} respectively, and references therein.} 

In this paper, we focus our attention on another, closely related class of $d=3$ infrared fixed points  -- the Gross-Neveu Yukawa (GNY) models \cite{Hasenfratz1991TheEO, Zinn-Justin:1991ksq}, which can also be studied in $d=4-\varepsilon$, as described by \cite{Moshe:2003xn, Herbut:2009vu, Fei:2016sgs} and references therein.\footnote{See \cite{Muta:1976js, Wetzel:1984nw, Gracey:1990wi, Zinn-Justin:1991ksq, Gracey:1991vy, Luperini:1991sv, Vasiliev:1992wr, Gracey:1992cp,  Kivel:1993wq, Gracey:1993kc,  Derkachov:1993uw, Gracey:2008mf, Raju:2015fza, Ghosh:2015opa, Manashov:2016uam, Gracey:2016mio, Giombi:2017rhm,  Zerf:2017zqi, Mihaila:2017ble, Ihrig:2018hho, Cresswell-Hogg:2022lgg, Erramilli:2022kgp} for computations in the Gross-Neveu (GN) and GNY models. Recently, it was shown that there is a melonic generalization of the GNY model in $d=3$ with a real spectrum \cite{Kim:2020jpz, Prakash:2022gvb}, which partially motivates this work.}  Can one generalize the vectorial GNY fixed points to large-$N$ fixed points dominated by planar diagrams, by promoting the fields from vectors to matrices? Two natural possibilities include a GNY model with matter transforming in the adjoint representation of $SU(N)$, and a GNY model with matter transforming in the symmetric-traceless rank-two ($S_2$) representation of $SO(N)$. We study both these models and find that the large-$N$ limits of both these GNY models are equivalent\footnote{See \cite{Dunne:2016nmc} and references therein for a review of large $N$ equivalence.} to all orders in perturbation theory. We also study a third GNY model with an equivalent large-$N$ limit, which contains bosons in the $S_2$ representation of $SO(N)$ but fermions in the adjoint, i.e. antisymmetric rank-two tensor ($A_2$), representation of $SO(N)$. 

We study these theories in $d=4-\varepsilon$ up to three loops and determine all fixed points as a function of $N$. We find that one can construct real interacting fixed points with and without supersymmetry at finite $N$ that possess large-$N$ limits dominated by planar diagrams in the large-$N$ limit. This is in contrast to the corresponding theories consisting only of scalars, for which no real fixed point exists for $N>3$.

The fermionic and bosonic degrees of freedom are equal in the $SU(N)$ adjoint model and the $SO(N)$ $S_2$ model, leading to the possibility of fixed points with emergent supersymmetry, and indeed both models possess one such fixed point for all $N$. The $SO(N)$ $S_2-A_2$ theory contains unequal bosonic and fermionic degrees of freedom for any finite $N$, but its large-$N$ limit includes a fixed point, that is perturbatively equivalent to the supersymmetric fixed point of the previous two models.

Our analysis shows that for, the adjoint model, for $N=3$ and $N\geq 7$, the fixed point is attractive to all supersymmetric and non-supersymmetric classically marginal deformations -- and hence possesses \textit{emergent} supersymmetry \cite{Thomas2005, Fei:2016sgs}.  For a fixed point to possess ``emergent supersymmetry'' it must be attractive to some or all (classically-marginal) non-supersymmetric deformations, indicating the existence of renormalization group flows from a quantum field theory described by a generic non-supersymmetric choice of coupling constants that end in a supersymmetric fixed point. Strikingly, this means that in order to reach this large-$N$ fixed point from a general parity-preserving non-supersymmetric point adjoint GNY theory, one needs to tune only one parameter -- the mass of the scalar field. This result is similar to that obtained for the $O(1)$ GNY model, consisting of a single real scalar and a single Majorana fermion, \cite{Fei:2016sgs}, which contains a fixed point that is attractive to the one non-supersymmetric deformation. The supersymmetric fixed point of the $SO(N)$-$S_2$ model is attractive to all marginal deformations for $N=3$ and $N \geq 9$. 

The supersymmetric fixed points of the adjoint model at finite $N$ have been studied extensively via $\mathcal N=1$ bootstrap in \cite{Rong:2019qer}, and the supersymmetric fixed point for $N=3$ is the subject of certain duality conjecture\footnote{Our three-loop results, as well as bootstrap results from \cite{Rong:2019qer}, appear to contradict the predictions of this duality.} given in \cite{Gaiotto:2018yjh, Benini:2018bhk}.  An $\mathcal N=1$ supersymmetric fixed point of the $SO(N)$-$S_2$ theory was studied in \cite{Liendo:2021wpo}. Our results supplement these studies by a systematic three-loop study of all fixed points of the $SU(N)$-adjoint and $SO(N)$-$S_2$ GNY models, without demanding supersymmetry, for arbitrary $N$. To our knowledge, the $SO(N)$ $S_2$ -$A_2$ GNY model has not appeared before in the literature. 

The large-$N$ supersymmetric fixed point is an example of an interacting CFT in $d=3$, dominated by planar diagrams, with one relevant time-reversal invariant operator -- the scalar mass. We find the model contains three other non-supersymmetric large-$N$ fixed points, but they contain additional relevant operators and possess a classical scalar potential that is not positive-definite in $d=4$. However, by generalizing the theories described above to include an $O(2N_f)$ flavor symmetry acting on fermions, we find that the stable supersymmetric fixed point is replaced by one which is non-supersymmetric, but is still attractive to all classically-marginal deformations and possesses a classical scalar potential that is positive definite. Therefore our construction gives rise to both supersymmetric and non-supersymmetric large $N$ planar fixed points in $d=3$.

Conjectures associated with the swampland program (see e.g., \cite{Agmon:2022thq} and references therein) suggest that in a consistent quantum theory of gravity, a stable $AdS$ vacuum requires both supersymmetry \cite{Ooguri:2016pdq} and (large) extra dimensions \cite{Montero:2022ghl}. One can ask whether the fixed points studied in this paper, which are strongly interacting planar large-$N$ CFT's, could be viewed as counter-examples to these conjectures, i.e., do they contain a gap? The answer appears to be no; we estimate the scaling dimensions of various easy-to-compute unprotected operators in $d=3$ via Pad\'e approximations. In particular, by computing $\Delta_\phi$ and $\Delta_\psi$ we can estimate the twist of the leading twist higher-spin operators $J_s$ in the limit $s\to \infty$, and we find no evidence of divergence. Based on the results we conclude that the fixed points in $d=3$ are likely dual to string theories with finite string tension.  

The theories defined here can be deformed in $d=3$ by including a Chern-Simons gauge field with level $k$. In the large-$N$ limit, the 't Hooft coupling for such a gauge field would be $\lambda=N/k$, which does not run under renormalization group flow, and can be varied continuously \cite{Giombi:2011kc, Aharony:2011jz}.  Each of the fixed points we study here is therefore the endpoint of a line of large-$N$ fixed points. We do not expect large $N$ equivalence to hold for these lines when $\lambda$ is nonzero. It would be interesting to know what happens to these fixed points when $\lambda$ is taken to strong coupling. While the theory is parity-violating at intermediate values of $\lambda$, it is possible that at very strong coupling, the theory again preserves parity.\footnote{Recall that, e.g., Chern-Simons theories coupled to a single fermion or a single boson \cite{Giombi:2011kc, Aharony:2011jz} preserve parity at both weak and strong coupling, as required for the bosonization duality \cite{Aharony:2012nh, GurAri:2012is, Aharony:2012ns, Jain:2013py, Takimi:2013zca, Bedhotiya:2015uga, Gur-Ari:2015pca, Minwalla:2015sca} to hold.} It is therefore conceivable that, when the 't Hooft coupling of the Chern-Simons gauge field is taken to strong coupling, the three theories admit simple (and possibly equivalent) holographic dual descriptions such as $\mathcal N=1$ supergravity coupled to a small number of fields. This conjecture is similar to conjectures about Chern-Simons theories coupled to bifundamental matter  \cite{Banerjee:2013nca, Gurucharan:2014cva, GuruCharan:2017ftx}, and appears difficult to test. 

Our computations are facilitated by two-loop results easily obtainable via the package RGBeta \cite{Thomsen:2021ncy} based on calculations in \cite{Poole:2019kcm, Bednyakov:2021qxa}. We supplemented them with three loop results from \cite{Jack:2023zjt} (see also \cite{Davies:2021mnc, Steudtner:2021fzs}).

Let us discuss some related work. \cite{Liendo:2021wpo} also presents a systematic study of fixed points with $\mathcal N=1$ supersymmetry, via one-loop calculations in $d=4-\varepsilon$. \cite{Jack:2023zjt} contains three-loop results for a general GNY model, and explores a wide class of fixed points for these theories. \cite{Benini:2018umh, Benini:2018bhk} perform some computations in the $\varepsilon$ expansion, in order to test certain proposed dualities between Wess-Zumino models in $d=3$ and super QED${}_3$. The existence of the fixed points that we study here follows from the general considerations given there. \cite{Gracey:2021ili} contains a study of a related but different class of GNY models. \cite{Pannell:2023tzc} appeared when the draft was in the final stages of preparation, and discusses various aspects of multicomponent GNY fixed-points.   

Let us also remark that large-$N$ equivalence has been used to construct several candidate non-supersymmetric large-$N$ CFTs via orbifolds/orientifolds of ABJM theory and $\mathcal N=4$ SYM. These include \cite{Duff:1984sv, Kachru:1998ys, Lawrence:1998ja, Seiberg:1999xz, Tseytlin:1999ii, Dymarsky:2005nc, Dymarsky:2005uh, Pomoni:2009joh}, which are are summarized in Appendix C of \cite{Giombi:2017mxl}. However\footnote{We thank Z. Komargodski for discussions on this point}, it is worth noting that \cite{Liendo:2011da} considers an orientifold of $\mathcal N=4$ SYM (similar to a construction for ABJM \cite{Armoni:2008kr}) for which double-trace operators are shown to have vanishing $\beta$-functions in the large-$N$ limit. Though \cite{Liendo:2011da} may give rise to a large-$N$ non-supersymmetric CFT with a gap, it does not appear to arise as the limit of a sequence of finite $N$ CFTs, as the arguments for conformality in \cite{Liendo:2011da} assume that $N$ is large. 

\cite{Herbut:2006cs, Herbut:2009qb, Herbut:2023xgz} and references therein discuss the application of GNY models to phase transitions in condensed matter physics.  See also,  \cite{Vojta:2000zz, Lee:2006if,Grover:2013rc,Shimada:2015gda}. In particular, it is conceivable that the fixed points we study here could emerge as a description of the boundary of some topological phase as in \cite{Grover:2013rc}.
\section{Preliminaries}
\label{sec:review}

\subsection{Theories}
\label{sec:theories}
We study Gross-Neveu Yukawa theories in $d=3$ spacetime dimensions with $O(N^2)$ real pseudo-scalars and $O(N^2)$ Majorana fermions, which both transform as matrices under a global symmetry group.\footnote{We can restrict to the singlet sector by gauging this global symmetry with a Chern-Simons gauge field, but we do not do this here.} We write the scalar and fermion fields as,
\begin{equation}
	\phi_{ij}=\phi_{a}\,{\mathcal T}^{a}_{ij}\hspace{2cm}\psi_{ij}=\psi_{a}\,{\mathcal T'}^{a}_{ij},
\end{equation}
where ${\mathcal T}^a$ and ${\mathcal T'}^a$ are bases for two (possibly different) sets of $N\times N$ matrices that each form an irreducible representation of the symmetry group. We require that both sets of matrices transform in the same way under the action of the symmetry group, so that they can be multiplied together.   

We will consider three choices for these bases.
\begin{itemize}
    \item $SU(N)$ adjoint model: ${\mathcal T}^{a}_{ij}={\mathcal T'}^{a}_{ij}=\Lambda^a_{ij}$, where $\Lambda^a_{ij}$ are a basis for the set of traceless Hermitian $N\times N$ matrices, normalized\footnote{These normalization conventions are chosen to match those used in RGBeta \cite{Thomsen:2021ncy}.} so that $\tr (\Lambda^a \Lambda^b) = \frac{1}{2} \delta^{ab}$, and $\Lambda^a_{ij} \Lambda^a_{kl}= \frac{1}{2}\delta_{ij}\delta_{kl} - \frac{1}{2N}\delta_{ij}\delta_{kl}$.
    \item $SO(N)$ $S_2$ model: ${\mathcal T}^{a}_{ij}={\mathcal T'}^{a}_{ij}=T^a_{ij}$, where $T^a_{ij}$ are a basis for the set of real, symmetric-traceless  $N\times N$ matrices, normalized so that $\tr (T^a T^b) = \delta^{ab}$, and $T^a_{ij} T^a_{kl}= \frac{1}{2}\left (\delta_{il}\delta_{jk} + \delta_{ik}\delta_{jl}\right) - \frac{1}{N}\delta_{ij}\delta_{kl}$. 
    \item $SO(N)$ $S_2$-$A_2$ model: ${\mathcal T}^{a}_{ij}=T^a_{ij}$, and  ${\mathcal T'}^{a}_{ij}=H^a_{ij}$, where $H^a_{ij}$ are a basis for real antisymmetric $N\times N$ matrices (i.e., generators of $SO(N)$), normalized so that $\tr (H^a H^b) = \delta^{ab}$, and $H^a_{ij} H^a_{kl} = \frac{1}{4} \left( \delta_{il}\delta_{jk} - \delta_{ik}\delta_{jl} \right)$. Although we are using different representations for fermions and bosons, there is no obstruction to obtaining $SO(N)$-invariant operators by taking the  trace of the product of any sequence of matrices, consisting of matrices from both representations.
\end{itemize}


We normalize the kinetic term in the action as follows:
\begin{equation}
    \frac{1}{2}(\partial \phi_a)^2 + \frac{i}{2} \bar{\psi}_{a'} \partial_\mu \gamma^\mu \psi_{a'}.
\end{equation}

We will study the theory via the $\varepsilon$ expansion in $d=4-\varepsilon$. The most general renormalizable action in $d=4$ contains the following three classically-marginal interactions\footnote{Conventionally, the quartic interactions should be normalized with a factor of $1/8$, as each possesses an automorphism symmetry group $D_4$ which has order $8$, but we omit this numerical factor for simplicity.} which become relevant in $d=3$:
\begin{align}
	V= & {g_1}\tr \phi^4 +g_2(\tr \phi^2)^2
 + \frac12 y \tr (\phi \bar{\psi}\psi). 
 \end{align}
The Yukawa coupling is proportional to $d^{abc} \phi_a \bar{\psi}_{b} \psi_{c}$ where $d^{abc}=\tr \left( {\mathcal T}^a \{ {\mathcal T'}^{b}, {\mathcal T'}^{c}\} \right)$. For the theories we consider, $d^{abc}$ vanishes for $N=2$. For $N=3$, there is only one independent quartic scalar coupling, which we denote as $g = g_2$. A field redefinition $\phi \to -\phi$ takes $y \to -y$, so it is convenient to specify fixed points by the value of $y^2$.
 
In addition, there are three relevant interactions: mass terms for the fermion and scalar, as well as a single-trace cubic scalar coupling.
 \be
 \frac{\chi}{2} \tr(\phi^2) + \frac{\rho}{2} \tr(\bar \psi \psi) + \alpha \tr(\phi^3).
 \ee
These couplings are constrained when we impose time-reversal symmetry in $d=3$. $\bar{\psi}\psi$ is odd under time-reversal symmetry, so $\phi$ must be a pseudo-scalar if the Yukawa interaction is to be time-reversal invariant. Demanding time-reversal invariance, then, forces both the fermion mass term and cubic scalar interaction to vanish -- leaving the scalar mass as the only classically-relevant parameter in the action that needs to be tuned to zero at a fixed point. 

We define the large-$N$ limit of the theory, by taking the limit $N \to \infty$ while keeping the following `t Hooft couplings constant: 
\begin{equation}\label{eq: t`hooft coupling}
    \lambda_y=y\sqrt{N}, \quad \lambda_1 = g_1 N, \quad \lambda_2 = N^2 g_2.
\end{equation}
The resulting theory is dominated by planar Feynman diagrams.

Below, we will study the theory in $d=4-\varepsilon$. This requires us to introduce an additional $O(N_f)$ flavour symmetry that acts only on the fermions, which are defined to transform as a vector under $O(N_f)$. Our $\beta$-functions are computed in $d=4$ via the MS-bar scheme, for a theory that  contains $N^2-1$ scalar fields and $N_f(N^2-1)$ Majorana fermions. When continuing results to $d=3$, we continue $N_f=\frac{1}{2}$, because Majorana spinors are two-dimensional in $d=3$, but four-dimensional in $d=4$. This is essentially the same approach used to study $\mathcal N=1$ supersymmetric quantum field theories in $d=3$ via $d=4-\varepsilon$ in earlier works such as \cite{Fei:2016sgs} and \cite{Jack:2023zjt}. However, it is important to note that there are subtleties associated with studying fermions in the epsilon expansion, but, for the observables we calculate, the simple method described here works up to three loops.\footnote{In particular, suppose we start with the $d=4$ theory with $SU(N)\times O(N_f)$ symmetry group described above. Upon rewriting the four-component Majorana spinors as two-component spinors, we would like to obtain a theory in $d=3$ with $SU(N)\times O(2N_f)$ symmetry, for the analytic continuation to $N_f=1/2$ to make sense and give rise to a supersymmetric theory in $d=3$. However, when the four-component Majorana spinors are reduced to pairs of two-component spinors, as described in Appendix A of \cite{Jack:2023zjt}, the symmetry group of the theory turns out instead to be $SU(N)\times \left( \left(O(N_f) \times O(N_f)\right) \rtimes \mathbb Z_2 \right)$.  While one may be able to continue $O(N_f)\times O(N_f) \rtimes \mathbb Z_2$ to $N_f=1/2$ following \cite{Binder:2019zqc}, such a theory would not be supersymmetric, and is not of interest to us in the present work. The difference between computations in theories with these two symmetry groups is closely tied to how one chooses to deal with traces of products of three or more gamma matrices in dimensional regularization, and luckily, as noted in \cite{Jack:2023zjt} and \cite{Steudtner:2021fzs}, the observables we compute here are identical for theories with both symmetry groups, up to three loops. There exist more sophisticated regulation schemes, such as ``DREG${}_3$'' in \cite{Zerf:2017zqi} or the scheme in \cite{Pannell:2023tzc}, that may perhaps allow one to perform higher-loop computations directly in the $SU(N)\times O(2N_f)$ theory. See  \cite{Erramilli:2022kgp} for detailed discussion of this in the context of the more familiar vectorial GNY model. We thank A. Stergiou for discussions on this point.}

\subsection{Review of large \texorpdfstring{$N$}{N} equivalence}
\label{sec:largeNequivalence}

Some examples of large-$N$ equivalence includes \cite{Eguchi:1982nm, Gonzalez-Arroyo:1982hyq, Kovtun:2007py}.
The theories we study turn out to be perturbatively equivalent in the large-$N$ limit.
We refer the reader \cite{Dunne:2016nmc} for a detailed review of large-$N$ equivalence. Some examples of large-$N$ equivalence include \cite{Kachru:1998ys, Bershadsky:1998cb,  Armoni:2003gp, Armoni:2003fb, Kovtun:2003hr, Armoni:2004ub,  Kovtun:2004bz,  Kovtun:2005kh, Unsal:2006pj, Bond:2019npq, Jepsen:2020czw}.
In particular, the perturbative equivalence of the theories we study here at large-$N$ follows from arguments given in \cite{Bershadsky:1998cb}. Let us briefly review how these arguments apply to our theories.

Consider a quantum field theory with a global symmetry group $G$, which we consider to be the parent theory. In our case, $G=SU(N)$, and all the matter fields transform in the adjoint representation of $SU(N)$.  This theory contains a discrete symmetry operation $C$ that maps $\Lambda^a \to (\Lambda^a)^*$, and satisfies $C^2=1$. Let $\Gamma = \{ 1, C\}$. We obtain an orbifold theory by projecting onto states that are invariant under the action of $\Gamma$. 

Let us explicitly construct a representation of $\Gamma$.  $SU(N)$ can be thought of as a subgroup of $SO(2N)$ as follows: we promote the hermitian traceless matrices $\Lambda^A$ forming the adjoint representation of $SU(N)$ to $2N\times 2N$ real symmetric matrices $\tilde{T}^A$, by making the replacements $1 \to \begin{pmatrix} 1 & 0 \\ 0 & 1 \end{pmatrix}$, and $i \to \begin{pmatrix} 0 & 1 \\ -1 & 0 \end{pmatrix}=i\sigma_y$.  The discrete symmetry operation $C$ is represented via conjugation by the traceless matrix $\gamma_C=1_{N\times N} \otimes \sigma_x$, so that  $\tilde{T}^A \to \gamma_C^{-1} \tilde{T}^A \gamma_C$. We define the orbifold theory of the parent field theory as the theory whose states are invariant under conjugation by all elements of $\Gamma$ in this representation. The states invariant under conjugation by $\gamma_C$ are clearly in one-to-one correspondence with real symmetric $N\times N$ matrices, so the orbifold theory for the parent $SU(N)$ adjoint GNY model is the $SO(N)$ $S_2$ GNY model. Note that the representation is \textit{regular}, i.e., $\tr \gamma_g = 0$ for all $g \neq 1$. 

The large-$N$ limit of the orbifold theory whose states are restricted to those invariant under conjugation by $\gamma_C$ is perturbatively equivalent to the original theory. To see this, define the projector $P$ which acts on fields as 
\begin{equation}
    P \phi^a \tilde{T}^a_{ij} = \frac{1}{|\Gamma|} \sum_{g \in \Gamma} \phi^a (\gamma_g^{-1})_{ik}\tilde{T}^a_{kl}(\gamma_g)_{lj},
\end{equation}
and projects fields in the parent theory onto the orbifold theory. The free propagator in the orbifold theory is related to the free propagator in the parent theory by a projection, which ensures that, in loops, we only sum over states in the orbifold theory: 
\begin{equation}\langle \phi_{ij} \phi_{kl} \rangle^{free}_{orbifold} = \frac{1}{|\Gamma|} \sum_{g \in \Gamma}  (\gamma_g^{-1})_{im}(\gamma_g)_{nj} \langle \phi_{mn} \phi_{kl} \rangle^{free}_{parent}.
\end{equation}

\begin{figure}
\begin{center}
    \includegraphics[width=4cm]{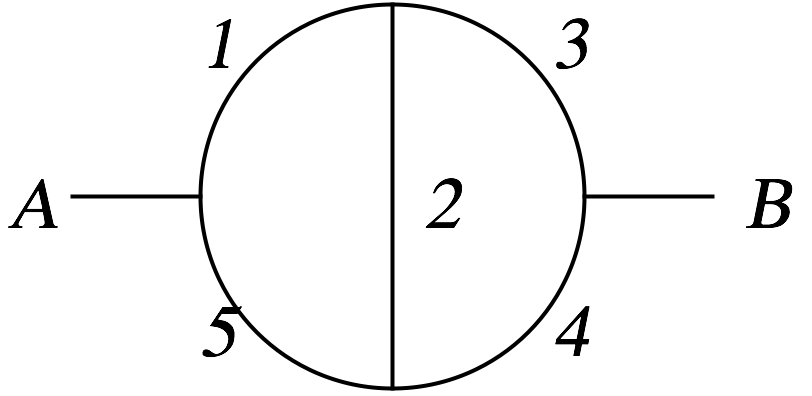}
    \caption{A planar two-loop correction to the propagator. One can show the value of an $L$-loop planar diagram in the orbifold theory is proportional to the corresponding diagram in the parent theory, with proportionality constant $1/|\Gamma|^L$. \label{two-loop}}
\end{center}
    \end{figure}

Consider any planar diagram in the orbifold theory, such as the two-loop correction to the scalar propagator shown in Figure \ref{two-loop}. 
This diagram is related to the corresponding diagram in the parent theory by the following factor:
\begin{equation}
    F^{AB}_{orbifold}(p_1)= \frac{1}{|\Gamma|^5}\sum_{g_1,~g_2,~g_3,~g_4,~g_5} \tr ( \gamma_{g_1}\gamma_{g_2} \gamma_{g_5})\tr (\gamma_{g_2} \gamma_{g_3} \gamma_{g_4} ) \tr (\tilde{T}^A \gamma_{g_1} \gamma_{g_3} \tilde{T}^B \gamma_{g_4} \gamma_{g_5} ) F^{AB}_{parent}
\end{equation}
with no sum on $A$ and $B$, and we use $g=g^{-1}$ for simplicity. Because $\tr \gamma_g=0$ unless $\gamma_g=1$, the only terms which are non-zero in this sum are those with $\gamma_{g_1}\gamma_{g_2} \gamma_{g_5}=1$, and $\gamma_{g_2} \gamma_{g_3} \gamma_{g_4}=1$. Using these constraints, we find that   \begin{equation}
    F^{AB}_{orbifold}(p_1)=\frac{1}{|\Gamma|^5} \sum_{g_1,~g_2,~g_3}\tr (\tilde{T}^A \gamma_{g_1} \gamma_{g_3} \tilde{T}^B \gamma_{g_3} \gamma_{g_1} ) F^{AB}_{parent} = \frac{1}{|\Gamma|^2}F^{AB}_{parent} ,
\end{equation}
if $\tilde{T}^A$ and $\tilde{T}^B$ are invariant under the projection. More, generally, in a planar diagram with $L$ loops, we would find $F^{AB}_{orbifold}(p_1)=\frac{1}{|\Gamma|^L}F^{AB}_{parent}$. This is because each loop gives a relation of the form $\Pi_i \gamma_i=1$, each of which can be used to eliminate one of the sums, and the planarity ensures that the final trace, containing external edges, will give rise to factors that cancel out if any external fields are invariant under the projections. On the other hand, if we considered a non-planar diagram, this procedure will clearly not work. For more details see \cite{Bershadsky:1998cb}. Thus, if we restrict attention to only planar diagrams, the two theories are equivalent with a rescaled coupling constant.

This argument applies to any symmetry operation. There is another symmetry operation $\gamma'=(-1)^F \gamma$, where $F$ is the fermion number operator. The orbifold (or orientifold) theory defined using $\gamma'$ in this case contains a scalar field in the symmetric-traceless-rank-two representation of $SO(N)$, and a Majorana fermion in the anti-symmetric rank-two representation of $SO(N)$. This theory contains $N(N+1)/2-1$ bosonic degrees of freedom, and $N(N-1)/2$ fermionic degrees of freedom, and is thus manifestly non-supersymmetric. However, in the large-$N$ limit, the planar diagrams contributing to observables common to both parent and orbifold theories will be equivalent, up to a re-scaling of the coupling constant. 

The above argument implies that, assuming the appropriate symmetry operations are unbroken, correlation functions of observables common to all three theories will coincide in the planar limit. 

What are the common observables in all three theories? From the perspective of holography, it is most interesting to restrict attention to the single-trace operators, composed of a finite number of fields. The single-trace operator content of the $SO(N)$ $S_2$ theory and the $SU(N)$ adjoint theory, differ from single-trace operator content of the $SO(N)$ $S_2$ $A_2$ theory. For the $SO(N)$ $S_2$ theory and the $SU(N)$ adjoint theory, the single-trace operators include any operator composed from a string of $\phi$ or $\psi$ fields, e.g., $\tr \phi^n$, $\tr \bar{\psi}\psi$ and $\tr \psi \bar{\psi}\psi$, possibly with derivatives included, such as the stress tensor and higher spin bilinears. Note that this list includes both bosonic and fermionic operators, depending on whether there are an even or odd number of $\psi$ fields in the trace. The $SO(N)$-$S_2$ theory and the $SU(N)$ adjoint theories contain an equal number of fermionic and bosonic single-trace operators, and could therefore have a supersymmetric holographic dual in the large $N$ limit. In the $SO(N)$ $S_2$-$A_2$ theory, however, many single-trace operators are missing. For instance, the only operators composed of a finite number of $\phi$ and $\psi$ fields (without any derivatives) are bosonic -- any operator containing an odd number of fermion fields, such as $\tr \psi^3 = \psi_{ij} \bar{\psi}_{jk} \psi_{ki}$ clearly vanishes, by the anti-symmetry of $\psi_{ij}$. (There are also fermionic operators, such as $\psi_{ij} \phi_{jk} \overleftrightarrow{\partial}_\mu \phi_{ki}$ present in the $S_2$-$A_2$ theory that vanish in the $S_2$ theory.) Therefore, a putative holographic dual description for fixed points of the large $N$ limit of the $SO(N)$ $S_2$-$A_2$ theory would necessarily be non-supersymmetric, although the masses of those fields common to the three holographic duals (i.e. anomalous dimensions of dual operators) would be identical to those in supersymmetric theories.

\subsection{\texorpdfstring{$\mathcal N=1$}{N = 1} supersymmetry in \texorpdfstring{$d=3$}{d=3}} 
\label{susy}

Both the $SU(N)$ adjoint model and the $SO(N)$-$S_2$ model contain an equal number of bosonic and fermionic degrees of freedom that transform in the same representation of the symmetry group. It is therefore possible to demand  $\mathcal N=1$ supersymmetry in $d=3$, which determines both of the quartic scalar couplings $g_1$ and $g_2$ in terms of the Yukawa coupling $y$. To see this, consider a theory with $N^2-1$ real scalar superfields $\Phi^a$ (using conventions in \cite{Gates:1983nr}) in $d=3$, transforming in the adjoint representation of $SU(N)$, and following relevant superpotential:
\begin{equation}
	f(\Phi) = \frac{1}{2}\,M_{ab}\, \Phi^a \Phi^b + \frac{1}{6}h_{abc} \Phi^a \Phi^b \Phi^c
\end{equation}
Rewriting this in components and integrating out the auxiliary field, we obtain:
\begin{align}
	V_{sup}(\psi,\phi) &= -\bigg(\frac{1}{8}\,h_{ab}{ }^{i}h_{cdi}\,\phi^{a}\phi^{b}\phi^{c}\phi^{d} + \frac{1}{2}\,h_{ab}{ }^{i}M_{ic}\,\phi^{a}\phi^{b}\phi^{c} + \frac{1}{2}\,M^{i}{ }_{a}M_{ib}\,\phi^{a}\phi^{b}\bigg)\nn\\
	&\hspace{2 cm}+ h_{aij}\,\phi^{a}\psi^{\mu i}\left(\frac{\delta_{\mu\nu}}{2}\right)\psi^{\nu j} + M_{ij}\psi^{\mu i}\left(\frac{\delta_{\mu \nu}}{2}\right) \psi^{\nu j}.
\end{align}
If we substitute 
\begin{equation}
    h_{abc}=y \frac{1}{2}\tr ({\Lambda }^a \{{\Lambda}^b, {\Lambda}^c\}),
\end{equation}
 we reproduce the action in the previous subsections, with the following relations between $g_1$, $g_2$ and $y$, 
\begin{equation}
	g_1 = \frac{y^2}{16}, ~ g_2= - \frac{y^2}{16 N} \label{susy-rel}.
\end{equation}
Similarly, for the $SO(N)$ $S_2$ model, using our conventions,  we find  $\mathcal N=1$ supersymmetry in $d=3$ implies,
\begin{equation}
    g_1= \frac{y^2}{8}, ~\quad g_{2} = - \frac{y^2}{8 N}.
\end{equation}
(For $N=3$, these results are modified. We find $g=\frac{y^2}{96}$ for the $SU(3)$ adjoint theory, and $g=\frac{y^2}{48}$ for the $SO(3)$ $S_2$ theory.)
We remark that we find that imposing $d=3$ supersymmetry is consistent with the three-loop $\beta$-functions we compute using four-dimensional theories with $SU(N)\times O(N_f)$ or $SO(N)\times O(N_f)$ symmetry groups, if and only if we set $N_f=1/2$.

Let us comment on the possibility of including classically marginal terms in the superpotential in $d=3$. These must be quartic in the superfield $\Phi_{ij}=\Phi^a {\mathcal T}^a_{ij}$. The only two possibilities invariant under $SU(N)$ are $\tr \Phi^4$ or $(\tr \Phi^2)^2$. However, such terms are forbidden by time-reversal invariance \cite{Gaiotto:2018yjh}, as we now review. The Grassman integration measure $d^2 \theta$ is odd under time-reversal.\footnote{Recall that $\bar{\chi}\chi$ is odd under time-reversal when $\chi$ is a Majorana fermion in $d=3$.} Therefore, if $d^2 \theta \tr \Phi^3$ is to be even under time-reversal, we require $\Phi$ to be a pseudo-scalar. We then also must demand the superpotential $W(\Phi)$ to be an odd function of $\Phi$. This means there can be no mass or quartic terms in the superpotential. Therefore, if we demand supersymmetry at the outset, flowing to a stable interacting fixed-point for the coupling $h$, requires \textit{zero} parameters to be tuned.

If we demand only time-reversal invariance, and not supersymmetry, marginal Yukawa interactions of the schematic form $\phi^2 \bar{\psi}\psi$ are forbidden, as is a fermion mass term. However, sextic scalar couplings and a scalar mass are allowed, though, we expect the sextic scalar couplings to become irrelevant at the interacting fixed points we find below.

\subsubsection{Scaling dimensions}
Below, we will compute the scaling dimensions of the following operators: $\psi^a$, $\phi^a$, $\tr \phi^2$, $\tr \psi^2$, $\tr \phi^3$, $\tr \phi \bar{\psi}\psi$, $\tr \phi^4$ and $(\tr \phi^2)^2$. Of these operators, $\tr \psi^2$ and $\tr \phi^3$ have the same classical scaling dimension in $d=4$ and therefore mix; we denote the two mixtures of these operators with well-defined scaling dimension as $\Delta_{(\bar{\psi}\psi,~\phi^3)_1}$ and $\Delta_{(\bar{\psi}\psi,~\phi^3)_2}$. The three operators, $\tr \phi \bar{\psi}\psi$, $\tr \phi^4$ and $(\tr \phi^2)^2$, that are classically-marginal in $d=4$ also mix, and we denote their mixtures as $\Delta_{(\phi \bar{\psi}\psi,~\phi^4)_i}$ for $i=1,~2,~3$. 

$\mathcal N=1$ supersymmetry in $d=3$ implies certain relations between scaling dimensions of these operators. These can be obtained by demanding that all components of certain composite operators formed from the superfield $\Phi^a$ have the same scaling dimension. In particular, consider the three composite superfield operators,
\begin{equation}\label{eq: Phi component expansion}
    \Phi^a (x,\theta) = \phi^a (x) + \theta^\alpha \psi^{a}{ }_\alpha - \theta^2 F^ a (x),
\end{equation}
\begin{equation}\label{eq: Phi^2 component expansion}
    \Phi^a \Phi^a = \phi^a \phi^a + 2 \theta^{ \mu} \psi^a{ }_\mu \phi^a + \theta^2 \left( h^{abc} \phi^a \phi^b \phi^c - \psi^2\right),
\end{equation}
and,
\begin{align}
    h_{abc} \Phi^a\Phi^b\Phi^c &= h_{abc} \left(\phi^a \phi^b \phi^c + 3 \theta^\alpha \phi^ a \phi^ b \psi^c{}_{\alpha}  + \theta^2 \left(\frac{3}{2} h^{aij}\phi^b \phi^c \phi^i \phi^j  -  3~ \phi^a \psi^{b \alpha} \psi^{c}{}_{\alpha} \right)\right).
\end{align}
Demanding all components of $\Phi^a$, \eqref{eq: Phi component expansion} have the same scaling dimension, yields, 
\begin{equation}
    \Delta_\psi = \Delta_\phi +1/2. \label{susy-1}
\end{equation}
Demanding all components of $\tr \Phi^2$, \eqref{eq: Phi^2 component expansion} have the same scaling dimension, yields
\begin{equation}
   \Delta_{(\bar{\psi}\psi,~\phi^3)_1} = 1 + \Delta_{\phi^2}. \label{susy-2} 
\end{equation}
Demanding all components of $\tr \Phi^3$ have the same scaling dimension, yields
\begin{equation}
     1 + \Delta_{(\bar{\psi}\psi,~\phi^3)_2} = \Delta_{(\phi\bar{\psi}\psi,~\phi^4)_1} \label{susy-3}.
\end{equation}
The two remaining operators which do not appear in the above relations, determine the scaling dimensions of $(\tr \Phi^2)^2$ and $\tr \Phi^4$ respectively, and are therefore independent. 

We will find that these relations hold up to three loops in the supersymmetric fixed points of the $SU(N)$ adjoint model and the $SO(N)$ $S_2$ model for all $N$. As required by large-$N$ equivalence, we also find that the $SO(N)$ $S_2$-$A_2$ model possesses a fixed point whose scaling dimensions also satisfy the above three relations, but only in the limit $N \to \infty$.

\subsubsection{Moduli space}

The scalar potential contains two terms -- the single trace quartic term and the double trace quartic term. Both terms are individually positive definite, but $g_1$ and/or $g_2$ may be negative for some fixed points. For what values of $g_1$, and $g_2$ is the classical scalar potential positive definite? $\phi$ is a traceless hermitian matrix, and can be diagonalized via a unitary transformation. We can therefore write $t_1=\tr \phi^4=\sum^N_{i=1} x_i^4$, and $t_2=(\tr \phi^2)^2=(\sum^N_{i=1}x_i^2)^2$, where $\sum_ i x_i=0$. One can show that 
\begin{equation}
     (t_1/t_2)_{min} \leq \frac{t_1}{t_2}\leq (t_1/t_2)_{max}
\end{equation}
 where
\begin{equation}
    1/N\leq \frac{t_1}{t_2}\leq \frac{N^2-3 N+3}{(N-1) N} 
\end{equation}
if $N$ is even, and
\begin{equation}
    \frac{N^2+3}{N^3-N} \leq \frac{t_1}{t_2} \leq \frac{N^2-3 N+3}{(N-1) N}
\end{equation}
if $N$ is odd.

This means that we require one of the following three conditions to hold: (i) both $g_1$ and $g_2$ to be positive, (ii) $g_2<0$ and $g_1>0$, with $(t_1/t_2)_{min}>-g_2/g_1$, or (iii) $g_1<0$ and $g_2>0$ with $(t_1/t_2)_{max}<-g_2/g_1$. Translating to $\lambda_1$ and $\lambda_2$, we have $\lambda_1>-\lambda_2$ in case (ii) and $\lambda_2>-N \lambda_1$ in case (iii).

Supersymmetry implies that $g_2 = - g_1/N$. For $N$ even, the discussion above implies that the potential possesses flat directions and that there exists a moduli space of classical vacuua. We discuss the moduli space in more detail, in Appendix \ref{moduli-space}. At a generic point in moduli space, other than $\phi = 0$, the $SU(N)$ symmetry is broken to $U(N/2)\times U(N/2)/U(1)$. The $SO(N)$ $S_2$ model also possesses a moduli space; at a generic point in the moduli space, the $SO(N)$ symmetry is broken to $O(N/2)\times O(N/2)/\mathbb Z_2$. 

\section{\texorpdfstring{$SU(N)$}{SU(N)} adjoint model}
\subsection{Fixed points for finite \texorpdfstring{$N>3$}{N>3}}\label{sec: finite N SU(N)}
We focus first on the adjoint model. We have computed the $\beta$-functions and anomalous dimensions at finite $N$ up to three loops. We computed up to two-loop\footnote{General results up to two loops are given in \cite{Machacek:1983tz, Machacek:1983fi, Machacek:1984zw, Jack:1990eb}. See also \cite{Luo:2002ti, Schienbein:2018fsw}. Note that \cite{Thomsen:2021ncy} also provides three-loop results for the Yukawa coupling which agreed with those computed from \cite{Jack:2023zjt}.} results via \cite{Thomsen:2021ncy}, and added three-loop corrections via   \cite{Jack:2023zjt}.  The $\beta-$functions at finite $N$, for $N$ greater than three, up to one loop\footnote{Finite $N$ three-loop results for each of the three theories studied in this paper are too lengthy to present here, but are available from the authors in electronic form upon request.} are,
\begin{align}
    \beta_{g_1} &= -\varepsilon~g_{1} + \frac{1}{\left(4 \pi \right)^2}\bigg[ \frac{1}{128 N} \bigg(512 g_{1}^2
   \left(-9+N^2\right)-\left(-32+N^2\right) y^4 N_f
   \nn\\
   & \quad +32 g_{1} \left(96 g_{2} N+ \left(-4+N^2\right) y^2 N_f \right)\bigg) \bigg]\\
    \beta_{g_2} &= -\varepsilon~g_{2} + \frac{1}{\left(4 \pi \right)^2}\bigg[ \frac{1}{128 N^2}  \bigg(768 g_{1}^2 \left(3+N^2\right)+256
   g_{2}^2 N^2 \left(7+N^2\right)\nn\\
   &\quad +512 g_{1}
   g_{2} N \left(-3+2 N^2\right)+32 g_{2} N
   \left(-4+N^2\right) y^2 N_f -\left(16+3
   N^2\right) y^4 N_f \bigg)\bigg]\\
   \beta_{y} &= -\frac{\varepsilon~y}{2} + \frac{1}{\left(4 \pi \right)^2}\bigg[ \frac{y^3 \left(N^2 (2+N_f )-4 (4+N_f )\right)}{16
   N } \bigg]
\end{align}
We present fixed point solutions of these $\beta$-functions for the case $N_f=1/2$, which allows for fixed points with emergent supersymmetry. Results for other values of $N_f>1/2$ are qualitatively similar. 

\begin{figure}[H]
    \centering
    \subfloat{{\includegraphics[width=4.6 cm]{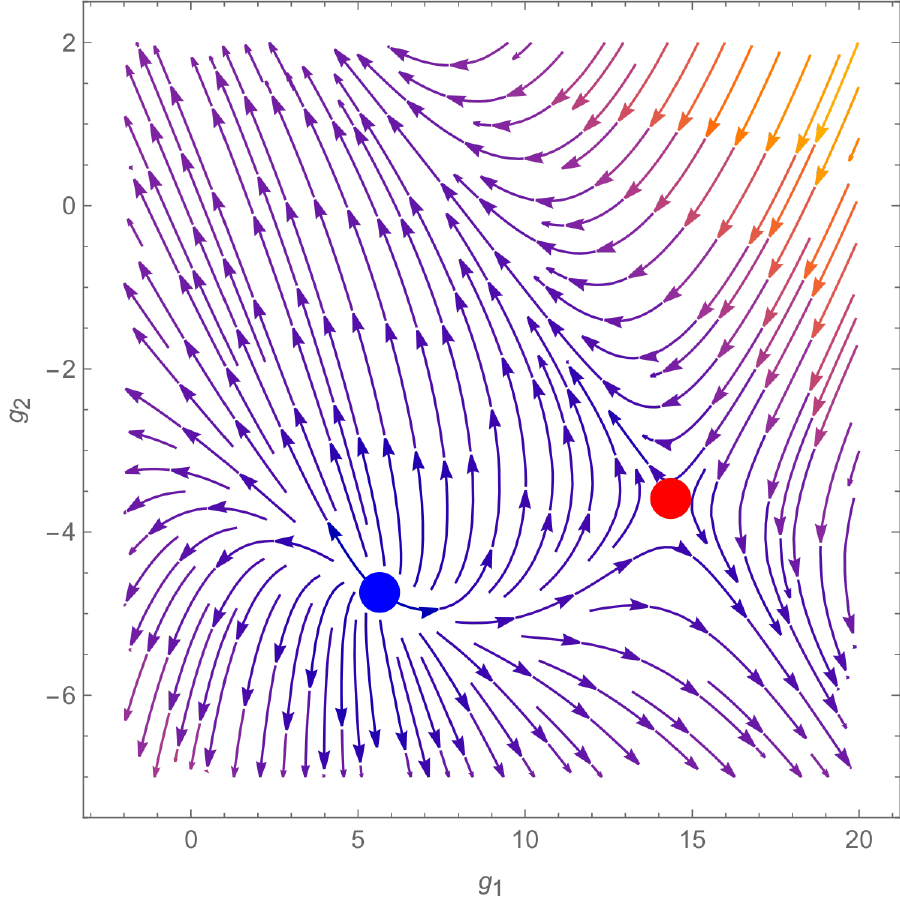} }}
    \quad
    \subfloat{{\includegraphics[width=4.6 cm]{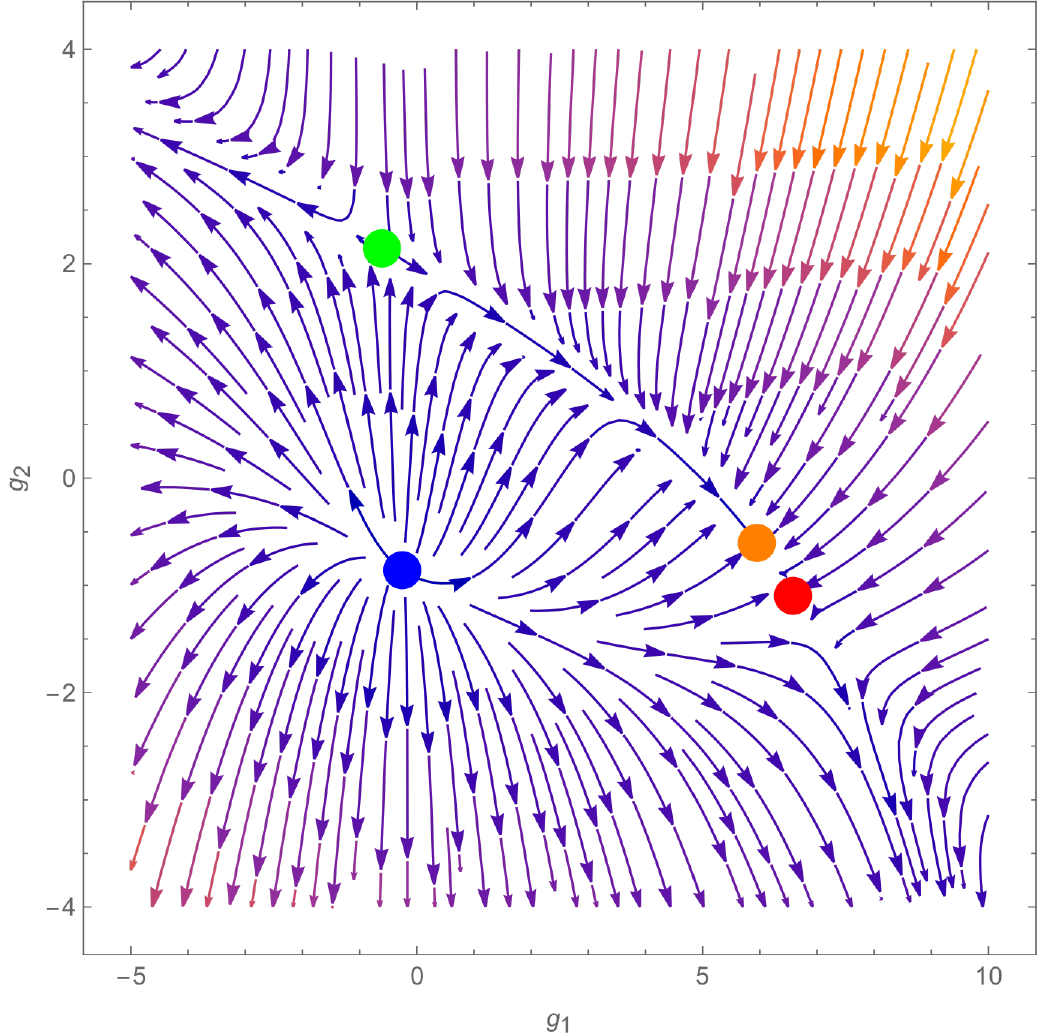} }}
    \quad
    \subfloat{{\includegraphics[width=4.6 cm]{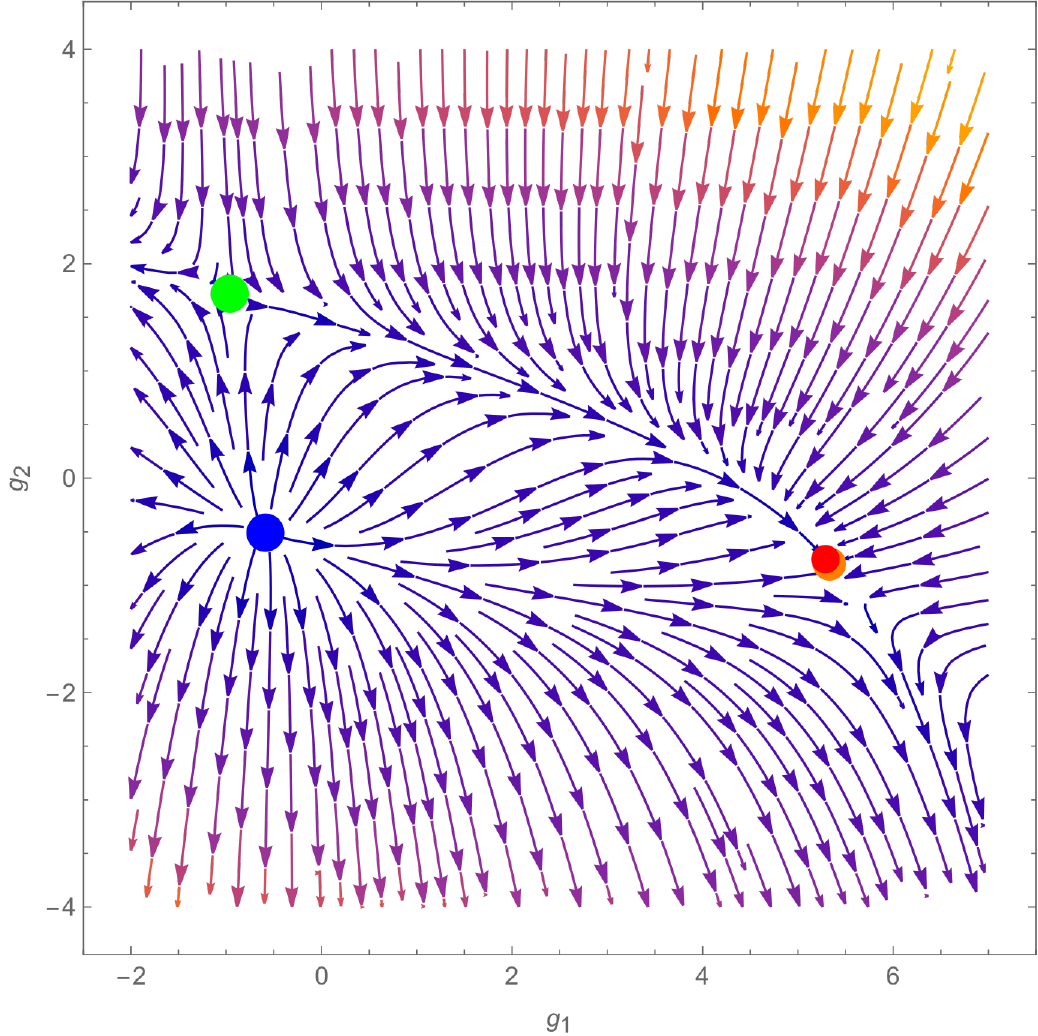} }}
    \caption{The figures show one-loop flows in the $g_1$-$g_2$ plane for $N=4$, $6$ and $7$ respectively, with, the Yukawa coupling tuned to criticality, $y=\frac{16 \pi  \sqrt{N}}{\sqrt{5 N^2-36}}\varepsilon$. The blue dot denotes the fixed point $[ns_+]$, and the red dot is the supersymmetric fixed point, $[susy]$, described in the text below. The orange and green dots are the fixed points $[ns_2]$ and $[ns_-]$ respectively, which are real only for $N>4$.}
    \label{figure: su(N) finite flow}
\end{figure}

Figure \ref{figure: su(N) finite flow} shows the fixed points and flows arising from the one-loop $\beta$-functions for various small values of $N>3$. For various values of $N$, there are up to four real fixed points with non-zero Yukawa couplings. One is the fixed point, denoted as $[susy]$, which can be presented in closed form, up to three loops for finite $N>3$, as follows:
\begin{align}
\frac{g_{1}^*}{\left(4 \pi\right)^2} &= \frac{ N \varepsilon }{5N^2-36}+\frac{3 N \left(3 N^4-48 N^2+208\right) \varepsilon ^2}{\left(5N^2-36\right)^3} + \frac{3 N  \varepsilon^3}{4 (-36 + 5 N^2)^5}\big(81 N^8\nn\\
&\quad  + 40 N^6 (-57 + 32 \zeta_{3}) - 32 N^4 (-738+ 1453 \zeta_{3})- 768 (-257 + 2178 \zeta_{3})\nn\\
  &\quad + 128 N^2 (-847
 + 3912 \zeta_{3})\big)\\
\frac{g_{2}^*}{\left(4 \pi\right)^2} &= -\frac{ \varepsilon }{5 N^2-36}-\frac{3 \left(3 N^4-48 N^2+208\right) \varepsilon ^2}{ \left(5 N^2-36\right)^3} - \frac{3 \varepsilon^3}{4 (-36 + 5 N^2)^5}\big(81 N^8\nn\\
  &\quad+ 40 N^6 (-57 + 32 \zeta_{3}) - 32 N^4 (-738 + 1453 \zeta_{3}) - 768 (-257 + 2178 \zeta_{3})\nn\\
  &\quad + 128 N^2 (-847 + 3912 \zeta_{3})\big)\\
\frac{\left(y^*\right)^2}{\left(4 \pi\right)^2} &= \frac{16 N \varepsilon }{5
		N^2-36}+\frac{48 N \left(3 N^4-48
		N^2+208\right) \varepsilon ^2}{\left(5
		N^2-36\right)^3} + \frac{12 N \varepsilon^3}{(-36 + 5 N^2)^5}\big(81 N^8\nn\\
  &\quad + 40 N^6 (-57 + 32 \zeta_{3}) - 32 N^4 (-738 + 1453 \zeta_{3}) - 768 (-257 + 2178 \zeta_{3})\nn\\
  &\quad + 128 N^2 (-847 + 3912 \zeta_{3})\big).
\end{align}
Much like the minimal $N=1$ GNY fixed point in \cite{Fei:2016sgs}, we find that up to three loops for all $N$, coupling constants at the fixed point $[susy]$ obey the conditions given in equation \eqref{susy-rel}, so we conjecture that the fixed point possesses $\mathcal N=1$ supersymmetry in $d=3$. The existence of a fixed point with supersymmetry in $d=3$ also provides a non-trivial check of our dimensional regularization procedure. 

The fixed point $[susy]$ is real for all $N$. The stability matrix at the fixed point $[susy]$ is given by,
\be
\left(
\begin{array}{ccc}
 1 & \frac{112 \sqrt{N} \pi }{\left(-36+5 N^2\right)^{3/2}} &
   -\frac{16 \left(3+N^2\right) \pi }{\sqrt{N} \left(-36+5
   N^2\right)^{3/2}} \\
 0 & \frac{68-5 N^2}{36-5 N^2} & \frac{4 \left(12+N^2\right)}{N
   \left(-36+5 N^2\right)} \\
 0 & \frac{24 N}{-36+5 N^2} & \frac{-12+N^2}{-36+5 N^2} \\
\end{array}
\right)\varepsilon + O(\varepsilon^2).
\ee
Two of the eigenvalues are positive for all values of $N>3$; the third eigenvalue is positive only after $N>2 \sqrt{\frac{1}{5} \left(28+\sqrt{889}\right)} \approx 6.7.$ As is clear from the discussion in section \ref{susy}, one linear combination of the coupling constants preserves supersymmetry while two linear combinations of the coupling constants break supersymmetry. For $N>6.7$, the supersymmetric fixed point is stable to all supersymmetric and non-supersymmetric classically-marginal deformations; while for smaller $N$, it is unstable in one of the non-supersymmetric directions. This implies, that, starting from a general non-supersymmetric time-reversal-invariant action, we only need to tune one parameter -- the mass of the scalar field -- to flow to the supersymmetric fixed point for any $N>6.7$.  We thus conclude that supersymmetry is emergent for $N$ greater than or equal to this ``critical'' value.

The anomalous dimensions at the $[susy]$ fixed point are given in eqns. \eqref{eq: gamma-phi-finiteN-susy-SU(N)}, \eqref{eq: gamma-phi^2-finiteN-susy-SU(N)}, \eqref{eq: gamma-phi^3-finiteN-susy-SU(N)} and  \eqref{eq: gamma-marg-finiteN-susy-SU(N)}. As one can see, the constraints on scaling dimensions following from supersymmetry given in equations \eqref{susy-1}, \eqref{susy-2} and \eqref{susy-3} are obeyed by scaling dimensions at the supersymmetric fixed point, up to three loops. 
\begin{align}\label{eq: gamma-phi-finiteN-susy-SU(N)}
&\gamma_{\phi} = \gamma_{\psi} = \frac{(-4 + N^2)  \varepsilon}{2 (-36 + 5 N^2)}
 + \frac{2 (-4 + N^2) (120 - 22 N^2 + N^4)  \varepsilon^2}{(-36 + 5 N^2)^3} \nn\\
& \quad  + \frac{(-4 + N^2)  \varepsilon^3}{2 (-36 + 5 N^2)^5}\bigg(211968- 84288 N^2 + 13328 N^4 - 972 N^6 + 27 N^8 \nn\\
& \quad - 1099008 \zeta_3 + 316800 N^2 \zeta_3 - 27552 N^4 \zeta_3 + 660 N^6 \zeta_3\bigg).
\end{align}
\begin{align}\label{eq: gamma-phi^2-finiteN-susy-SU(N)}
&\gamma_{\phi^2} = \gamma_{\left(\Bar{\psi}\psi,\phi^3\right)_1} + \varepsilon = \frac{3 (-4 + N^2) \varepsilon}{-36 + 5 N^2}
 + \frac{(-576 - 240 N^2 + 148 N^4 - 13 N^6) \varepsilon^2}{(-36 + 5 N^2)^3}\nonumber\\
& \quad + \frac{\varepsilon^3 }{2 (-36 + 5 N^2)^5}\bigg(138240 - 417024 N^2 + 177792 N^4 - 25696 N^6 + 1340 N^8 \nn\\
& \quad - 13 N^{10} + 3981312 \zeta_3 + 1105920 N^2 \zeta_3 - 1388160 N^4 \zeta_3 + 301632 N^6 \zeta_3 \nn\\
& \quad - 23880 N^8 \zeta_3 + 600 N^{10} \zeta_3\bigg).
\end{align}
\begin{align}\label{eq: gamma-phi^3-finiteN-susy-SU(N)}
& \gamma_{\left(\Bar{\psi}\psi,\phi^3\right)_2} = \gamma_{\left(g_1,g_2,y\right)_{2}} - \varepsilon = - \frac{3 (208 - 48 N^2 + 3 N^4) \varepsilon^2}{(-36 + 5 N^2)^2} + \frac{3 \varepsilon^3 }{2 (-36 + 5 N^2)^4} \bigg(321792 \nn\\
& \quad - 131200 N^2 + 19008 N^4 - 1176 N^6 + 27 N^8 + 1672704 \zeta_3 - 500736 N^2 \zeta_3 \nn\\
& \quad + 46496 N^4 \zeta_3 - 1280 N^6 \zeta_3\bigg).
\end{align}
\begin{align}\label{eq: gamma-marg-finiteN-susy-SU(N)}
\gamma_{\left(g_1,g_2,y\right)_{1,3}} &= \frac{40 - 3 N^2 \pm 2 \sqrt{484 - 4 N^2 + N^4}}{36 - 5 N^2}\varepsilon \pm \frac{\varepsilon^2}{\left(-36+5 N^2\right)^3 \sqrt{484-4 N^2+N^4}}\times\nn\\
& \quad \bigg[17 N^8+N^6 \left(538 \mp 43 \sqrt{484-4 N^2+N^4}\right) - 3568 N^2 \left(82 \pm 5 \sqrt{484-4
   N^2+N^4}\right) \nn\\
   & \quad +4 N^4 \left(1598 \pm 343 \sqrt{484-4 N^2+N^4}\right) + 192 \left(7678 \pm 377
   \sqrt{484-4 N^2+N^4}\right)\bigg] + \order{\varepsilon^3}
\end{align}

There are three more non-supersymmetric fixed points with non-zero Yukawa couplings, which we denote as $[ns_+]$, $[ns_2]$, and $[ns_-]$. For these three fixed points, the Yukawa coupling is given by
\begin{equation}
    y^* = \frac{16 \pi  \sqrt{N}}{\sqrt{5 N^2-36}}\varepsilon + O(\varepsilon^2),
\end{equation}
at one loop. The values of the quartic couplings, and anomalous dimensions, at one-loop for these fixed points, for small values of $N$ are given in Table \ref{table: finite N-non susy-fixed point}-\ref{table: finite N-[ns-]-gamma marginal}. Analytical expressions for arbitrary $N$ are possible to obtain but are tedious to present here. The fixed points $[ns_2]$ and $[ns_-]$ are real only for $N>5.4$. For $N=6$, $[ns_2]$ is stable in all three directions, and $[susy]$ can flow to $[ns_2]$. 

For $N\geq 4$, there are also two real fixed points when the Yukawa coupling is zero -- the Gaussian (free) fixed point, which has three unstable directions, and the critical  $O(N^2-1)$ vector model fixed point, denoted as $[vec]$, for which $g_1=0$ and
\begin{align}
    \frac{g_{2}^*}{\left(4 \pi\right)^2} &= \frac{  \varepsilon}{2(7+N^2)} + \frac{3 \left(11+3 N^2\right)  \varepsilon^2}{2\left(7+N^2\right)^3} + \frac{- \varepsilon^3}{16 \left(7+N^2\right)^5} \bigg(-2927-1441 N^2-209 N^4\nn\\
    &\quad +33 N^6+11424
   \zeta_3+4992 N^2 \zeta_3+480 N^4
   \zeta_3\bigg).
\end{align}

%
The vector model fixed point has two unstable directions and can flow to the supersymmetric fixed point with non-zero Yukawa coupling. There is no real fixed point with $y=0$ but $g_1\neq 0$ for $N\geq 4$. There are two fixed points, denoted as $[adj_{\pm}]$, that are complex for all values of $N > \frac{3 \sqrt{1 + \sqrt{3}}}{\sqrt{2}} + O(\varepsilon) \approx 3.5$,
\begin{align}
    \frac{g_{1}^*}{\left(4 \pi\right)^2} &= \frac{ 9 N-20 N^3+N^5 \mp 6 \sqrt{81 N^2+18 N^4-2 N^6} }{4 \left(567+315 N^2-35 N^4+N^6\right)} \varepsilon
    \\
    \frac{g_{2}^*}{\left(4 \pi\right)^2} &= \frac{ \left(108 N+21 N^3-N^5\mp (9 - N^2) \sqrt{81 N^2+18 N^4-2 N^6} \right) \varepsilon }{4N \left(567+315 N^2-35 N^4+N^6\right)}.
\end{align}

\begin{figure}[H]
    \centering
    \includegraphics[scale=0.8]{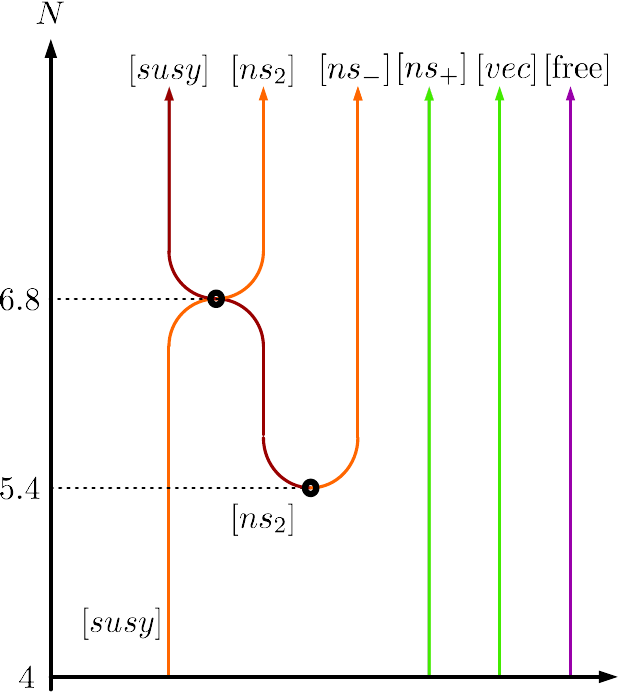}
    \caption{This figure illustrates the real fixed points at finite $N\geq 4$ of the adjoint GNY model at one-loop. Each line's color (red, orange, green, and violet) indicates the number of marginally unstable directions (0, 1, 2, and 3, respectively). Black dots denote mergers. }
    \label{mergers}
\end{figure}

Figure \ref{mergers} summarizes the stability and existence of the various fixed points of the adjoint GNY model as a function of $N\geq 4$.

\begin{table}[H]
	\centering
 \resizebox{\columnwidth}{!}{
	\begin{tabular}{cLLLLLL}
		\toprule
		\multicolumn{1}{c}{} & \multicolumn{2}{c}{\textbf{$[ns_+]$}} & \multicolumn{2}{c}{\textbf{$[ns_2]$}} & \multicolumn{2}{c}{\textbf{$[ns_-]$}}\\
		\cmidrule(rl){2-3} \cmidrule(rl){4-5} \cmidrule(rl){6-7}
		$N$ & {g_1 } & {g_2 } & {g_1 } & {g_2 } & {g_1 } & {g_2 }  \\
		\midrule
		 4 & 5.647 \varepsilon +1.147 \varepsilon ^2 & -4.739 \varepsilon +0.203 \varepsilon ^2 &
   (-15.315+4.996 i) \varepsilon  & (9.485\, -0.796
   i) \varepsilon  & (-15.315-4.996 i) \varepsilon
    & (9.485\, +0.796 i) \varepsilon  \\
 5 & 0.974 \varepsilon +0.375 \varepsilon ^2 & -1.758 \varepsilon +0.079 \varepsilon ^2 & (1.45\,
   +5.472 i) \varepsilon  & (2.494\, -2.135 i)
   \varepsilon  & (1.45\, -5.472 i) \varepsilon
    & (2.494\, +2.135 i) \varepsilon  \\
6 & -0.256 \varepsilon +0.353 \varepsilon ^2 & -0.858 \varepsilon +0.024 \varepsilon ^2 & 5.948
   \varepsilon +4.142 \varepsilon ^2 & -0.603 \varepsilon -1.609 \varepsilon ^2 & -0.61
   \varepsilon +1.027 \varepsilon ^2 & 2.143 \varepsilon +0.041 \varepsilon ^2 \\
 7 & -0.588 \varepsilon +0.327 \varepsilon ^2 & -0.508 \varepsilon +0.011 \varepsilon ^2 & 5.331
   \varepsilon +1.973 \varepsilon ^2 & -0.8 \varepsilon -0.427 \varepsilon ^2 & -0.962
   \varepsilon +0.62 \varepsilon ^2 & 1.721 \varepsilon +0.06 \varepsilon ^2 \\
		\bottomrule
	\end{tabular}}
 \caption{Quartic scalar couplings for the three non-supersymmetric fixed points of the $SU(N)$ adjoint theory for small values of $N$. \label{table: finite N-non susy-fixed point}}
\end{table}
\begin{table}[H]
	\centering
 \resizebox{1.1 \columnwidth}{!}{\hspace{-.1\columnwidth}
		\begin{tabular}{cLLLLLL}
			\toprule
			\multicolumn{1}{c}{} & \multicolumn{2}{c}{$[ns_+]$} & \multicolumn{2}{c}{$[ns_2]$} & \multicolumn{2}{c}{$[ns_-]$}\\
			\cmidrule(rl){2-3} \cmidrule(rl){4-5} \cmidrule(rl){6-7}
			$N$ & \gamma_{\phi}  & \gamma_{\psi} & \gamma_{\phi} & \gamma_{\psi} & \gamma_{\phi} & \gamma_{\psi}  \\
			\midrule
			4 & 0.136 \varepsilon -0.017 \varepsilon ^2 & 0.136 \varepsilon +0.001 \varepsilon ^2 & 0.136 \varepsilon +(0.04\, -0.007 i) \varepsilon ^2 & 0.136 \varepsilon +(0.039\, +0.003 i) \varepsilon ^2 & 0.136 \varepsilon +(0.04\, +0.007 i) \varepsilon ^2 & 0.136 \varepsilon +(0.039\, -0.003 i) \varepsilon ^2 \\
5 & 0.118 \varepsilon -0.01 \varepsilon ^2 & 0.118 \varepsilon +0.01 \varepsilon ^2 & 0.118 \varepsilon +(0.004\, -0. i) \varepsilon ^2 & 0.118 \varepsilon +(0.024\, -0.003 i) \varepsilon ^2 & 0.118 \varepsilon +(0.004\, +0. i) \varepsilon ^2 & 0.118 \varepsilon +(0.024\, +0.003 i) \varepsilon ^2 \\
6  & 0.111 \varepsilon -0.009 \varepsilon ^2 & 0.111 \varepsilon +0.011 \varepsilon ^2 & 0.111 \varepsilon +0.012 \varepsilon ^2 & 0.111 \varepsilon +0.015 \varepsilon ^2 & 0.111 \varepsilon +0.003 \varepsilon ^2 & 0.111 \varepsilon +0.019 \varepsilon ^2 \\
7 & 0.108 \varepsilon -0.009 \varepsilon ^2 & 0.108 \varepsilon +0.011 \varepsilon ^2 & 0.108 \varepsilon +0.014 \varepsilon ^2 & 0.108 \varepsilon +0.014 \varepsilon ^2 & 0.108 \varepsilon -0. \varepsilon ^2 & 0.108 \varepsilon +0.016 \varepsilon ^2 \\
			\bottomrule
		\end{tabular}
	}
	\caption{Anomalous dimensions $\phi$ and $\psi$ at the three non-supersymmetric fixed points of the $SU(N)$ adjoint theory for small values of $N$. \label{table: finite N-non susy-gamma phi}}
\end{table}
\begin{table}[H]
\centering
 \resizebox{.6\columnwidth}{!}{
\begin{tabular}{cLLLLLLLLL}
    \toprule
    $N$ & \gamma_{\phi^2} & \gamma_{\left(\bar\psi\psi,\phi^3\right)_{1}} & \gamma_{\left(\bar\psi\psi,\phi^3\right)_{2}} \\
    \midrule
    4 & -0.229 \varepsilon +0.194 \varepsilon ^2 & -0.915 \varepsilon -0.064 \varepsilon ^2 & -0.182
   \varepsilon -0.175 \varepsilon ^2 \\
 5 & -0.227 \varepsilon +0.166 \varepsilon ^2 & -1.139 \varepsilon -0.074 \varepsilon ^2 & -0.292
   \varepsilon -0.037 \varepsilon ^2 \\
6 &  -0.217 \varepsilon +0.163 \varepsilon ^2 & -1.28 \varepsilon -0.026 \varepsilon ^2 & -0.333
   \varepsilon -0.005 \varepsilon ^2 \\
7 & -0.208 \varepsilon +0.161 \varepsilon ^2 & -1.353 \varepsilon -0.001 \varepsilon ^2 & -0.354
   \varepsilon +0.006 \varepsilon ^2 \\
\bottomrule
		\end{tabular}}
	\caption{Anomalous dimensions of classically-relevant operators at the  $[ns_+]$ fixed point of the $SU(N)$ adjoint theory for small $N$. \label{table: finite N-[ns+]-gamma relevant}}
\end{table}
\begin{table}[H]
\centering
\resizebox{\columnwidth}{!}{
\begin{tabular}{cLLLLLLLLL}
    \toprule
    $N$ & \gamma_{\phi^2} & \gamma_{\left(\bar\psi\psi,\phi^3\right)_{1}} & \gamma_{\left(\bar\psi\psi,\phi^3\right)_{2}} \\
    \midrule
    4 & (0.909\, +0.287 i) \varepsilon -(0.431\, -0.207 i) \varepsilon ^2 & (-1.825+0.414 i)
   \varepsilon -(0.606\, +0.203 i) \varepsilon ^2 & -0.182 \varepsilon +(0.043\, +0.02 i) \varepsilon ^2 \\
 5 & (1.23\, -0.051 i) \varepsilon -(0.033\, +0.375 i) \varepsilon ^2 & (-0.747+0.628 i)
   \varepsilon +(0.984\, -0.066 i) \varepsilon ^2 & -0.292 \varepsilon +(0.036\, +0.012 i) \varepsilon ^2 \\
 6 & 0.806 \varepsilon -0.446 \varepsilon ^2 & -0.333 \varepsilon -0.17 \varepsilon ^2 & -0.083
   \varepsilon +0.027 \varepsilon ^2 \\
 7 & 0.625 \varepsilon -0.209 \varepsilon ^2 & -0.354 \varepsilon -0.129 \varepsilon ^2 & 0.006
   \varepsilon -0.336 \varepsilon ^2 \\
\bottomrule
		\end{tabular}
	}
	\caption{Anomalous dimensions of classically-relevant operators at the $[ns_2]$ fixed points of the $SU(N)$ adjoint theory for small $N$. \label{table: finite N-[ns2]-gamma relevant}}
\end{table}
\begin{table}[H]
\centering
\resizebox{\columnwidth}{!}{
\begin{tabular}{cLLLLLLLLL}
    \toprule
    $N$ & \gamma_{\phi^2} & \gamma_{\left(\bar\psi\psi,\phi^3\right)_{1}} & \gamma_{\left(\bar\psi\psi,\phi^3\right)_{2}} \\
    \midrule
    4 &  (0.909\, -0.287 i) \varepsilon -(0.431\, +0.207 i) \varepsilon ^2 & (-1.825-0.414 i)
   \varepsilon -(0.606\, -0.203 i) \varepsilon ^2 & -0.182 \varepsilon +(0.043\, -0.02 i) \varepsilon ^2 \\
 5 & (1.23\, +0.051 i) \varepsilon -(0.033\, -0.375 i) \varepsilon ^2 & (-0.747-0.628 i)
   \varepsilon +(0.984\, +0.066 i) \varepsilon ^2 & -0.292 \varepsilon +(0.036\, -0.012 i) \varepsilon ^2 \\
 6 & 1.137 \varepsilon -0.062 \varepsilon ^2 & -1.12 \varepsilon +0.053 \varepsilon ^2 & -0.333
   \varepsilon +0.048 \varepsilon ^2 \\
 7 & 1.14 \varepsilon -0.076 \varepsilon ^2 & -1.271 \varepsilon +0.056 \varepsilon ^2 & -0.354
   \varepsilon +0.039 \varepsilon ^2 \\
\bottomrule
		\end{tabular}
	}
	\caption{Anomalous dimensions of classically-relevant operators at the $[ns_-]$ fixed points of the $SU(N)$ adjoint theory for small $N$. \label{table: finite N-[ns-]-gamma relevant}}
\end{table}

\begin{table}[H]
\centering
 \resizebox{.5\columnwidth}{!}{
\begin{tabular}{cLLLLLLLLL}
\toprule
$N$ & \gamma_{\left(g_1,g_2,y\right)_{1}} & \gamma_{\left(g_1,g_2,y\right)_{2}} & \gamma_{\left(g_1,g_2,y\right)_{3}} \\
			\midrule
4 &  -1.932 \varepsilon -0.069 \varepsilon ^2 & -0.92 \varepsilon +0.156 \varepsilon ^2 & \varepsilon -0.279 \varepsilon ^2 \\
5 & -1.673 \varepsilon +0.191 \varepsilon ^2 & -0.685 \varepsilon -0.035 \varepsilon ^2 & \varepsilon -0.321 \varepsilon ^2 \\
6 & -1.577 \varepsilon +0.249 \varepsilon ^2 & -0.732 \varepsilon -0.011 \varepsilon ^2 & \varepsilon -0.322 \varepsilon ^2 \\
7 & -1.521 \varepsilon +0.27 \varepsilon ^2 & -0.789 \varepsilon +0.022 \varepsilon ^2 & 
   \varepsilon -0.318 \varepsilon ^2 \\
\bottomrule
\end{tabular}
}
 \caption{ Anomalous dimensions of classically-marginal operators at the $[ns_+]$ fixed points of the $SU(N)$ adjoint theory at small $N$. \label{table: finite N-[ns+]-gamma marginal}}
\end{table}

\begin{table}[H]
\centering
\resizebox{\columnwidth}{!}{
\begin{tabular}{cLLLLLLLLL}
\toprule
$N$ & \gamma_{\left(g_1,g_2,y\right)_{1}} & \gamma_{\left(g_1,g_2,y\right)_{2}} & \gamma_{\left(g_1,g_2,y\right)_{3}} \\
			\midrule
4 &  (0.119\, +0.547 i) \varepsilon +(0.208\, +0.091 i) \varepsilon ^2 & \varepsilon
   -(0.559\, +0.022 i) \varepsilon ^2 & (1.769\, +0.229 i) \varepsilon -(1.596\, +0.163
   i) \varepsilon ^2 \\
5 & (0.1\, +0.23 i) \varepsilon -(0.151\, +0.327 i) \varepsilon ^2 & \varepsilon -(0.441\,
   -0.027 i) \varepsilon ^2 & (1.824\, -0.095 i) \varepsilon -(0.718\, +0.61 i) \varepsilon
   ^2 \\
6 & 0.073 \varepsilon -0.106 \varepsilon ^2 & \varepsilon -0.355 \varepsilon ^2 & 1.155
   \varepsilon -1.1 \varepsilon ^2 \\
7 & -0.016 \varepsilon -0.067 \varepsilon ^2 & 0.997 \varepsilon -0.653 \varepsilon ^2 & 1.
   \varepsilon -0.346 \varepsilon ^2 \\
\bottomrule
\end{tabular}
	}
 \caption{ Anomalous dimensions of classically-marginal operators at the $[ns_2]$ fixed points of the $SU(N)$ adjoint theory at small $N$. \label{table: finite N-[ns2]-gamma marginal}}
\end{table}

\begin{table}[H]
\centering
\resizebox{\columnwidth}{!}{
\begin{tabular}{cLLLLLLLLL}
\toprule
$N$ & \gamma_{\left(g_1,g_2,y\right)_{1}} & \gamma_{\left(g_1,g_2,y\right)_{2}} & \gamma_{\left(g_1,g_2,y\right)_{3}} \\
			\midrule
4 &  (0.119\, -0.547 i) \varepsilon +(0.208\, -0.091 i) \varepsilon ^2 & \varepsilon
   -(0.559\, -0.022 i) \varepsilon ^2 & (1.769\, -0.229 i) \varepsilon -(1.596\, -0.163
   i) \varepsilon ^2 \\
5 & (0.1\, -0.23 i) \varepsilon -(0.151\, -0.327 i) \varepsilon ^2 & \varepsilon -(0.441\,
   +0.027 i) \varepsilon ^2 & (1.824\, +0.095 i) \varepsilon -(0.718\, -0.61 i) \varepsilon
   ^2 \\
6 & -0.342 \varepsilon -0.027 \varepsilon ^2 & \varepsilon -0.39 \varepsilon ^2 & 1.573
   \varepsilon -0.631 \varepsilon ^2 \\
7 & -0.55 \varepsilon +0.047 \varepsilon ^2 & \varepsilon -0.364 \varepsilon ^2 & 1.506
   \varepsilon -0.544 \varepsilon ^2 \\
\bottomrule
\end{tabular}
	}
 \caption{ Anomalous dimensions of classically-marginal operators at the $[ns_-]$ fixed points of the $SU(N)$ adjoint theory at small $N$. \label{table: finite N-[ns-]-gamma marginal}}
\end{table}
Let us comment on whether the scalar potential is positive definite or not at these fixed points, at finite $N$:
\begin{itemize}
    \item For $[susy]$, when $N$ is odd, the classical potential is positive definite. When $N$ is even, it possesses flat directions.
    \item For $[ns_+]$, $g_1>0$ for $N\in (3,5.65)$ and $g_1<0$ for $N>5.65$; while $g_2<0$ for $N>3$. The classical scalar potential is positive definite for $N=5$ only. 
    \item For $[ns_2]$, $g_1>0$ and $g_2<0$ for $N>5.4$. The classical scalar potential is positive definite only for $N = 6$ and $7$.
    \item For $[ns_-]$, $g_1 >0$  for $N\in (5.4,5.65)$ and $g_1<0$ for $N>5.65$; while $g_2>0$ for $N>5.4$. The classical scalar potential is positive-definite only for $N=6$, $7$ , $8$, $9$, $10$, $11$, and $12$.
\end{itemize}
At sufficiently large-$N$, we find that the $[susy]$ fixed point is the only fixed point with non-zero Yukawa coupling that possesses a non-negative classical scalar potential. 

\subsection{Fixed points at large \texorpdfstring{$N$}{N}}
\begin{figure}[H]
    \centering
    \subfloat{{\includegraphics[width = 7cm]{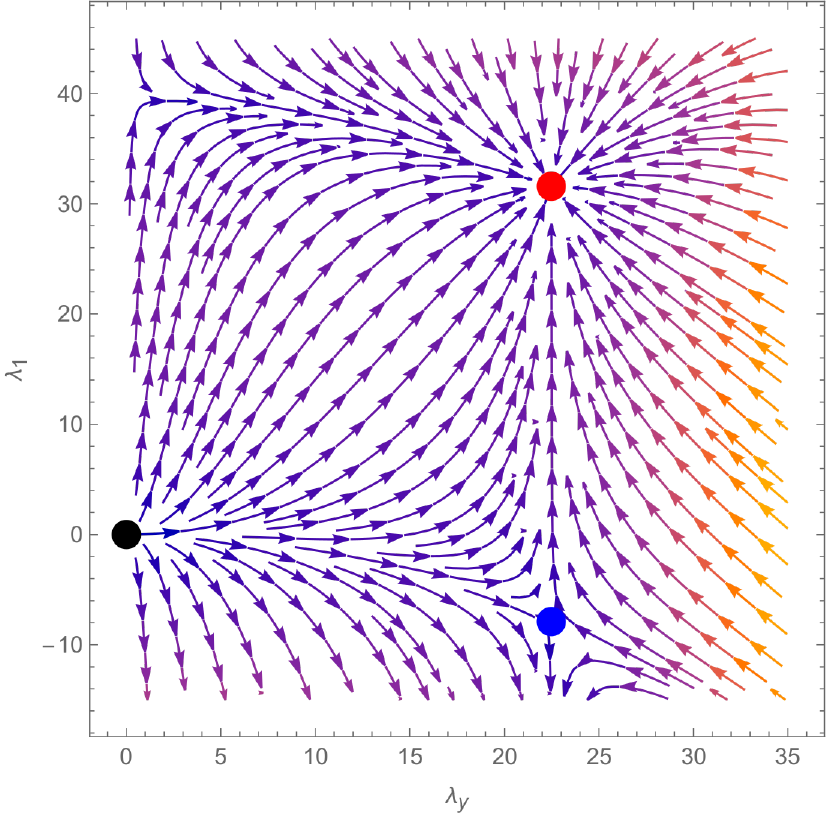}} }
    \quad
    \subfloat{{\includegraphics[width=7
 cm]{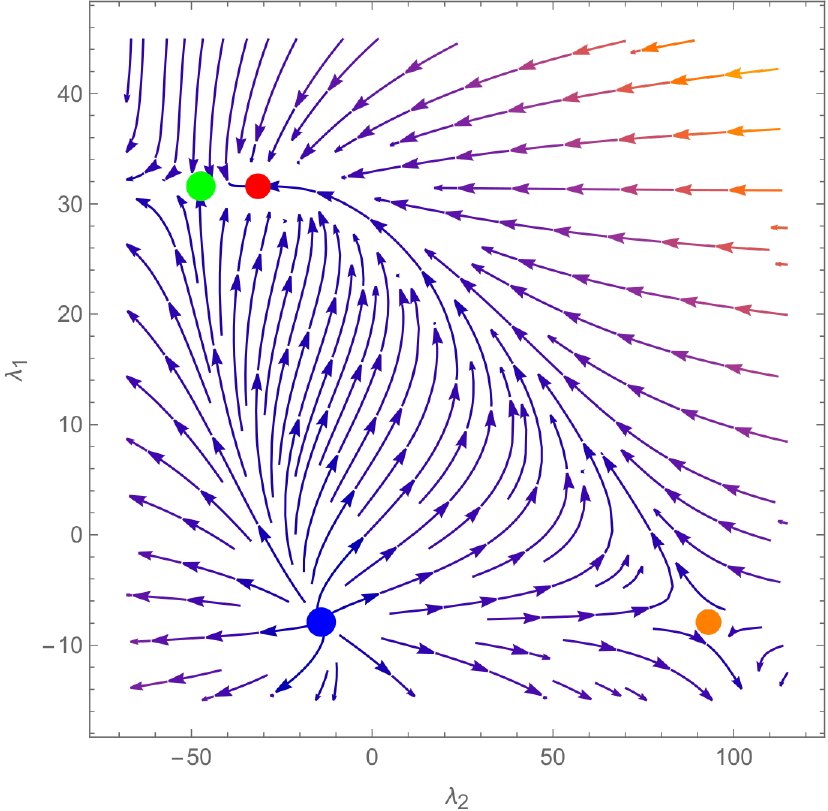} }}
    \caption{Flows arising from the one-loop $\beta$-functions at large-$N$. Flows are plotted in the $\lambda_1$-$\lambda_y$ (left) and $\lambda_1-\lambda_2$ plane at  (right). For the plot on the left, note that $\beta_{\lambda_1}$ and $\beta_{\lambda_y}$ are independent of $\lambda_2$. Each dot in the left plot denotes two fixed points, which differ only in the value of $\lambda_2$. The red dot denotes $[susy]$ and $[ns_2]$, the blue dot corresponds to $[ns_\pm]$ and the black dot corresponds to $[free]$ and $[vec]$. In the right plot, $\lambda_y^2$ is tuned to criticality, via \eqref{eq: lambda-y fixed point}.   \label{flowsLargeN}}
\end{figure}

Let us now discuss the limiting behavior of the fixed points at large-$N$. In the large-$N$ limit, the one-loop $\beta$-functions for the 't Hooft couplings for the theory are
\begin{align}\label{eq: marginal Beta function SU(N)}
    \beta_{\lambda_{y}} & = -\frac{\varepsilon}{2}\,\lambda_{y} + \frac{1}{\left(4 \pi\right)^2}\bigg[\frac{1}{16 } \lambda_y^3 (2 + N_f )\bigg],\\
    \beta_{\lambda_{1}} &= -\varepsilon\,\lambda_{1} + \frac{1}{\left(4 \pi\right)^2}\bigg[4 \lambda_1^2
 + \frac{1}{4} \lambda_1 \lambda_y^2 N_f
 -  \frac{1}{128} \lambda_y^4 N_f\bigg] ,\\
    \beta_{\lambda_{2}} & = -\varepsilon\,\lambda_{2} + \frac{1}{\left(4 \pi\right)^2}\bigg[6 \lambda_1^2
 + 8 \lambda_1 \lambda_2
 + 2 \lambda_2^2
 + \frac{1}{4} \lambda_2 \lambda_y^2 N_f
 -  \frac{3}{128} \lambda_y^4 N_f\bigg] .
\end{align}
Below, we set $N_f=1/2$. We plot the fixed points and flows for the theory arising from the large-$N$ one-loop $\beta$-functions in Figure \ref{flowsLargeN}.

The large-$N$ limit of $[ns_+]$ and $[ns_-]$, have the same Yukawa and $\lambda_1$ quartic coupling, up to three loops,
\begin{equation}
    \frac{\left(\lambda_{y}^*\right)^2}{\left(4 \pi\right)^2} = \frac{16 \varepsilon }{5} + \frac{119  \varepsilon^2}{125} + \frac{11501 \varepsilon ^3}{25000}, \quad
    \frac{\lambda_{1}^*}{\left(4 \pi\right)^2} = -\frac{ \varepsilon }{20}+\frac{17 \varepsilon ^2}{1000} + \frac{7411 \varepsilon ^3}{400000}.
\end{equation}
The $\lambda_2$ quartic coupling at the $[ns_{\pm}]$ fixed points is given by,
\begin{align}
\frac{\lambda_{2}^*}{\left(4 \pi\right)^2} &= \frac{1}{20}\left(5 \mp \sqrt{46}\right)  \varepsilon -\frac{3 \left(230 \mp 43 \sqrt{46} \right) \varepsilon ^2}{92000} \nn\\
    &\quad -\frac{ \left(552000 \left(23 \mp \sqrt{46}\right)
   \zeta_{3} \mp 2728111 \sqrt{46}-18255790\right)
   \varepsilon ^3}{846400000},
\end{align}

The Yukawa and $\lambda_1$ quartic coupling for the large-$N$ limit of $[ns_2]$ and $[susy]$ are given by,
\begin{equation}\label{eq: lambda-y fixed point}
         \frac{\left(\lambda_{y}^*\right)^2}{\left(4 \pi\right)^2} = \frac{16  }{5}\varepsilon + \frac{144}{125}\varepsilon^2 + \frac{972}{3125}\varepsilon^{3}, \quad \frac{\lambda_{1}^*}{\left(4 \pi\right)^2} = \frac{ \varepsilon }{5} + \frac{9  \varepsilon ^2}{125} + \frac{243  \varepsilon ^3}{12500}
\end{equation}
The $\lambda_2$ quartic coupling at the $[ns_2]$ fixed point is,
\begin{align}
    \frac{\lambda_{2}^*}{\left(4 \pi\right)^2} &= -\frac{3 \varepsilon }{10} + \frac{3 \varepsilon ^2}{250} - \frac{ (2250 \zeta_{3}-331) \varepsilon
   ^3}{25000},
\end{align}
and the $\lambda_2$ quartic coupling at the $[susy]$ fixed point is,
\begin{eqnarray}\label{eq: susy fixed point}
    \frac{\lambda_{2}^*}{\left(4 \pi\right)^2} &=& -\frac{ 1}{5}\varepsilon  -\frac{9}{125}\varepsilon^2 - \frac{243 }{12500}\varepsilon^{3}.
\end{eqnarray}
The stability matrix $(\partial_I \beta_J)$, evaluated at the supersymmetric fixed point, where $I=\lambda_y,~\lambda_1,~\lambda_2$, is
\be
\partial_{I}\beta_{J} = 
\begin{bmatrix}
\varepsilon-\frac{9 \varepsilon
   ^2}{25}+\frac{51 \varepsilon ^3}{1250} & \quad\frac{6 \pi  \varepsilon ^{5/2}}{25
   \sqrt{5}}  & \quad-\frac{16 \pi  \varepsilon ^{3/2}}{5 \sqrt{5}}+\frac{394 \pi  \varepsilon ^{5/2}}{125
   \sqrt{5}}\\
    \frac{3 \varepsilon ^{5/2}}{50 \sqrt{5} \pi } & \quad\varepsilon-\frac{12 \varepsilon ^2}{25}+\frac{94
   \varepsilon ^3}{625} & \quad\frac{4 \varepsilon
   }{5}-\frac{214 \varepsilon ^2}{125}+\frac{3 (1700 \zeta_{3}+581) \varepsilon^3}{3125} \\
   \quad0 & \quad0 & \quad\frac{\varepsilon }{5}-\frac{26 \varepsilon ^2}{125}+\frac{(600 \zeta_{3}-13) \varepsilon
   ^3}{3125}
\end{bmatrix}
\ee
For small $\varepsilon$, all three eigenvalues of this matrix are positive, so the fixed point is attractive to all deformations, including the two supersymmetry-breaking directions in the large-$N$ limit. These eigenvalues imply the following scaling dimensions of the classically marginal operators:
\begin{align}
    \Delta_{\left(\lambda_1,\lambda_y\right)_1} &= 4-\frac{12 \varepsilon ^2}{25}+\frac{79 \varepsilon ^3}{625},\\
    \Delta_{\left(\lambda_1,\lambda_y\right)_2} &= 1 + \Delta_{(\bar\psi\psi,\phi^3)_{2}}\\
    \Delta_{\lambda_2}  &= 4-\frac{4 \varepsilon }{5}-\frac{26 \varepsilon ^2}{125}+\frac{\varepsilon ^3 (-13+600 \zeta_{3})}{3125}.
\end{align}  
The scaling dimensions for $\phi$ and $\psi$ at the supersymmetric fixed point \eqref{eq: susy fixed point} are
\begin{equation}
    \Delta_{\phi} = \frac{1}{2} (2
 -  \varepsilon)
 + \frac{1}{10} \varepsilon
 + \frac{2}{125} \varepsilon^2
 + \frac{27}{6250} \varepsilon^3,\quad\Delta_{\psi} = \Delta_{\phi}+\frac{1}{2}.
\end{equation}
The scaling dimensions of the classically relevant operators are given by
\begin{align}
\Delta_{\phi^2} &= 2 - \frac{2 \varepsilon }{5}-\frac{13 \varepsilon ^2}{125}+\frac{\varepsilon ^3 (-13+600 \zeta_{3})}{6250} ,\\
\Delta_{(\bar\psi\psi,\phi^3)_{1}} &= 1+\Delta_{\phi^2}\\
\Delta_{(\bar\psi\psi,\phi^3)_{2}} &= 3 -\frac{9 \varepsilon ^2}{25} + \frac{81 \varepsilon   ^3}{1250}.
\end{align}
These scaling dimensions obey the constraints given in \eqref{susy-1}, \eqref{susy-2}, and \eqref{susy-3}. We list the scaling dimensions of the other non-supersymmetric fixed points in Appendix \ref{App: non-supersymmetric fp large N}. 

Let us briefly discuss the fixed points with zero Yukawa coupling. The large $N$ limit of the fixed point $[vec]$ is given by $\lambda_y=\lambda_1=0$, and $\lambda_2=8\pi^2\varepsilon$, and is the large $N$ critical $O(N)$ vector model. The large $N$ limit of the two complex fixed points, $[adj{}_\pm]$ is given by the following coupling constants:
\begin{align}
    \frac{\lambda_{1}^*}{\left(4 \pi\right)^2} &= \frac{\varepsilon}{4} + \frac{3 \varepsilon^2}{32} + \frac{5 \varepsilon^3}{256}\\
    \frac{\lambda_{2}^*}{\left(4 \pi\right)^2} &= \frac{-\varepsilon(1 + i\sqrt{2})}{4} + \frac{\varepsilon^2}{32}(-1+2\sqrt{2}i) + \frac{\varepsilon^3}{512}\big[-74 + 29\sqrt{2}i(1-2\zeta_{3})\big],
\end{align}
and its complex conjugate.

It is interesting that, unlike the case of purely bosonic theories, there is no obstruction to finding real fixed points at large-$N$ with adjoint matter and non-vanishing single-trace couplings, with and without supersymmetry, in GNY models. However, the quartic scalar potential is not positive definite at the non-supersymmetric fixed points $[ns_\pm]$ and $[ns_2]$, so these fixed points are not physical in $d=4$; however, this does not mean that the fixed point in $d=3$ or $d=2$ is necessarily unphysical, as discussed in \cite{Fei:2016sgs} and \cite{Nakayama:2022svf}.

In the discussion above we set $N_f=1/2$, to allow for emergent $\mathcal N=1$ supersymmetry in 
$d=3$. There is no difficulty in solving the $\beta$-functions for arbitrary values of $N_f$, such as $N_f=1$, corresponding to two flavors of Majorana fermions in $d=3$.  Results are summarized in Appendix \ref{App: stable-large-N-nonsusy}. We find that the fixed point $[susy]_{N_f}$ generalized to arbitrary $N_f>1/2$ (which is not supersymmetric if $N_f \neq 1/2$) remains stable and gives rise to a classical scalar potential that is positive definite for $N_f>1/2$. Therefore, the above construction gives rise to an interacting non-supersymmetric large-$N$ fixed point in $d=3$ dominated by planar diagrams.

\subsection{Fixed points at \texorpdfstring{$N=3$}{N=3}}
As noted earlier, the case of $N=3$ needs to be treated separately, as the two scalar quartic couplings are not independent. We denote $g=g_2$, and set $g_1=0$.

\begin{figure}[H]
    \centering
    \includegraphics[width = 8 cm]{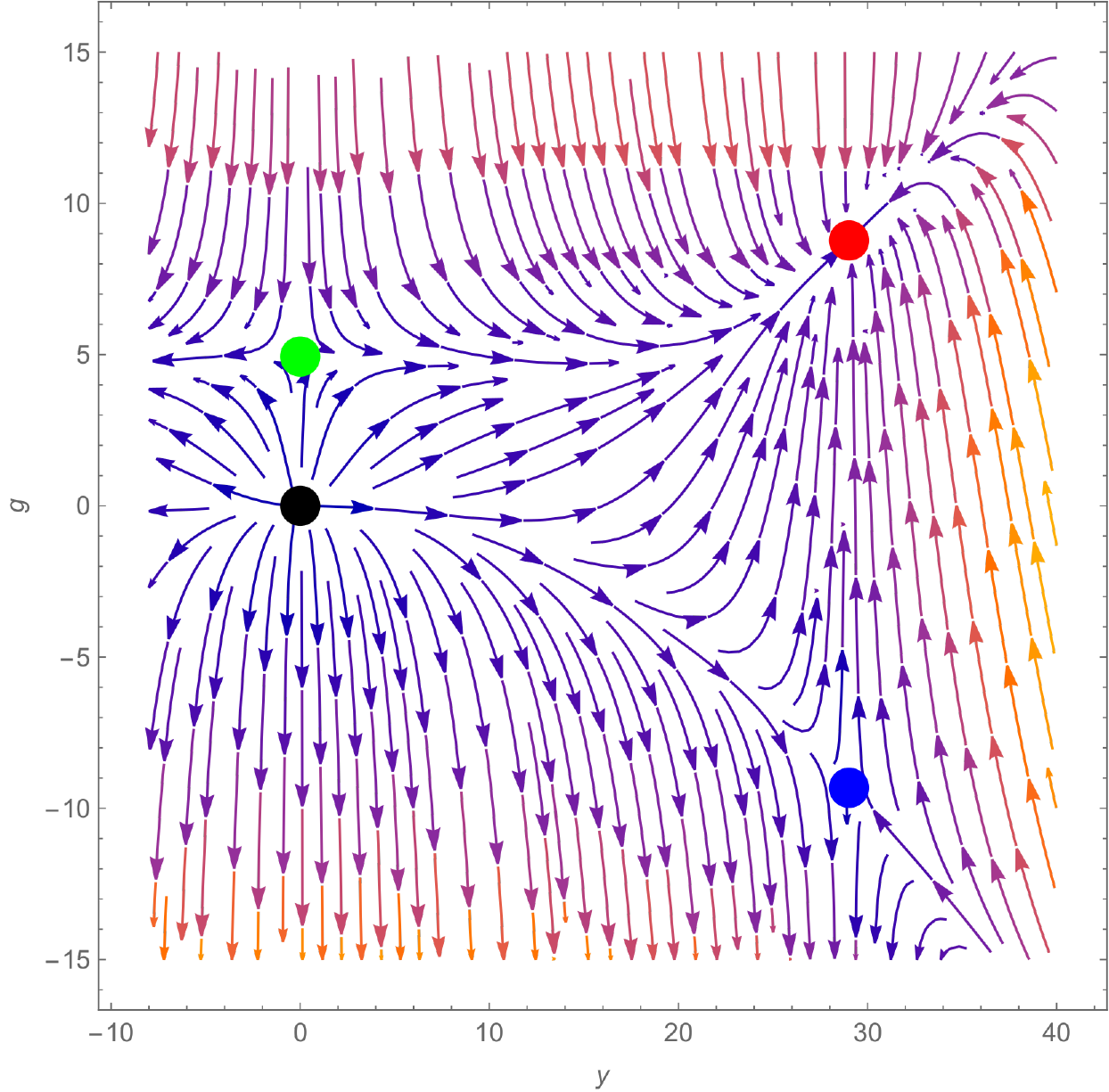}
    \caption{The figure shows flows in the $g$-$y$ plane for $N=3$. The blue dot denotes the fixed point $[ns]$ and the red dot is the supersymmetric fixed point. The black dot is the free fixed point and the green dot is the critical $O(8)$ vector model fixed point.  \label{SU3plot}}
\end{figure}

The one-loop $\beta$-functions for $g$ and $y$ are
\be
\beta_g = -\varepsilon g + \frac{1}{\left(4 \pi\right)^2}\bigg[\frac{73728 g^2+N_f\left(960 g\,y^2- 17 y^4\right)}{2304}\bigg],\quad \beta_y = -\frac{\varepsilon y}{2} + \frac{1}{\left(4 \pi\right)^2}\bigg[\frac{(5 N_f +2) y^3}{48}\bigg].
\ee
We now set $N_f=1/2$. The model contains 8 real pseudo-scalars and 8 Majorana fermions. The plot of fixed points and flows in the $g-y$ plane are shown in Figure \ref{SU3plot}. Overall the structure is very similar to that of the $N=1$ Gross-Neveu Model consisting of a  single Majorana fermion and a single real scalar studied in \cite{Fei:2016sgs}.
There are two fixed points with non-zero Yukawa coupling. We find one stable, $\mathcal N=1$ supersymmetric fixed point,
\begin{align}
\frac{g^{*}}{\left(4 \pi\right)^2} &= \frac{ \varepsilon }{18}+\frac{19
   \varepsilon ^2}{486} + \frac{ \left(288  \zeta_{3} + 1283 \right) \varepsilon ^3}{52488}
\\
\frac{(y^{*})^2}{\left(4 \pi\right)^2} &= \frac{16 \varepsilon }{3} + \frac{304 \varepsilon ^2}{81} + \frac{4 \left(288 \zeta_{3}+1283 \right) \varepsilon ^3}{2187}.
\end{align}
There is also one unstable non-supersymmetric fixed point, which we denote as $[ns]$,
\begin{align}
    \frac{g^{*}}{\left(4 \pi\right)^2} &= -\frac{17 \varepsilon
   }{288} + \frac{34183  \varepsilon ^2}{497664}+ \frac{\left(34875432 \zeta_{3} - 34011811 \right) \varepsilon
   ^3}{859963392}
   \\
   \frac{(y^{*})^2}{\left(4 \pi\right)^2} &= \frac{16 \varepsilon }{3} - \frac{1175 \varepsilon ^2}{1296} + \frac{\left(589824  \zeta_{3} + 10969801 \right) \varepsilon
   ^3}{1119744}.
\end{align}
The scaling dimensions of $\phi$ and $\psi$ at the supersymmetric fixed point are
\be
\Delta_{\phi} = \frac{2-\varepsilon }{2}+\frac{10
   \varepsilon ^2}{243}+\frac{5 \varepsilon }{18}+\frac{5 (180 \zeta_{3}+167) \varepsilon ^3}{13122},\quad \Delta_\psi = \frac12+\Delta_\phi.
\ee
While at $[ns]$ the scaling dimensions of $\phi$ and $\psi$ are
\begin{align}
\Delta_{\phi} &= \frac{2-\varepsilon }{2}+\frac{5 \varepsilon }{18}-\frac{6145 \varepsilon
   ^2}{31104}+\frac{5 \varepsilon ^3 (353693+46080 \zeta_{3})}{3359232}\\
\Delta_{\psi} &= \frac{3-\varepsilon }{2}+\frac{5 \varepsilon }{18}-\frac{25075 \varepsilon
   ^2}{124416}+\frac{5 \varepsilon ^3 (12857293+1474560 \zeta_{3})}{107495424}.
\end{align}
The scaling dimensions of composite operators at the supersymmetric fixed point are
\begin{align}
\Delta_{\phi^2} &= 2+\frac{2 \varepsilon }{3}-\frac{25 \varepsilon ^2}{81}+\frac{5 \varepsilon ^3   (577+1008 \zeta_{3})}{4374}\\
\Delta_{\left(\bar\psi\psi,\phi^3\right)_1} &= 3-\frac{19 \varepsilon ^2}{27}+\frac{\varepsilon ^3 (161-288 \zeta_3)}{1458}\\
\Delta_{\left(\bar\psi\psi,\phi^3\right)_2} &= 1 + \Delta_{\phi^2},
\end{align}
and at the $[ns]$ fixed point,
\begin{align}
\Delta_{\phi^2} &=2-\frac{13 \varepsilon }{8}+\frac{14255 \varepsilon ^2}{13824}+\frac{5   \varepsilon ^3 (-10602847+5003496 \zeta_{3})}{23887872}\\
\Delta_{\left(\bar\psi\psi,\phi^3\right)_1} &= 3-\frac{11 \varepsilon }{8}-\frac{8563 \varepsilon   ^2}{677376}+\frac{\varepsilon ^3 (-2597982067613-2469733926408   \zeta_{3})}{2810384252928}\\
\Delta_{\left(\bar\psi\psi,\phi^3\right)_2} &= 3+\frac{2 \varepsilon }{3}-\frac{2173315 \varepsilon ^2}{1016064}+\frac{5   \varepsilon ^3 (2303839181197+485745426432 \zeta_{3})}{2107788189696}.
\end{align}
 At the $[susy]$ fixed point, the scaling dimension for the classically marginal couplings are,
\begin{align}
    \Delta_{\left(\phi^4,\phi\bar\psi\psi\right)_1} &= 1 + \Delta_{\left(\bar\psi\psi,\phi^3\right)_1}\\
    \Delta_{\left(\phi^4,\phi\bar\psi\psi\right)_2} &= 4+\frac{8 \varepsilon }{3}-\frac{268 \varepsilon ^2}{81}+\frac{\varepsilon ^3 (15847+12024 \zeta_{3})}{2187},
\end{align}
and at $[ns]$,
\begin{align}
    \Delta_{\left(\phi^4,\phi\bar\psi\psi\right)_1} &=4-\frac{14 \varepsilon }{3}+\frac{2723 \varepsilon ^2}{10368}+\frac{\varepsilon ^3 (-1733727883-865785600 \zeta_{3})}{143327232}\\
    \Delta_{\left(\phi^4,\phi\bar\psi\psi\right)_2} &= 4+\frac{1175 \varepsilon ^2}{6912}+\frac{\varepsilon ^3 (-166800959-9437184 \zeta_{3})}{47775744}.
\end{align}
In addition, there are two fixed points with zero Yukawa coupling. These are the Gaussian (free) fixed point, which is unstable in all directions in $d=4-\varepsilon$, and the $O(8)$ vector model, denoted by $([wf])$, which has one unstable direction and can flow to the supersymmetric fixed point, as is clear from Figure \ref{SU3plot}.  The quartic coupling at the $[wf]$ fixed point is
\be
\frac{g^*}{\left(4 \pi\right)^2} = \frac{ \varepsilon }{32} + \frac{57 \varepsilon ^2}{4096} - \frac{\varepsilon ^3 (-137+1488 \zeta_3)}{262144}.
\ee
For $N=3$, there is no distinction between $[adj]$ and $[vec]$, as there is only one quartic scalar coupling.
\section{\texorpdfstring{$SO(N)$}{SO(N)} symmetric traceless rank-2 model} 

We now turn to the model in which fields are real, symmetric traceless matrices.
\subsection{Fixed points for finite \texorpdfstring{$N>3$}{N>3}}
\begin{figure}[H]
	\centering
	\subfloat{{\includegraphics[width=4.7 cm]{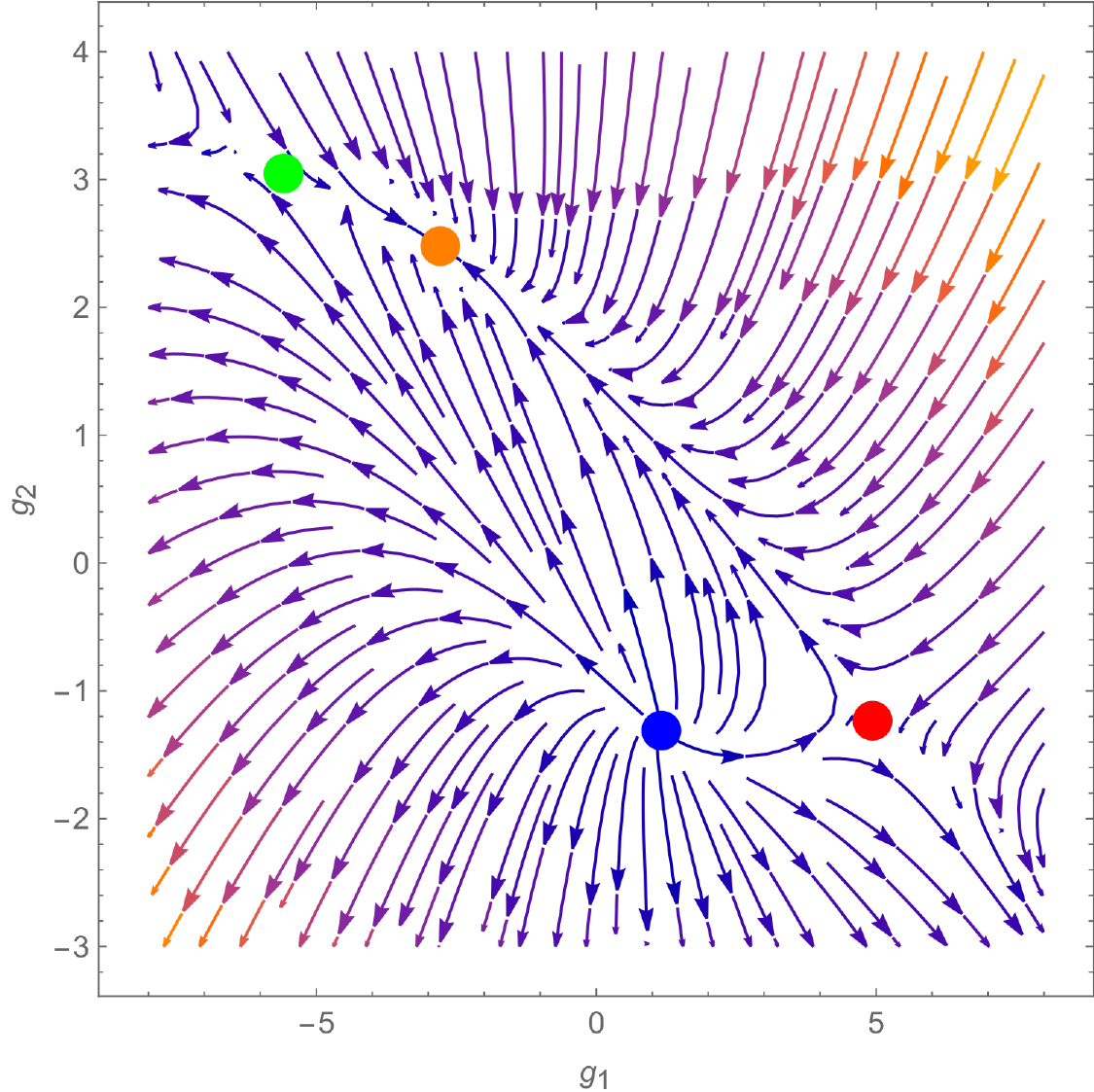} }}
	\quad
	\subfloat{{\includegraphics[width=4.7 cm]{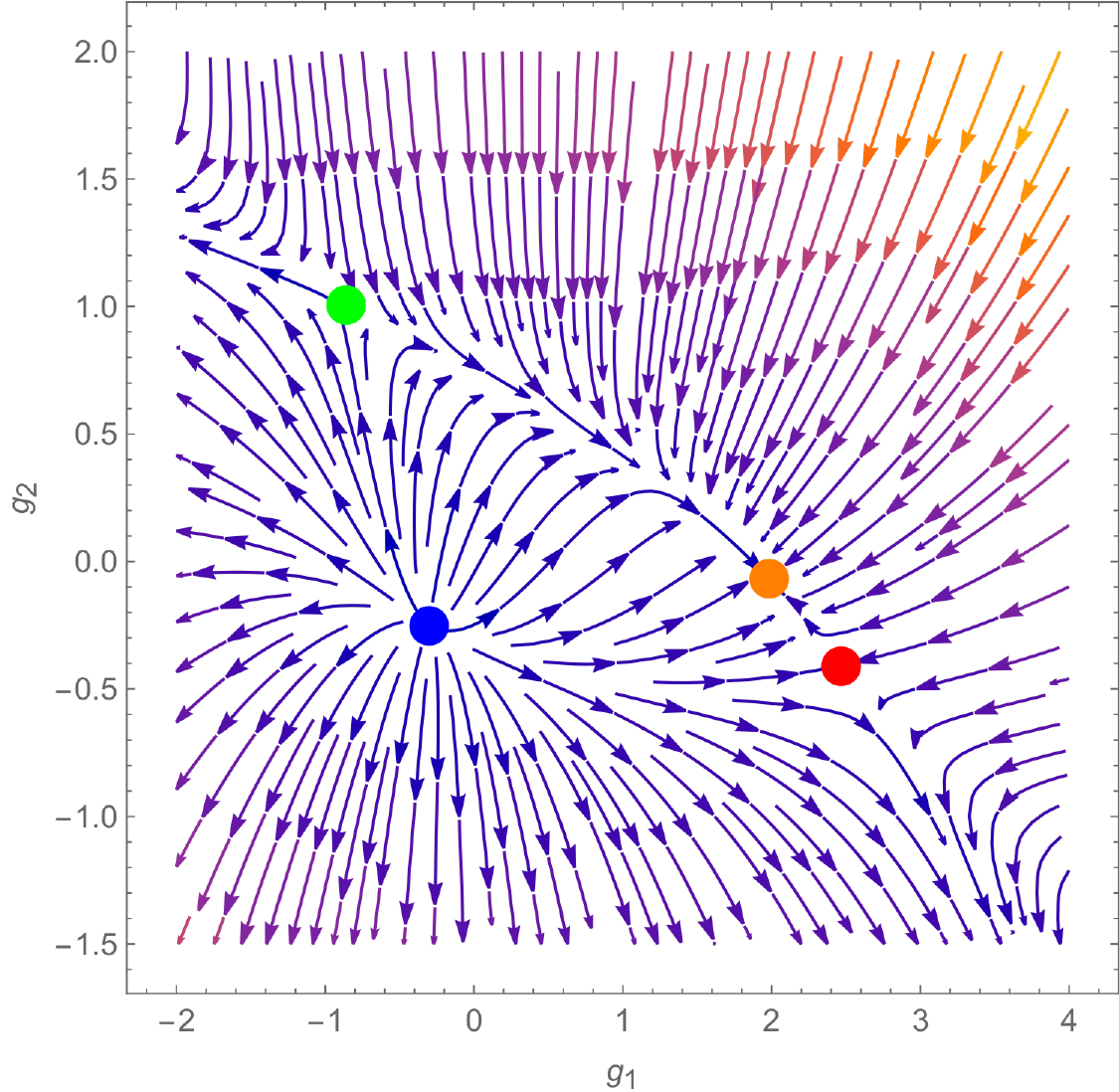} }}
	\quad
	\subfloat{{\includegraphics[width=4.7 cm]{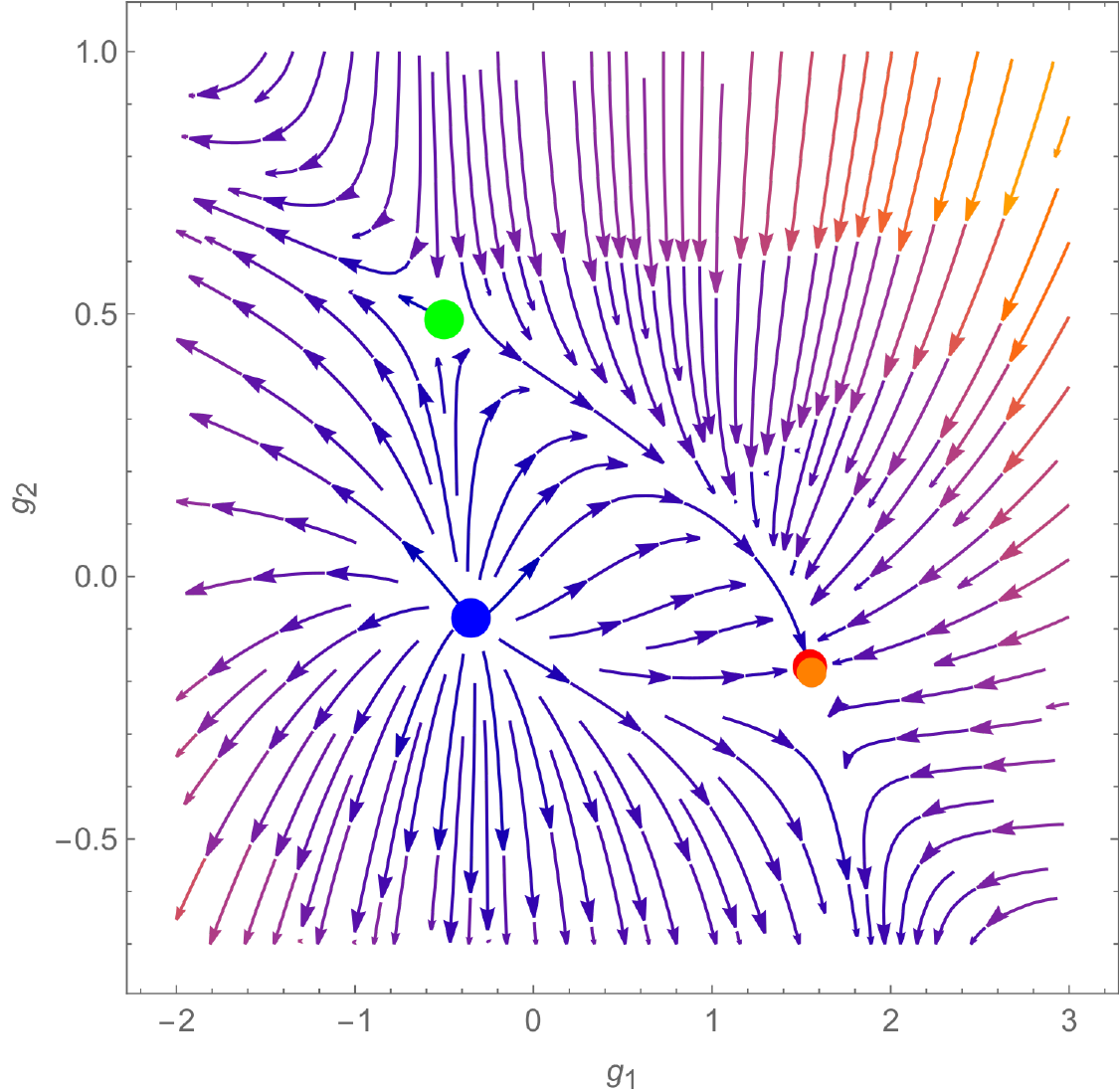} }}
	\caption{The figures show flows in the $g_1$-$g_2$ plane for $N=4$, $6$ and $9$ respectively. Here, the Yukawa coupling is tuned to criticality, $y=\frac{8 \pi }{\sqrt{14-\frac{72}{N}+5 N}} \varepsilon$. The blue dot denotes the fixed point $[ns_+]$ and the red dot is the supersymmetric fixed point. The orange and green dots are the fixed points $[ns_2]$ and $[ns_-]$ respectively.}
	\label{figure: SO(N) finite flow}
\end{figure}

 The $\beta$-functions at finite $N$, up to one loop, are,
\begin{align}
	\beta_{g_1} &= -\varepsilon  g_1 + \frac{1}{\left(4 \pi\right)^2}\bigg[\frac{1}{16 N }\bigg(64 g_1^2
	\left(-36+9 N+2 N^2\right)-\left(-64+8
	N+N^2\right) y^4 N_{f} \nn\\
	&\quad +16 g_1 \left(96
	g_2 N+\left(-8+2 N+N^2\right) y^2 N_{f}
	\right)\bigg)\bigg] , \\
	\beta_{g_2} &= -\varepsilon  g_2 + \frac{1}{\left(4 \pi\right)^2}\bigg[\frac{1}{16 N^2 }\bigg(192 g_1^2
	\left(6+N^2\right)+64 g_2^2 N^2
	\left(14+N+N^2\right)\bigg]\nn \\
	&\quad +128 g_1
	g_2 N \left(-6+3 N+2 N^2\right)+16
	g_2 N \left(-8+2 N+N^2\right) y^2 N_{f}\nn
	\\
	&\quad-\left(32+3 N^2\right) y^4 N_{f} \bigg) , \\
	\beta_{y} &= \frac{-\varepsilon}{2} y + \frac{1}{\left(4 \pi\right)^2}\bigg[\frac{y^3 \left(N^2
		(2+N_{f} )+2 N (3+N_{f} )-8 (4+ N_f )\right)}{4 N }\bigg].
\end{align}
We set $N_f=1/2$ below.
Plots of the flows and fixed points for small values of $N>3$ are shown in Figure \ref{figure: SO(N) finite flow}.

Again, there exist up to four real fixed points with non-zero Yukawa coupling, which we refer to as $[susy]$, $[ns_+]$, $[ns_2]$ and $[ns_-]$. The supersymmetric fixed point, $[susy]$ has the following form for $N>3$:
\begin{align}
    \frac{\left(y^*\right)^2}{\left(4 \pi\right)^2} &= \frac{4 N \varepsilon }{-72+14 N+5 N^2} + \frac{12 N
    \left(832-304 N-68 N^2+18 N^3+3 N^4\right)
    \varepsilon ^2}{\left(-72+14 N+5 N^2\right)^3} \nn\\
    & \quad + \frac{3 N  \varepsilon^3}{\left(-72+14 N+ 5N^2\right)^5} \bigg(81 N^8+N^7 (894-690 \zeta_3)-8 N^6 (111+494 \zeta_3) \nn \\
    & \quad -512 N^2 (296+583 \zeta_3)-12288 (-257+2178 \zeta_3)+8 N^5 (-3827+5825 \zeta_3)\nn \\
    & \quad +16 N^4 (-157+8304\zeta_3)+1024 N (-2389+16980 \zeta_3)-64 N^3 (-6573+23386 \zeta_3)\bigg) ,\\
    \frac{g^*_1}{\left(4 \pi\right)^2} &= \frac{ N \varepsilon }{2\left(-72+14 N+5 N^2\right)} + \frac{3 N \left(832-304 N-68 N^2+18 N^3+3 N^4\right) \varepsilon ^2}{2 \left(-72+14 N+5 N^2\right)^3}\nn
	\\
    & \quad + \frac{3 N  \varepsilon^3}{ 8\left(-72+14 N+ 5N^2\right)^5} \bigg(81 N^8+N^7 (894-690 \zeta_3)-8 N^6 (111+494 \zeta_3) \nn \\
    & \quad -512 N^2 (296+583 \zeta_3)-12288 (-257+2178 \zeta_3)+8 N^5 (-3827+5825 \zeta_3)\nn \\
    & \quad +16 N^4 (-157+8304
   \zeta_3)+1024 N (-2389+16980
   \zeta_3)-64 N^3 (-6573+23386
   \zeta_3)\bigg)
    \\
    \frac{g^*_2}{\left(4 \pi\right)^2} &= -\frac{  \varepsilon }{2 \left(-72+14 N+5 N^2\right)} - \frac{3 \left(832-304 N-68 N^2+18 N^3+3 N^4\right)
    \varepsilon ^2}{2 \left(-72+14 N+5 N^2\right)^3} \nn\\
    & \quad - \frac{3~\varepsilon^3}{8 \left(-72+14 N+ 5N^2\right)^5} \bigg(81 N^8+N^7 (894-690 \zeta_3)-8 N^6 (111+494 \zeta_3) \nn \\
    & \quad -512 N^2 (296+583 \zeta_3)-12288 (-257+2178 \zeta_3)+8 N^5 (-3827+5825 \zeta_3)\nn \\
    & \quad +16 N^4 (-157+8304 \zeta_3)+1024 N (-2389+16980 \zeta_3)-64 N^3 (-6573+23386  \zeta_3)\bigg).
\end{align}
The anomalous dimension of $\phi$ and $\psi$ at $[susy]$ fixed point is given in equation \eqref{eq: gamma-phi-susy-SO(N)}; the anomalous dimensions of the relevant couplings are given in the equation \eqref{eq: gamma-phi^2-susy-SO(N)}; and the anomalous dimensions of the marginal couplings are given in the equation \eqref{eq: gamma-marg-susy-SO(N)}.
\begin{align}\label{eq: gamma-phi-susy-SO(N)}
&\gamma_{\phi} = \gamma_{\psi} = \frac{(-8 + 2 N + N^2) \varepsilon}{2 (-72 + 14 N + 5 N^2)}
 + \frac{(-8 + 2 N + N^2) (960 - 328 N - 60 N^2 + 15 N^3 + 2 N^4) \varepsilon^2}{(-72 + 14 N + 5 N^2)^3}\nonumber\\
& \quad + \frac{(-8 + 2 N + N^2) \varepsilon^3 }{2 (-72 + 14 N + 5 N^2)^5} \bigg(3391488 - 2370048 N - 50176 N^2 + 286112 N^3 - 7536 N^4 \nn\\
& \quad - 15340 N^5 - 220 N^6 + 351 N^7 + 27 N^8 - 17584128 \zeta_3 + 11326464 N \zeta_3 - 247296 N^2 \zeta_3 \nn\\
& \quad - 932352 N^3 \zeta_3 + 88704 N^4 \zeta_3 + 27312 N^5 \zeta_3 - 2634 N^6 \zeta_3 - 405 N^7 \zeta_3\bigg).
\end{align}
\begin{align}\label{eq: gamma-phi^2-susy-SO(N)}
&\gamma_{\phi^2} = \gamma_{\left(\Bar{\psi}\psi,\phi^3\right)_1} + \varepsilon = \frac{3 (-8 + 2 N + N^2) \varepsilon}{-72 + 14 N + 5 N^2} + \frac{\varepsilon^2}{(-72 + 14 N + 5 N^2)^3}\bigg[-4608 + 768 N \nn\\
& \quad - 640 N^2 + 776 N^3 + 168 N^4 - 76 N^5 - 13 N^6\bigg]
 + \frac{(-8 + 2 N + N^2) \varepsilon^3}{2 (-72 + 14 N + 5 N^2)^5}\times  \nn \\
& \quad \bigg[-110592 (5 + 144 \zeta_3) + 9216 N (113 + 408 \zeta_3) + N^8 (-13 + 600 \zeta_3)  \nn \\
& \quad  - 1536 N^2 (-224 + 2343 \zeta_3) - 4 N^6 (-62 + 4305 \zeta_3) + N^7 (-374 + 6780 \zeta_3) \nn\\
& \quad  + 16 N^4 (-47 + 17292 \zeta_3) - 8 N^5 (-2899
 + 33711 \zeta_3)
 + 64 N^3 (-6061
 + 42936 \zeta_3)\bigg]
\end{align}
\begin{align}
&\gamma_{\left(\Bar{\psi}\psi,\phi^3\right)_2} = \gamma_{\left(g_1,g_2,y\right)_1} -  \varepsilon = - \frac{3 (832 - 304 N - 68 N^2 + 18 N^3 + 3 N^4) \varepsilon^2}{(-72 + 14 N + 5 N^2)^2}\nonumber\\
& + \quad \frac{3 \varepsilon^3 }{2 (-72 + 14 N + 5 N^2)^4} \bigg[5148672 - 3623936 N - 97280 N^2 + 434880 N^3 - 13424 N^4 \nn\\
& \quad  - 20648 N^5 - 120 N^6 + 402 N^7 + 27 N^8 + 26763264 \zeta_3 - 17387520 N \zeta_3 + 298496 N^2 \zeta_3 \nn\\
& \quad + 1496704 N^3 \zeta_3 - 132864 N^4 \zeta_3 - 46600 N^5 \zeta_3 + 3952 N^6 \zeta_3 + 690 N^7 \zeta_3\bigg].
\end{align}
\begin{equation}\label{eq: gamma-marg-susy-SO(N)}
    \gamma_{\left(g_1,g_2,y\right)_{1,3}} = \frac{3 N (4+N)-2 \left(40 \pm \sqrt{1936+N (-536+N (41+N
   (14+N)))}\right)}{-72+N (14+5 N)}\varepsilon + \order{\varepsilon^2}
\end{equation}

\begin{figure}[H]
    \centering
  \includegraphics[scale=0.8]{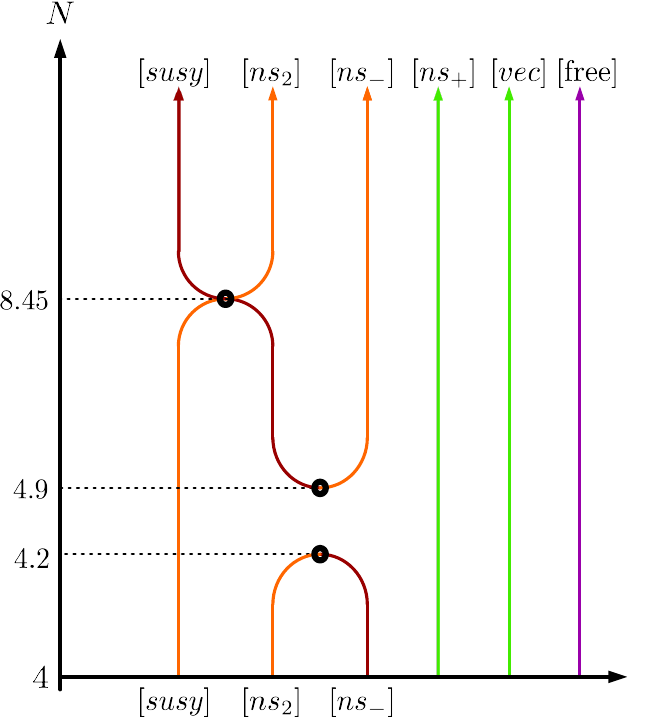}
    \caption{This figure illustrates the real fixed points at finite $N\geq 4$ of the $SO(N)$-$S_2$ GNY model at one-loop. Each line's color (red, orange, green, and violet) indicates the number of marginally unstable directions (0, 1, 2, and 3, respectively). Black dots denote mergers.}
    \label{SO(N)-mergers}
\end{figure}

At the non-supersymmetric fixed points, the Yukawa coupling is given by, 
\begin{equation}
    y = \frac{8 \pi }{\sqrt{14-\frac{72}{N}+5 N}} \varepsilon,
\end{equation}
at one loop. The values of the scalar couplings, and anomalous dimensions, up to two loops for non-supersymmetric fixed points, for small values of $N$ is given in Table \ref{table: SO(N)-finiteN-non susy-fixed point}-\ref{table: SO(N)-finite N-non susy - gamma - marginal}.
For $N\geq 4$, there are also two real fixed points when the Yukawa coupling is zero -- the free fixed point and the critical  $O(N(N+1)/2-1)$ vector model fixed point, denoted as $[vec]$, for which $g_1=0$ and
\begin{align}
    \frac{g^*_1}{\left(4 \pi\right)^2} &= \frac{ \varepsilon}{4 \left(14+N+N^2\right)} + \frac{3 \left(22+3 N+3 N^2\right) \varepsilon^2}{2 \left(14+N+N^2\right)^3}  - \frac{ \varepsilon^3}{16 (14 + N + N^2)^5} \bigg( -23416 + 99 N^5 \nn\\ 
    & \quad  + 33 N^6  + 91392 \zeta_{3} + N^4 (-319 + 960 \zeta_{3}) + N^3 (-803 + 1920 \zeta_{3}) + 4 N (-1441  \nn\\ 
    & \quad + 4992 \zeta_{3}) + N^2 (-6182 + 20928 \zeta_{3})\bigg).
\end{align}
For $N>4$, we do not find any real fixed point for which $y=0$ but $g_1 \neq 0$.  There are two fixed points which we denote as $[mat_\pm]$, that are complex for all values of $N> 3.6$
\begin{align}
\frac{g^*_1}{\left(4 \pi\right)^2} &=  N\varepsilon \left(N^2+N-10\right) (N (2 N+9)-36)\bigg[4 N \big(3888+1890 N -1215 N^2 -115 N^3+40 N^4+4   N^5\big)\nn\\
&\quad - 48
   \left(216 \pm \sqrt{-(N (2 N+9)-36)^2 (N (N (8 N
   (N+3)-153)+216)-1296)}\right)\bigg]^{-1} + \order{\varepsilon^2}
   \\
   \frac{g^*_2}{\left(4 \pi\right)^2} &= \bigg[8 \bigg(18144+4230
   N^2-1359 N^3-163 N^4+40 N^5+4 N^6\bigg)\bigg]^{-1} \varepsilon\bigg[1728-324 N+195 N^2\nn\\
   & \quad -12 N^3-4 N^4 \mp \sqrt{-\left(-36+9 N+2 N^2\right)^2 \left(-1296+216
   N-153 N^2+24 N^3+8 N^4\right)} \bigg] + \order{\varepsilon^2}.
\end{align}
Figure \ref{SO(N)-mergers} illustrates the stability and existence of various fixed points of $SO(N)-S_2$ GNY model as a function of $N \geq 4 .$
\begin{table}[H]
\centering
\resizebox{1.1\columnwidth}{!}{ \hspace{-.1\columnwidth}
\begin{tabular}{LLLLLLL}
\toprule
\multicolumn{1}{c}{} & \multicolumn{2}{c}{\textbf{$[ns_+]$}} & \multicolumn{2}{c}{\textbf{$[ns_2]$}} & \multicolumn{2}{c}{\textbf{$[ns_-]$}}\\
\cmidrule(rl){2-3} \cmidrule(rl){4-5} \cmidrule(rl){6-7}
N & g_1 & g_2 & g_1 & g_2 & g_1 & g_2  \\
\midrule
4 & 1.163 \varepsilon +0.255 \varepsilon ^2 & -1.31 \varepsilon +0.183 \varepsilon ^2 & -2.785 \varepsilon -3.438 \varepsilon ^2 & 2.479 \varepsilon +1.808 \varepsilon ^2 & -5.584 \varepsilon -2.683 \varepsilon ^2 & 3.047 \varepsilon +1.297 \varepsilon ^2 \\
5 & -0.03 \varepsilon +0.175 \varepsilon ^2 & -0.494 \varepsilon +0.065 \varepsilon ^2 & 0.345 \varepsilon -0.28 \varepsilon ^2 & 0.878 \varepsilon +0.387 \varepsilon ^2 & -0.888 \varepsilon +2.92 \varepsilon ^2 & 1.311 \varepsilon -0.615 \varepsilon ^2 \\
6 & -0.302 \varepsilon +0.175 \varepsilon ^2 & -0.252 \varepsilon +0.029 \varepsilon ^2 & 1.984 \varepsilon +1.495
   \varepsilon ^2 & -0.068 \varepsilon -0.494 \varepsilon ^2 & -0.86 \varepsilon +0.514 \varepsilon ^2 & 1.006
   \varepsilon +0.055 \varepsilon ^2 \\
 7 & -0.362 \varepsilon +0.165 \varepsilon ^2 & -0.155 \varepsilon +0.017 \varepsilon ^2 & 1.937 \varepsilon +0.892
   \varepsilon ^2 & -0.202 \varepsilon -0.216 \varepsilon ^2 & -0.697 \varepsilon +0.316 \varepsilon ^2 & 0.767
   \varepsilon +0.048 \varepsilon ^2 \\
 8 & -0.365 \varepsilon +0.152 \varepsilon ^2 & -0.107 \varepsilon +0.011 \varepsilon ^2 & 1.74 \varepsilon +0.658
   \varepsilon ^2 & -0.204 \varepsilon -0.101 \varepsilon ^2 & -0.583 \varepsilon +0.24 \varepsilon ^2 & 0.605
   \varepsilon +0.031 \varepsilon ^2 \\
 9 & -0.351 \varepsilon +0.14 \varepsilon ^2 & -0.079 \varepsilon +0.007 \varepsilon ^2 & 1.558 \varepsilon +0.548
   \varepsilon ^2 & -0.183 \varepsilon -0.052 \varepsilon ^2 & -0.502 \varepsilon +0.198 \varepsilon ^2 & 0.489
   \varepsilon +0.019 \varepsilon ^2 \\
		\bottomrule
	\end{tabular}}
	\caption{Quartic couplings at the three non-supersymmetric fixed points of the $SO(N)$-$S_2$ theory for small $N$. \label{table: SO(N)-finiteN-non susy-fixed point}}
\end{table}
\begin{table}[H]
	\centering
	\resizebox{\columnwidth}{!}{
		\begin{tabular}{cLLLLLL}
			\toprule
			\multicolumn{1}{c}{} & \multicolumn{2}{c}{$[ns_+]$} & \multicolumn{2}{c}{$[ns_2]$} & \multicolumn{2}{c}{$[ns_-]$}\\
			\cmidrule(rl){2-3} \cmidrule(rl){4-5} \cmidrule(rl){6-7}
			$N$ & \gamma_{\phi} & \gamma_{\psi} & \gamma_{\phi} & \gamma_{\psi} & \gamma_{\phi} & \gamma_{\psi}  \\
			\midrule
			4 & 0.125 \varepsilon -0.021 \varepsilon ^2 & 0.125 \varepsilon -0.003
			\varepsilon ^2 & 0.125 \varepsilon +0.024 \varepsilon ^2 & 0.125
			\varepsilon +0.031 \varepsilon ^2 & 0.125 \varepsilon +0.027
			\varepsilon ^2 & 0.125 \varepsilon +0.024 \varepsilon ^2 \\
			5  & 0.11 \varepsilon -0.013 \varepsilon ^2 & 0.11 \varepsilon +0.005
			\varepsilon ^2 & 0.11 \varepsilon +0.01 \varepsilon ^2 & 0.11
			\varepsilon +0.021 \varepsilon ^2 & 0.11 \varepsilon +0.011
			\varepsilon ^2 & 0.11 \varepsilon +0.021 \varepsilon ^2 \\
			6 & 0.104 \varepsilon -0.012 \varepsilon ^2 & 0.104 \varepsilon +0.007
			\varepsilon ^2 & 0.104 \varepsilon +0.013 \varepsilon ^2 & 0.104
			\varepsilon +0.017 \varepsilon ^2 & 0.104 \varepsilon +0.005
			\varepsilon ^2 & 0.104 \varepsilon +0.016 \varepsilon ^2 \\
			7 & 0.101 \varepsilon -0.011 \varepsilon ^2 & 0.101 \varepsilon +0.007
			\varepsilon ^2 & 0.101 \varepsilon +0.015 \varepsilon ^2 & 0.101
			\varepsilon +0.016 \varepsilon ^2 & 0.101 \varepsilon +0.002
			\varepsilon ^2 & 0.101 \varepsilon +0.014 \varepsilon ^2 \\
			8 & 0.1 \varepsilon -0.011 \varepsilon ^2 & 0.1 \varepsilon +0.007
			\varepsilon ^2 & 0.1 \varepsilon +0.016 \varepsilon ^2 & 0.1
			\varepsilon +0.016 \varepsilon ^2 & 0.1 \varepsilon -0.001
			\varepsilon ^2 & 0.1 \varepsilon +0.012 \varepsilon ^2 \\
			9 & 0.099 \varepsilon -0.011 \varepsilon ^2 & 0.099 \varepsilon +0.007
			\varepsilon ^2 & 0.099 \varepsilon +0.016 \varepsilon ^2 & 0.099
			\varepsilon +0.016 \varepsilon ^2 & 0.099 \varepsilon -0.002
			\varepsilon ^2 & 0.099 \varepsilon +0.011 \varepsilon ^2 \\
			\bottomrule
		\end{tabular}
	}
	\caption{Anomalous dimensions of $\phi$ and $\psi$ at the three non-supersymmetric fixed points of the  $SO(N)$-$S_2$ theory at small $N$.\label{table: SO(N)-finite N-non susy - gamma - phi}}
\end{table}

\begin{table}[H]
	\centering
	\resizebox{1.2\columnwidth}{!}{ \hspace{-.2\columnwidth}
		\begin{tabular}{cLLLLLLLLL}
			\toprule
			\multicolumn{1}{c}{} & \multicolumn{3}{c}{$[ns_+]$} & \multicolumn{3}{c}{$[ns_2]$} & \multicolumn{3}{c}{$[ns_-]$}\\
			\cmidrule(rl){2-4} \cmidrule(rl){5-7} \cmidrule(rl){8-10}
			$N$ & \gamma_{\phi^2} & \gamma_{\left(\bar\psi\psi,\phi^3\right)_{1}} & \gamma_{\left(\bar\psi\psi,\phi^3\right)_{2}} &  \gamma_{\phi^2} & \gamma_{\left(\bar\psi\psi,\phi^3\right)_{1}} & \gamma_{\left(\bar\psi\psi,\phi^3\right)_{2}} &  \gamma_{\phi^2} & \gamma_{\left(\bar\psi\psi,\phi^3\right)_{1}} & \gamma_{\left(\bar\psi\psi,\phi^3\right)_{2}}  \\
			\midrule
			4 & -0.2 \varepsilon +0.198 \varepsilon ^2 & -1.17 \varepsilon +0.019
			\varepsilon ^2 & -0.25 \varepsilon -0.151 \varepsilon ^2 & 0.962
			\varepsilon -0.124 \varepsilon ^2 & -1.218 \varepsilon -0.449
			\varepsilon ^2 & -0.25 \varepsilon +0.052 \varepsilon ^2 & 0.604
			\varepsilon -0.234 \varepsilon ^2 & -1.896 \varepsilon -0.116
			\varepsilon ^2 & -0.25 \varepsilon +0.005 \varepsilon ^2 \\
			5 & -0.19 \varepsilon +0.174 \varepsilon ^2 & -1.334 \varepsilon +0.005
			\varepsilon ^2 & -0.341 \varepsilon -0.036 \varepsilon ^2 & 1.034
			\varepsilon -0.018 \varepsilon ^2 & -0.757 \varepsilon -0.385
			\varepsilon ^2 & -0.341 \varepsilon +0.113 \varepsilon ^2 & 1.017
			\varepsilon +0.122 \varepsilon ^2 & -1.15 \varepsilon +0.942
			\varepsilon ^2 & -0.341 \varepsilon +0.065 \varepsilon ^2 \\
			6 & -0.18 \varepsilon +0.167 \varepsilon ^2 & -1.425 \varepsilon +0.028
			\varepsilon ^2 & -0.375 \varepsilon -0.009 \varepsilon ^2 & 0.837
			\varepsilon -0.285 \varepsilon ^2 & -0.375 \varepsilon -0.186
			\varepsilon ^2 & -0.152 \varepsilon +0.081 \varepsilon ^2 & 1.024
			\varepsilon +0.005 \varepsilon ^2 & -1.339 \varepsilon +0.227
			\varepsilon ^2 & -0.375 \varepsilon +0.05 \varepsilon ^2 \\
			7 & -0.173 \varepsilon +0.164 \varepsilon ^2 & -1.471 \varepsilon +0.042
			\varepsilon ^2 & -0.391 \varepsilon +0.002 \varepsilon ^2 & 0.699
			\varepsilon -0.213 \varepsilon ^2 & -0.391 \varepsilon -0.118
			\varepsilon ^2 & -0.037 \varepsilon -0.245 \varepsilon ^2 & 1.046
			\varepsilon -0.029 \varepsilon ^2 & -1.401 \varepsilon +0.153
			\varepsilon ^2 & -0.391 \varepsilon +0.043 \varepsilon ^2 \\
			8 & -0.169 \varepsilon +0.161 \varepsilon ^2 & -1.496 \varepsilon +0.05
			\varepsilon ^2 & -0.4 \varepsilon +0.008 \varepsilon ^2 & 0.621
			\varepsilon -0.135 \varepsilon ^2 & -0.4 \varepsilon -0.105
			\varepsilon ^2 & -0.006 \varepsilon -0.341 \varepsilon ^2 & 1.065
			\varepsilon -0.051 \varepsilon ^2 & -1.437 \varepsilon +0.124
			\varepsilon ^2 & -0.4 \varepsilon +0.039 \varepsilon ^2 \\
			9 & -0.167 \varepsilon +0.16 \varepsilon ^2 & -1.511 \varepsilon +0.054
			\varepsilon ^2 & -0.405 \varepsilon +0.011 \varepsilon ^2 & 0.574
			\varepsilon -0.082 \varepsilon ^2 & -0.405 \varepsilon -0.1
			\varepsilon ^2 & 0.004 \varepsilon -0.373 \varepsilon ^2 & 1.08
			\varepsilon -0.068 \varepsilon ^2 & -1.461 \varepsilon +0.107
			\varepsilon ^2 & -0.405 \varepsilon +0.036 \varepsilon ^2 \\
			\bottomrule
		\end{tabular}
	}
	\caption{  Anomalous dimensions of classically-relevant operators at the three non-supersymmetric fixed points of the $SO(N)$-$S_2$ theory for small $N$. \label{table: SO(N)-finite N-non susy - gamma - phi^2}}
\end{table}

\begin{table}[H]
	\centering
	\resizebox{1.2\columnwidth}{!}{ \hspace{-.2\columnwidth}
		\begin{tabular}{cLLLLLLLLL}
			\toprule
			\multicolumn{1}{c}{} & \multicolumn{3}{c}{$[ns_+]$} & \multicolumn{3}{c}{$[ns_2]$} & \multicolumn{3}{c}{$[ns_-]$}\\
			\cmidrule(rl){2-4} \cmidrule(rl){5-7} \cmidrule(rl){8-10}
			$N$ & \gamma_{\left(g_1,g_2,y\right)_{1}} & \gamma_{\left(g_1,g_2,y\right)_{2}} & \gamma_{\left(g_1,g_2,y\right)_{3}}  & \gamma_{\left(g_1,g_2,y\right)_{1}} & \gamma_{\left(g_1,g_2,y\right)_{2}} & \gamma_{\left(g_1,g_2,y\right)_{3}} & \gamma_{\left(g_1,g_2,y\right)_{1}} & \gamma_{\left(g_1,g_2,y\right)_{2}} & \gamma_{\left(g_1,g_2,y\right)_{3}}  \\
			\midrule
			4 &  -1.955 \varepsilon +0.019 \varepsilon ^2 & -1.068 \varepsilon +0.139
			\varepsilon ^2 & \varepsilon -0.224 \varepsilon ^2 & 0.385
			\varepsilon -0.279 \varepsilon ^2 & \varepsilon -0.495 \varepsilon
			^2 & 1.925 \varepsilon -1.471 \varepsilon ^2 & -0.516 \varepsilon
			+0.161 \varepsilon ^2 & \varepsilon -0.445 \varepsilon ^2 &
			1.667 \varepsilon -1.329 \varepsilon ^2 \\
			5 & -1.687 \varepsilon +0.21 \varepsilon ^2 & -0.873 \varepsilon +0.032
			\varepsilon ^2 & \varepsilon -0.268 \varepsilon ^2 & 0.097
			\varepsilon -0.394 \varepsilon ^2 & \varepsilon -0.413 \varepsilon
			^2 & 1.684 \varepsilon -0.997 \varepsilon ^2 & -0.125 \varepsilon
			+0.602 \varepsilon ^2 & \varepsilon -0.408 \varepsilon ^2 &
			1.66 \varepsilon -0.815 \varepsilon ^2 \\
			6 & -1.583 \varepsilon +0.249 \varepsilon ^2 & -0.897 \varepsilon +0.058
			\varepsilon ^2 & \varepsilon -0.274 \varepsilon ^2 & 0.139
			\varepsilon -0.119 \varepsilon ^2 & \varepsilon -0.373 \varepsilon
			^2 & 1.376 \varepsilon -1.194 \varepsilon ^2 & -0.489 \varepsilon
			+0.258 \varepsilon ^2 & \varepsilon -0.361 \varepsilon ^2 &
			1.523 \varepsilon -0.757 \varepsilon ^2 \\
			7 & -1.523 \varepsilon +0.264 \varepsilon ^2 & -0.934 \varepsilon +0.085
			\varepsilon ^2 & \varepsilon -0.274 \varepsilon ^2 & 0.072
			\varepsilon -0.072 \varepsilon ^2 & \varepsilon -0.363 \varepsilon
			^2 & 1.238 \varepsilon -0.98 \varepsilon ^2 & -0.661 \varepsilon
			+0.217 \varepsilon ^2 & \varepsilon -0.336 \varepsilon ^2 &
			1.46 \varepsilon -0.665 \varepsilon ^2 \\
			8 & -1.485 \varepsilon +0.271 \varepsilon ^2 & -0.964 \varepsilon +0.103
			\varepsilon ^2 & \varepsilon -0.274 \varepsilon ^2 & 0.019
			\varepsilon -0.024 \varepsilon ^2 & \varepsilon -0.361 \varepsilon
			^2 & 1.182 \varepsilon -0.843 \varepsilon ^2 & -0.765 \varepsilon
			+0.195 \varepsilon ^2 & \varepsilon -0.321 \varepsilon ^2 &
			1.424 \varepsilon -0.597 \varepsilon ^2 \\
			9 & -1.457 \varepsilon +0.275 \varepsilon ^2 & -0.986 \varepsilon +0.114
			\varepsilon ^2 & \varepsilon -0.274 \varepsilon ^2 & -0.019
			\varepsilon +0.015 \varepsilon ^2 & \varepsilon -0.361 \varepsilon
			^2 & 1.156 \varepsilon -0.768 \varepsilon ^2 & -0.834 \varepsilon
			+0.181 \varepsilon ^2 & \varepsilon -0.311 \varepsilon ^2 &
			1.401 \varepsilon -0.546 \varepsilon ^2 \\
			\bottomrule
		\end{tabular}
	}
	\caption{ Anomalous dimensions of classically marginal operators at the three non-supersymmetric fixed points of the  $SO(N)$-$S_2$ theory at small $N$. \label{table: SO(N)-finite N-non susy - gamma - marginal}}
\end{table}
Let us discuss the positivity of the quartic scalar potential at these fixed points. As in the $SU(N)$ adjoint theory, the supersymmetric fixed point gives rise to a classical scalar potential that has a unique minimum when $N$ is odd, but possesses flat directions when $N$ is even, giving rise to a moduli space as discussed above. For the other, non-supersymmetric fixed points, at finite $N$ we have the following results.
\begin{itemize}
\item For $[ns_+]$ fixed point $g_1>0$ for $N\in(3,4.95)$, $g_1<0$ for $N>4.95$ and $g_2<0$ for all values of $N$. The classical scalar potential is not positive for any value of $N$.
\item For the $[ns_2]$ fixed point $g_1>0 \, g_2>0$ for $N\in(4.95,5.8)$ and $g_1>0\,g_2<0$ for $N>5.8$. The classical scalar potential is positive for $N=5,6,7$ and $8$.
\item For $[ns_-]$ fixed point $g<0,g_2>0$ for $N\in(3,4.2)\cup(4.9,\infty)$. The classical scalar potential is positive for all integers in the range $N\in[3,10]$.
\end{itemize}
For sufficiently large-$N$, the classical scalar potential is not positive definite for any of the non-supersymmetric fixed points.
\subsection{Fixed points at large \texorpdfstring{$N$}{N}}
In the large-$N$ limit, of the one-loop $\beta$-functions for the 't Hooft couplings for the theory are,
\begin{align}
    \beta_{\lambda_{y}} & = -\frac{\varepsilon}{2}\,\lambda_{y} + \frac{1}{\left(4 \pi\right)^2}\bigg[\frac{\lambda_y^3 (2+N_f )}{4}\bigg] ,\\
    \beta_{\lambda_{1}} &= -\varepsilon\,\lambda_{1} + \frac{1}{\left(4 \pi\right)^2}\bigg[\frac{128 \lambda_1^2+16 \lambda_1 \lambda_{y}^2 N_f -\lambda_y^4 N_f }{16}\bigg] , \\
    \beta_{\lambda_{2}} & = -\varepsilon\,\lambda_{2} + \frac{1}{\left(4 \pi\right)^2}\bigg[\frac{192 \lambda_{1}^2+256 \lambda_{1} \lambda_{2}+64 \lambda_{2}^2+16 \lambda_{2} \lambda_{y}^2 N_f -3 \lambda_{y}^4  N_f }{16}\bigg].
\end{align}
As expected from the general arguments reviewed in section \ref{sec:largeNequivalence}, these $\beta$-functions are equivalent to the $\beta$-function of marginal couplings for the case of $SU(N)$ adjoint theory, given in eqn. \eqref{eq: marginal Beta function SU(N)} using the following re-scaling of the `t Hooft couplings:
\begin{equation}
\lambda_y^{SO(N)~S_2} \mapsto \frac{1}{2} \lambda_y^{SU(N)},\quad \lambda_1^{SO(N)~S_2} \mapsto \frac{1}{2} \lambda_1^{SU(N)}, \quad \lambda_2^{SO(N)~S_2} \mapsto \frac{1}{2} \lambda_2^{SU(N)}.
\end{equation}

The equivalence of $\beta$-functions for the two theories under the above map holds in the large-$N$ limit, for all $N_f$, but at finite $N$, the $\beta$-functions for the $SO(N)~S_2$ and $SU(N)$ adjoint theories are not equivalent. Therefore, the fixed points and anomalous dimensions of all operators we calculated are identical for both models in the large-$N$ limit. Note, however, that $1/N$ corrections to these quantities differ in each theory.

\subsection{Fixed points at \texorpdfstring{$N=3$}{N=3}}
For $N=3$, the two quartic couplings are not independent, and we set $g_1=0$ and define $g=g_2$ as before. For this theory, there are five real pseudo-scalars, and five Majorana fermions.

\begin{figure}[H]
    \centering
    \includegraphics[width = 8 cm]{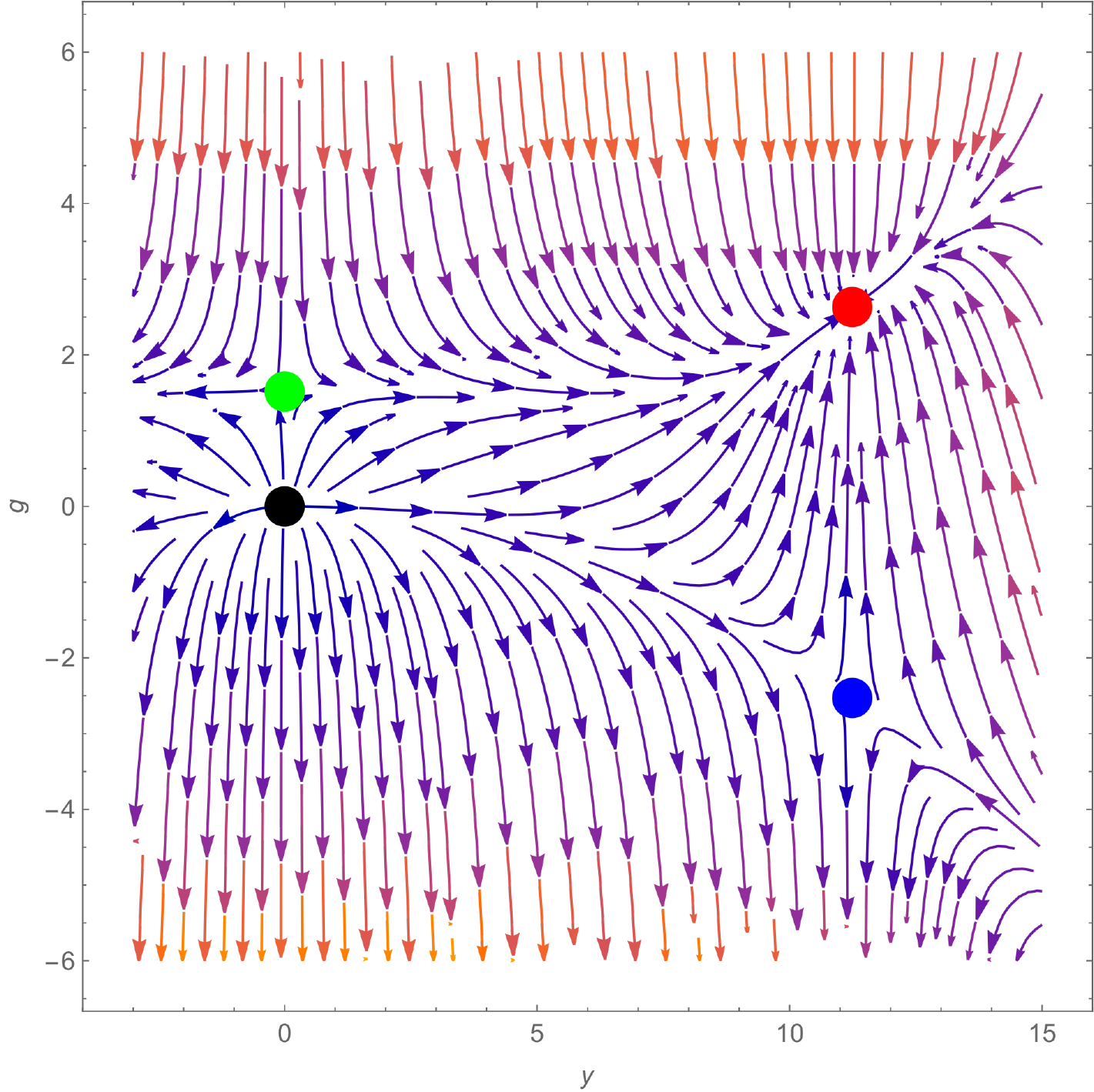}
    \caption{The figure shows flows in the $g$-$y$ plane for $N=3$. The blue dot denotes the fixed point $[ns]$ and the red dot is the supersymmetric fixed point. The black dot is the free fixed point and the green dot is the critical $O(5)$ vector model fixed point.  \label{SO3plot}}
\end{figure}

The $\beta$-functions for the two classically-marginal couplings when $N=3$ are
\begin{equation}
\beta_g =-\varepsilon ~ g + \frac{1}{\left(4 \pi\right)^2}\bigg[\frac{29952 \, g^2 + 672 g y^2 N_f - 25 y^4 N_f}{288}\bigg] ,\quad \beta_y = -\frac{\varepsilon~ y}{2} + \frac{1}{\left(4 \pi\right)^2}\bigg[\frac{y^3 (4 + 7 N_f)}{12 }\bigg].
\end{equation}
We present the fixed points after continuing to $N_f=1/2$. The plot for the fixed points and flows are shown in Figure \ref{SO3plot}
There are two fixed points with non-zero Yukawa coupling. We find one stable, $\mathcal N=1$ supersymmetric fixed point,
\begin{align}
\frac{g^*}{\left(4 \pi\right)^2} &= \frac{ \varepsilon }{60}+\frac{37 \varepsilon ^2}{4500} + \frac{\varepsilon ^3 \left(3097 -810  \zeta_3\right)}{1350000}
\\
\frac{\left(y^*\right)^2}{\left(4 \pi\right)^2} &= \frac{4 \varepsilon }{5} + \frac{148  \varepsilon ^2}{375} + \frac{ \varepsilon ^3  \left(3097 - 810 \zeta_3\right)}{28125}.
\end{align}
There is also one unstable non-supersymmetric fixed point, which we denote as $[ns]$,
\begin{align}
    \frac{g^*}{\left(4 \pi\right)^2} &= -\frac{5 \varepsilon}{312}  + \frac{90403  \varepsilon^2}{3954600} + \frac{\varepsilon ^3 \left(-2025716417  + 970207524  \zeta_3\right)}{80199288000}
   \\
   \frac{\left(y^*\right)^2}{\left(4 \pi\right)^2} &= \frac{4 \varepsilon }{5}-\frac{758
   \varepsilon ^2}{2535}-\frac{ \varepsilon ^3
   \left(-1187710595 + 18507528 
   \zeta_3\right)}{642622500}.
\end{align}
The scaling dimensions of $\phi$ and $\psi$ at the supersymmetric fixed point are
\be
\Delta_{\phi} = 1-\frac{4 \varepsilon }{15}+\frac{7 \varepsilon ^2}{1125}+\frac{7 \varepsilon ^3 (128+1335
   \zeta_3)}{168750},\quad \Delta_\psi = \frac12+\Delta_\phi
\ee
While at $[ns]$ the scaling dimensions of $\phi$ and $\psi$ are
\begin{align}
\Delta_{\phi} &= 1-\frac{4 \varepsilon }{15}-\frac{7546 \varepsilon ^2}{38025}+\varepsilon ^3
   \left(\frac{418479173}{771147000}+\frac{623 \zeta_3}{11250}\right)\\
\Delta_{\psi} &= \frac{3}{2}-\frac{4 \varepsilon }{15}-\frac{29827 \varepsilon ^2}{152100}+\frac{7
   \varepsilon ^3 (661963565+61006296 \zeta_3)}{7711470000}.
\end{align}
The scaling dimensions of composite operators at the supersymmetric fixed point are
\begin{align}
\Delta_{\phi^2} &=2+\frac{2 \varepsilon }{5}-\frac{161
   \varepsilon ^2}{375}+\frac{7 \varepsilon ^3
   (577+3040 \zeta_{3})}{18750}\\
\Delta_{\left(\bar\psi\psi,\phi^3\right)_1} &= 3-\frac{37 \varepsilon   ^2}{75}+\frac{\varepsilon ^3 (793+270
   \zeta_{3})}{3750}\\
\Delta_{\left(\bar\psi\psi,\phi^3\right)_2} &= 1 + \Delta_{\phi^2}
\end{align}
and at the non-supersymmetric fixed point,
\begin{align}
\Delta_{\phi^2} &= 2-\frac{93 \varepsilon }{65}+\frac{47432
   \varepsilon ^2}{54925}+\frac{7 \varepsilon
   ^3 (-755731285+350786696 \zeta_{3})}{2784697500} \\
   \Delta_{\left(\bar\psi\psi,\phi^3\right)_1} &= 3-\frac{102 \varepsilon }{65}+\frac{6444059
   \varepsilon ^2}{10545600}-\frac{\varepsilon
   ^3 (550342950430945+501910168141824
   \zeta_{3})}{364995870720000} \\
\Delta_{\left(\bar\psi\psi,\phi^3\right)_2} &= 3+\frac{2 \varepsilon }{5}-\frac{1514023
   \varepsilon ^2}{811200}+\varepsilon ^3
   \left(\frac{27309580648609}{561532108
   8000}+\frac{2128 \zeta_{3}}{1875}\right).
\end{align}
We also computed the anomalous dimension for marginal couplings. At the supersymmetric fixed point, the scaling dimension for these marginal couplings are
\begin{align}
    \Delta_{\left(\phi^4,\phi\bar\psi\psi\right)_1} &= 1 + \Delta_{\left(\bar\psi\psi,\phi^3\right)_1}\\
    \Delta_{\left(\phi^4,\phi\bar\psi\psi\right)_2} &= 4+\frac{12 \varepsilon }{5}-\frac{1526
   \varepsilon ^2}{375}+\frac{\varepsilon ^3
   (207611+197520 \zeta_{3})}{28125},
\end{align}
and at $[ns]$,
\begin{align}
    \Delta_{\left(\phi^4,\phi\bar\psi\psi\right)_1} &= 4-\frac{22 \varepsilon }{5}+\frac{10483
   \varepsilon ^2}{12675}-\frac{\varepsilon ^3
   (3346564865+2129410608 \zeta_{3})}{321311250}\\
    \Delta_{\left(\phi^4,\phi\bar\psi\psi\right)_2} &= 4+\frac{379 \varepsilon ^2}{1014}+\varepsilon
   ^3
   \left(-\frac{72168823}{17136600}+\frac{9 \zeta_{3}}{125}\right).
\end{align}

In addition, there are two fixed points with zero Yukawa coupling. These are the Gaussian (free) fixed point, which is unstable in all directions in $d=4-\varepsilon$, and the $O(5)$ vector model, denoted by $([wf])$, which has one unstable direction and can flow to the supersymmetric fixed point, as is clear from Figure \ref{SO3plot}. The quartic coupling at $[wf]$ is
\be
\frac{g^*}{\left(4 \pi\right)^2} = \frac{\varepsilon
   }{104} + \frac{87 \varepsilon
   ^2}{17576} + \frac{\varepsilon ^3
   \left(11969 - 58656 \zeta_{3}\right)}{23762752}.
\ee
For $N=3$, there is no distinction between $[adj]$ and $[vec]$, as there is only one quartic scalar coupling.
\section{\texorpdfstring{$SO(N)$ $S_2$-$A_2$}{SO(N) S2-A2} model}
We now turn our attention to a GNY model in which scalars transform as a real, symmetric matrix and the fermions transform as a real anti-symmetric matrix under $SO(N)$ global symmetry. This theory is manifestly non-supersymmetric as there are an unequal number of bosonic and fermionic degrees of freedom at finite $N$. Nevertheless, at large-$N$, the model possesses a fixed point for which correlation functions of single-trace operators are identical those computed in the supersymmetric large $N$ fixed point studied in the previous sections. This is a consequence of large-$N$ equivalence which we explicitly verify using the three-loop beta functions below.

\subsection{Fixed points for finite \texorpdfstring{$N>3$}{N>3}}
\begin{figure}[H]
	\centering
	\subfloat{{\includegraphics[width=7.2 cm]{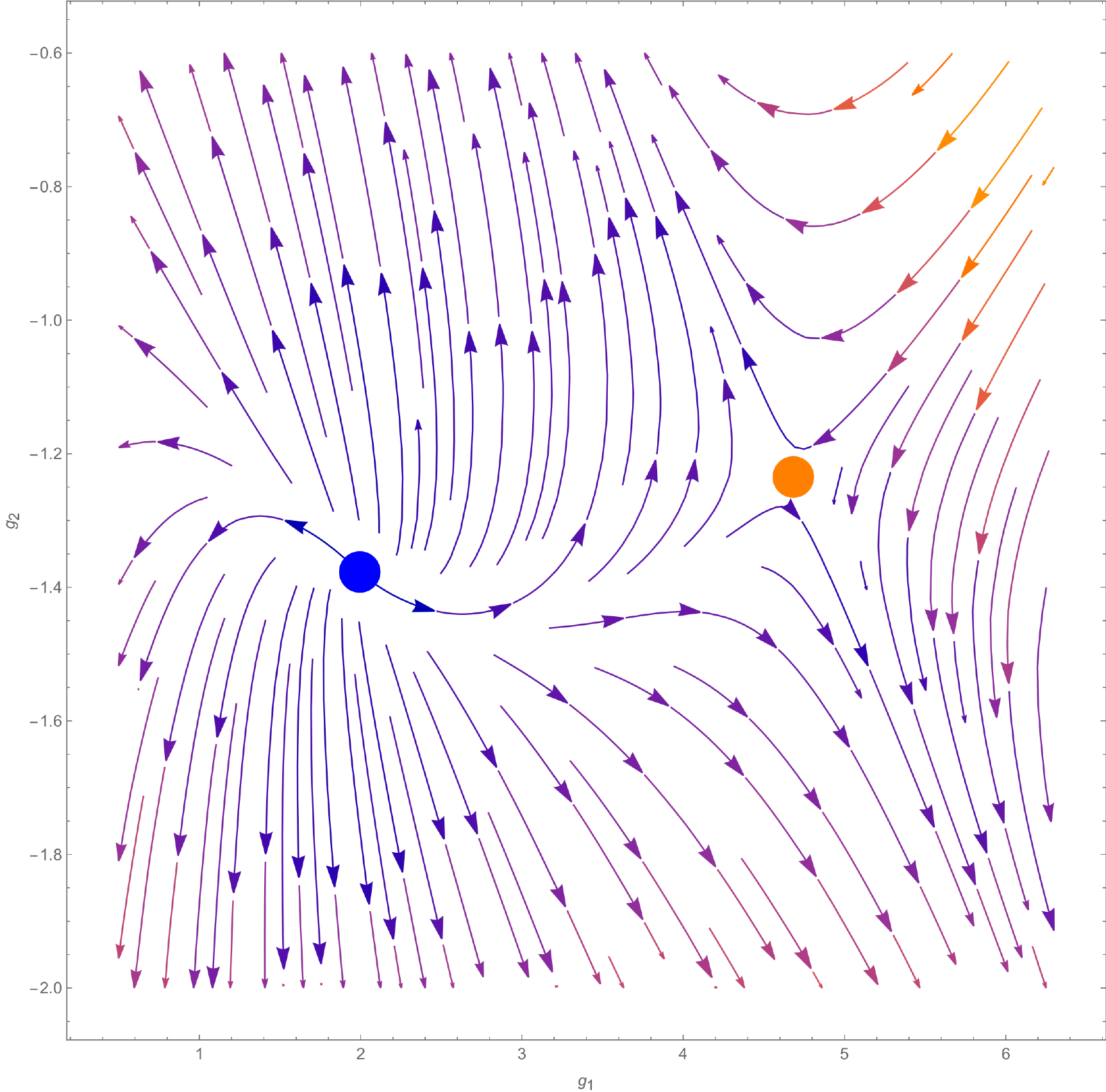} }}
	\quad
	\subfloat{{\includegraphics[width=7.2 cm]{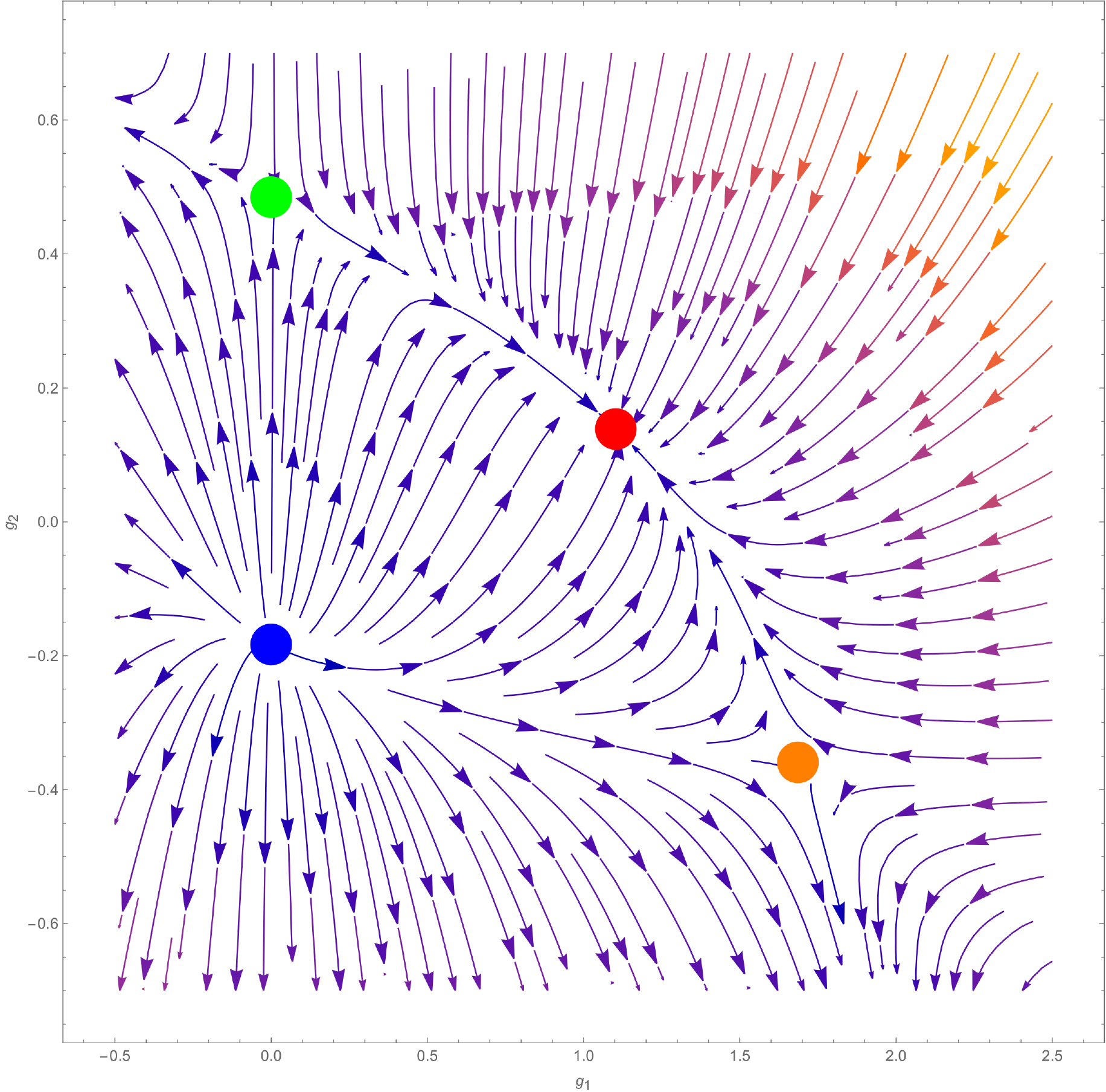} }}
	\caption{The figures show flows in the $g_1$-$g_2$ plane for $N=4$ and $8$ respectively. Here, the Yukawa coupling is tuned to criticality. The blue dot denotes the fixed point $[\widetilde{ns}_+]$ and the red dot is the $[\widetilde{susy}]$. The orange and green dots are the fixed points $[\widetilde{ns}_2]$ and $[\widetilde{ns}_-]$ respectively.}
	\label{figure: S2-A2 finite flow}
\end{figure}

\begin{figure}[H]
    \centering
  \includegraphics[scale=0.8]{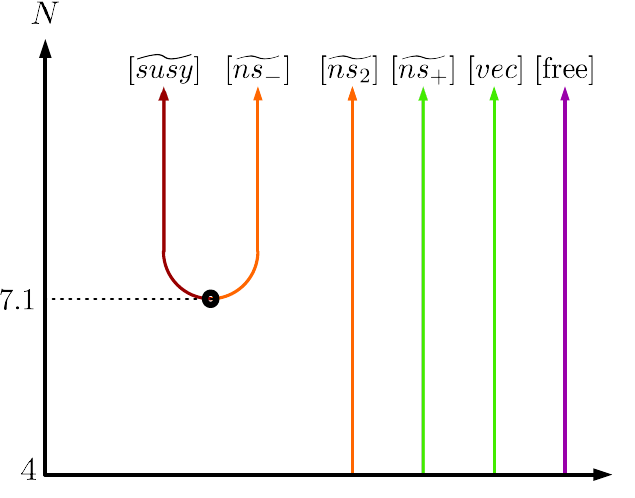}
    \caption{This figure illustrates the real fixed points at finite $N\geq 4$ of the $S_2-A_2$ GNY model at one-loop. Each line's color (red, orange, green, and violet) indicates the number of marginally unstable directions (0, 1, 2, and 3, respectively). Black dots denote mergers. }
    \label{S2-A2(N)-mergers}
\end{figure}

The $\beta$-functions at finite $N$, to one loop, are,
\begin{align}
	\beta_{g_1} &= - \varepsilon~g_1 + \frac{1}{\left(4 \pi\right)^2}\bigg[\frac{1}{256 N }\bigg(1024 g_1^2 (-36 + 9 N + 2 N^2) -  (-8 + N) N y^4 N_f \nn\\
	& \quad + 64 g_1 N \bigl(384 g_2 + (-2 + N) y^2 N_f \bigr)\bigg)\bigg]\\
	\beta_{g_2} &= - \varepsilon~g_2 + \frac{1}{\left(4 \pi\right)^2}\bigg[\frac{1}{256 N^2} \bigg(3072 g_1^2 (6 + N^2) + 2048 g_1 g_2 N (-6 + 3 N + 2 N^2) \nn\\
	& \quad + N^2 \bigl(1024 g_2^2 (14 + N + N^2) + 64 g_2 (-2 + N) y^2 N_f - 3 y^4 N_f \bigr)\bigg)\bigg]\\
	\beta_{y} &= - \frac{\varepsilon}{2}~y + \frac{1}{\left(4 \pi\right)^2}\bigg[\frac{y^3 \bigl(-12 - 2 N N_f + N^2 (2 + N_f)\bigr)}{16 N }\bigg]
\end{align}
Plots of the flows and fixed points for small values of $N$ are shown in Figure \ref{figure: S2-A2 finite flow}.
There exist up to four real fixed points with non-zero Yukawa coupling, which we refer to as $[\widetilde{ns}_+]$, $[\widetilde{ns}_-]$, $[\widetilde{ns}_2]$ and $[\widetilde{susy}]$. The last fixed point is not-supersymmetric, but we continue to refer to it using similar nomenclature as for the other two theories. The existence of these fixed points for various $N$ is illustrated in Figure \ref{S2-A2(N)-mergers}. 

\begin{figure}[H]
    \centering
    \subfloat{{\includegraphics[width=7 cm]{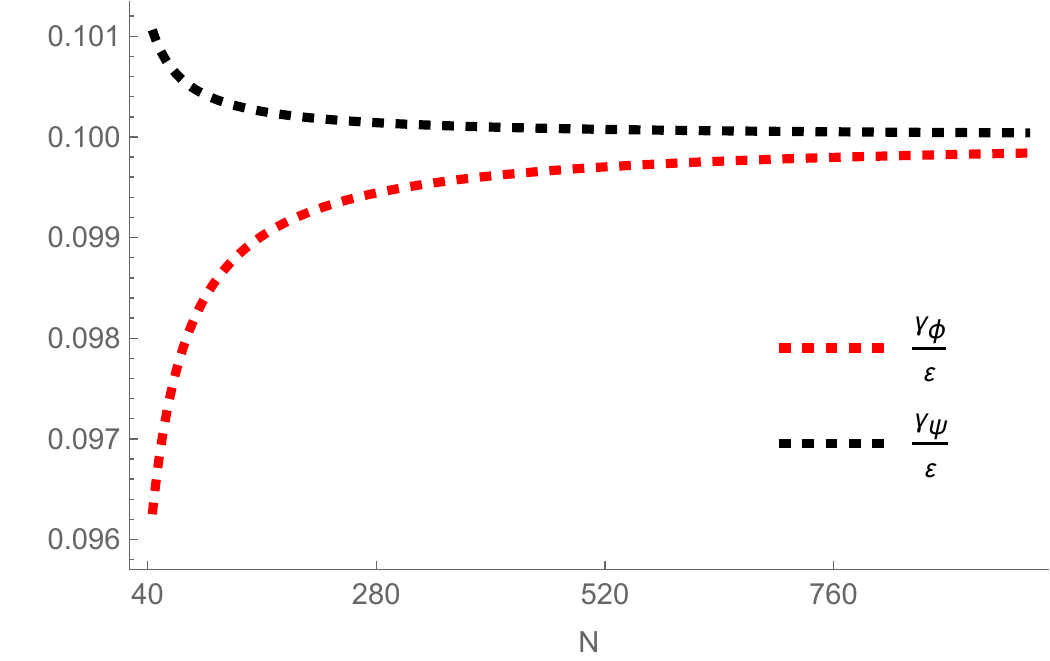} }}
    \quad
    \subfloat{{\includegraphics[width=7 cm]{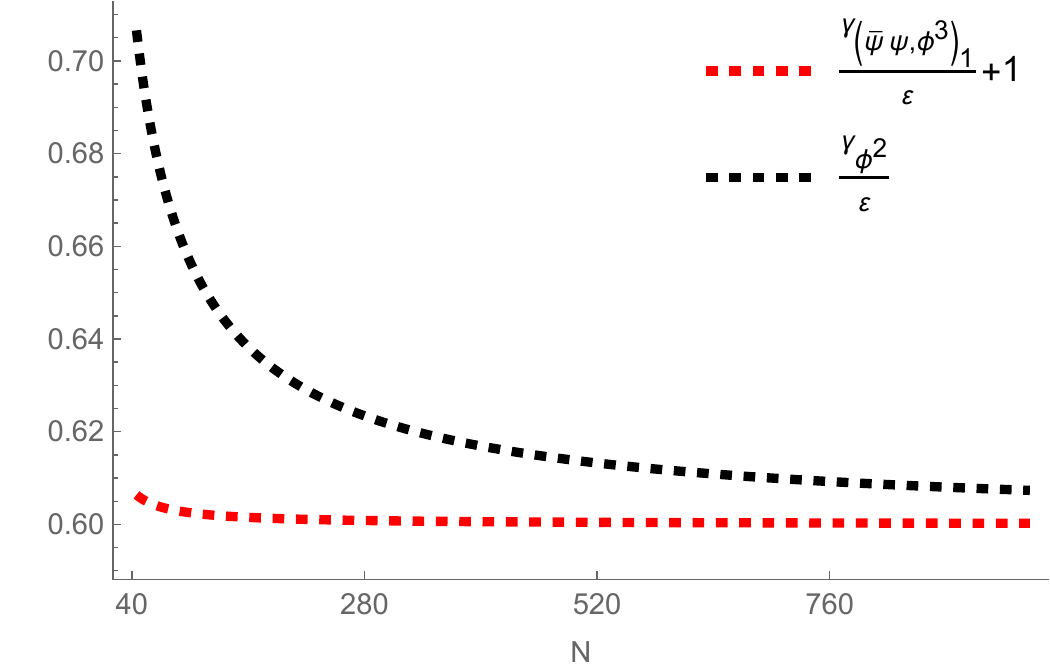} }}
    \quad
    \subfloat{{\includegraphics[width=7 cm]{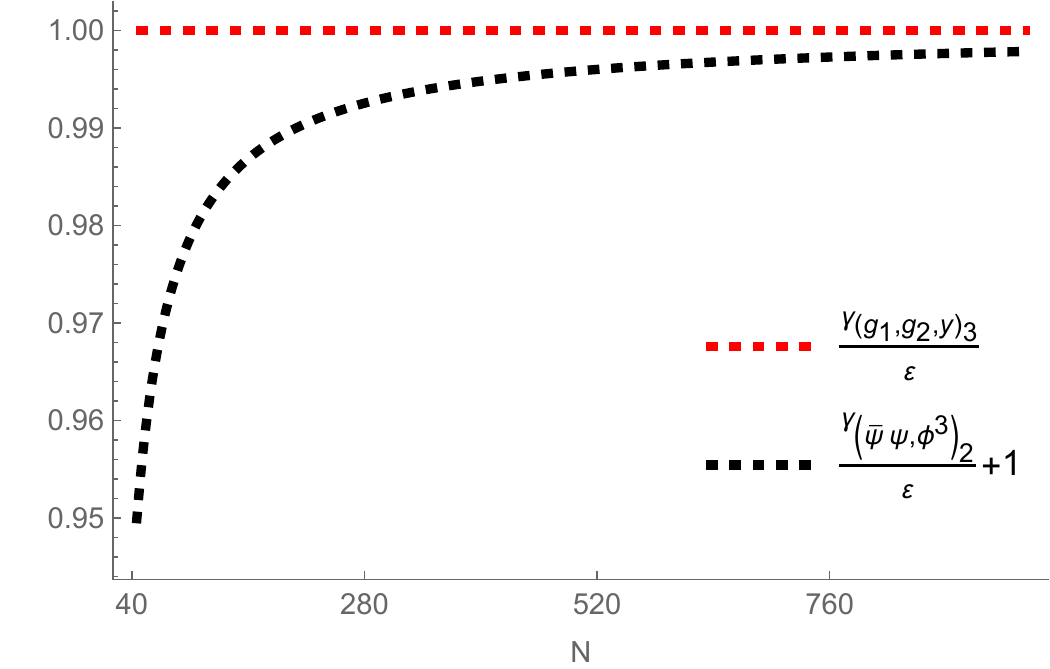} }}
    \caption{Plots of the scaling dimensions as a function of $N$ in the $[\widetilde{susy}]$ fixed point of the $SO(N)~S_2-A_2$ theory. As $N \to \infty$, they obey the relations \eqref{susy-1}, \eqref{susy-2} and \eqref{susy-3} that follow from supersymmetry. \label{gamma-So(N) S2A2-[susy]}}
\end{figure}

Let us examine the large-$N$ limit of the fixed point, $[\widetilde{susy}]$ in this model. At any $N$, the fixed point is not supersymmetric.  However, as mentioned above, in the large-$N$ limit, all observables in sectors common to the supersymmetric theories above, most notably correlation functions of single-trace operators, obey the constraints implied by supersymmetry. The anomalous dimensions at the $[\widetilde{susy}]$ fixed point are plotted as a function of $N$ in Figure \ref{gamma-So(N) S2A2-[susy]}, and we observe that scaling dimensions for large values of $N$ approach those of the $[susy]$ as obtained in the case of adjoint and $SO(N) - S_2$ GNY model, and hence obey the constraints of equations \eqref{susy-1}, \eqref{susy-2} and \eqref{susy-3} in the limit of large $N$.   

\subsection{Fixed points at large \texorpdfstring{$N$}{N}}
The large-$N$ limit of the one-loop $\beta$-functions for the 't Hooft couplings are,
\begin{align}
    \beta_{\lambda_{y}} & = -\frac{\varepsilon}{2}\,\lambda_{y} + \frac{1}{\left(4 \pi\right)^2}\bigg[\frac{\lambda_{y}^3 (2+N_f )}{16} - \frac{\lambda_y^3 N_f }{8 N}\bigg]\\
    \beta_{\lambda_{1}} &= -\varepsilon\,\lambda_{1} + \frac{1}{\left(4 \pi\right)^2}\bigg[\frac{2048 \lambda_{1}^2+64 \lambda_{1} \lambda_{y}^2 N_f -\lambda_{y}^4 N_f }{256} + \frac{1152 \lambda_1^2 - 16 \lambda_1 \lambda_y^2 N_f + \lambda_y^4 N_f}{32 N }\bigg]
	\\
	\beta_{\lambda_{2}} & = -\varepsilon\,\lambda_{2} + \frac{1}{\left(4 \pi\right)^2}\bigg[\frac{3072 \lambda_{1}^2+4096 \lambda_{1} \lambda_{2}+1024 \lambda_{2}^2+64 \lambda_{2} \lambda_{y}^2 N_f -3 \lambda_{y}^4 N_f }{256} \nn\\
 & \quad + \frac{6144 \lambda_1 \lambda_2 + 1024 \lambda_2^2 - 128 \lambda_2 \lambda_y^2 N_f}{256 N }\bigg]
\end{align}
We can relate the $\beta$-function of marginal couplings for the case of $SU(N)$ adjoint theory with the $\beta$-function of the marginal  couplings in the case of $SO(N) ~ S_2-A_2$ by following  re-scaling of the `t Hooft couplings 
\begin{equation}
    \lambda_y^{SO(N) S_2-A_2} \mapsto  \lambda_y^{SU(N)}, \quad \lambda_1^{SO(N) S_2-A_2} \mapsto  \frac{1}{2}\lambda_1^{SU(N)}, \quad \lambda_2^{SO(N) S_2-A_2} \mapsto  \frac{1}{2}\lambda_2^{SU(N)},
\end{equation}
$ $ and keeping the marginal quartic `t Hooft couplings unchanged.
The equivalence of $\beta$-functions for the two theories under the above map holds in the large-$N$ limit, for all $N_f$; and to all orders in perturbation theory, although we have only explicitly checked it up to three loops. The fixed points and anomalous dimensions of all operators we calculated are identical for both the adjoint and $SO(N)$-$S_2$ model in the large-$N$ limit. However, $1/N$ corrections differ. In particular, the scaling dimensions at fixed point $[\widetilde{susy}]$, which are listed up to one loop with $1/N$ corrections in \eqref{eq: 1/N correction field}, \eqref{eq: 1/N correction relevant}, and \eqref{eq: 1/N correction marginal},
can be seen to satisfy the constraints given in eqns. \eqref{susy-1}, \eqref{susy-2} and \eqref{susy-3}, at leading order in the large-$N$ expansion, but not at the level of $1/N$ corrections.

\begin{align}\label{Susy-largeN-SO(N)-S2 A2}
    \frac{\lambda^*_1}{\left(4 \pi\right)^2} = \left(\frac{1}{10} - \frac{11 }{25 N}\right)\varepsilon , \quad
    \frac{\lambda^*_2}{\left(4 \pi\right)^2} = \left(-\frac{1}{10} + \frac{67 }{25 N}\right)\varepsilon, \quad \frac{\left(\lambda_y^*\right)^2}{\left(4 \pi\right)^2} =  \left(\frac{16}{5} + \frac{32}{25 N}\right)\varepsilon
\end{align}
\be\label{eq: 1/N correction field}
\Delta_{\phi} = 1-\frac{2 \varepsilon }{5}-\frac{4   \varepsilon }{25 N}
, \quad 
\Delta_{\psi} = \frac{3}{2}-\frac{2 \varepsilon }{5}+\frac{\varepsilon }{25 N}
\ee

\be\label{eq: 1/N correction relevant}
\Delta_{\phi^{2}} = 2-\frac{2 \varepsilon   }{5}+\frac{192 \varepsilon }{25   N}
, \quad
\Delta_{\left(\bar\psi\psi,\phi^{3}\right)_{1}} = 3-\frac{2 \varepsilon   }{5}+\frac{6 \varepsilon }{25 N}
, \quad 
\Delta_{\left(\bar\psi\psi,\phi^{3}\right)_{2}} = 3-\frac{54 \varepsilon }{25 N}
\ee

\be\label{eq: 1/N correction marginal}
\Delta_{\left(\phi^2\right)^2} = 4-\frac{4 \varepsilon   }{5}+\frac{384 \varepsilon }{25   N}
, \quad 
\Delta_{\left(\phi^4,\phi\bar\psi\psi\right)_{1}} = 4
, \quad
\Delta_{\left(\phi^4,\phi\bar\psi\psi\right)_{2}} = 4-\frac{12 \varepsilon   }{25 N}
\ee


\subsection{Fixed points at \texorpdfstring{$N=3$}{N=3}}
\begin{figure}[H]
    \centering
    \includegraphics[width = 8 cm]{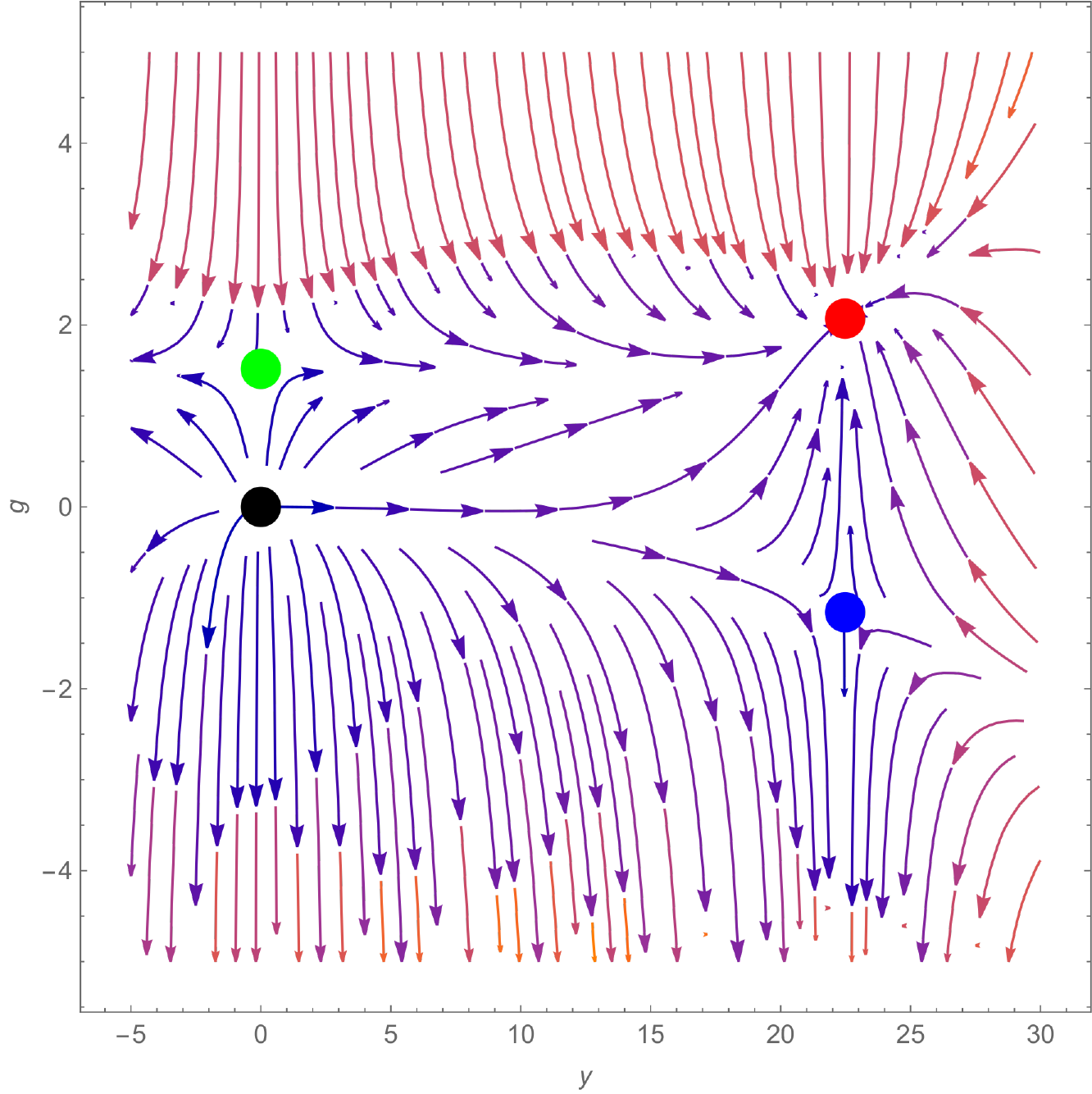}
    \caption{The figure shows flows in the $g$-$y$ plane for $N=3$. The blue dot denotes the unstable fixed point $[ns_-]$ and the red dot is  the stable $[ns_+]$ fixed point. The black dot is the free fixed point and the green dot is the critical $O(5)$ vector model fixed point.  \label{S2-A23plot}}
\end{figure}

For $N=3$, the $\beta$-function for $g$ and $y$ are
\begin{equation}
    \beta_{g} = - \varepsilon~g + \frac{1}{\left(4 \pi\right)^2}\bigg[\frac{53248 g^2 + 128 g y^2 N_f -  y^4 N_f}{512}\bigg], \quad \beta_{y} = -\frac{\varepsilon}{2}~ y + \frac{1}{\left(4 \pi\right)^2}\bigg[\frac{y^3 (2 + N_f)}{16}\bigg]
\end{equation}
We present the fixed points after continuing to $N_f=1/2$. There are 5 real pseudo-scalars and 3 Majorana fermions. The plot for the fixed points and flows are shown in Figure \ref{S2-A23plot}
There are two fixed points with non-zero Yukawa coupling, denoted as $[ns_\pm]$.  $[ns_+]$ is stable and $[ns_-]$ is unstable.
\begin{align}
    \frac{g^{*}}{\left(4 \pi\right)^2} &= \frac{3 \pm \sqrt{113}}{1040} \varepsilon +\frac{\left(661841  \mp 14981 \sqrt{113} \right) \varepsilon ^2}{114582000} \nn\\
    & \quad +\frac{ \varepsilon ^3 }{3938710417200000} \bigg[-885818563403  \pm 430102347235 \sqrt{113} \nn\\
    & \quad + 2838154137900  \zeta_{3} \mp 912675495900 \sqrt{113}  \zeta_{3}\bigg]
    \\
    \frac{\left(y^{*}\right)^2}{\left(4 \pi\right)^2} &= \frac{16 \varepsilon }{5}+\frac{4 \left(2693 \pm 343 \sqrt{113} \right) \varepsilon ^2}{21125} \nn\\
    & \quad - \frac{7 \varepsilon ^3 \left(-8097806339  \pm 684445563 \sqrt{113}  + 7447830000  \zeta_{3}\right)}{41894043750}.
\end{align}
The scaling dimensions of $\phi$ and $\psi$ at the $[ns_{\pm}]$ fixed point are
\begin{align}
    \Delta_{\phi} &= \frac{\varepsilon }{10}+\frac{\left(-3139 \pm 336 \sqrt{113}\right)
   \varepsilon ^2}{126750} \nn\\
   & \quad + \frac{\varepsilon ^3
   \left(10423469057 \mp 1283090319 \sqrt{113}-7894699800
   \zeta_{3}\right)}{335152350000} \\
   \Delta_{\psi} &= \frac{\varepsilon }{6}+\frac{\left(-3751 \pm 1029 \sqrt{113}\right) \varepsilon ^2}{304200} \nn\\
   & \quad + \frac{\varepsilon ^3 \left(25317258229 \mp 2423092203 \sqrt{113} - 15789399600 \zeta_{3}\right)}{402182820000}
\end{align}
The scaling dimensions of classically relevant operators  at the $[ns_{\pm}]$ fixed point are
\begin{align}
    \Delta_{\phi^2} &=\frac{47 \pm 7 \sqrt{113}}{130} \varepsilon + \frac{\left(1480413 \mp 289961 \sqrt{113}\right) \varepsilon ^2}{14322750} \nn\\
    & \quad  + \frac{\varepsilon ^3 }{246169401075000} \bigg[-40501165392615 \pm 4849253416469 \sqrt{113} \nn\\
    & \quad + 39528571923000 \zeta_{3} \pm 252007379400 \sqrt{113} \zeta_{3}\bigg]\\
    \Delta_{\left(\bar\psi\psi,\phi^3\right)_1} &= \frac{3 \left(-23 \pm \sqrt{113}\right) \varepsilon}{65}  + \frac{\left(246895847 \pm 15108003 \sqrt{113}\right) \varepsilon ^2}{42968250 \left(-23 \pm \sqrt{113}\right)} \nn\\
    & \quad + \frac{\varepsilon ^3 }{553881152418750 \left(-23 \pm \sqrt{113}\right)^3} \bigg[1552248488286510403 \nn\\
    & \quad \mp 211297594055407869 \sqrt{113} - 1076504544581729400 \zeta_{3} \nn\\
    & \quad \pm 60896334116608200 \sqrt{113} \zeta_{3}\bigg]\\
    \Delta_{\left(\bar\psi\psi,\phi^3\right)_2} &= \frac{\left(579295 \mp 66477 \sqrt{113}\right) \varepsilon ^2}{76050 \left(-23 \pm \sqrt{113}\right)} + \frac{\varepsilon ^3}{75409278750 \left(-23 \pm \sqrt{113}\right)^3} \times \nn\\
    & \quad \bigg[-659845212941173 \pm 75953536701051 \sqrt{113} - 294403186677600 \zeta_{3} \nn\\
    & \quad \pm 25069395780000 \sqrt{113} \zeta_{3}\bigg]
\end{align}
The scaling dimension of the classically marginal operators at the $[ns_\pm]$ fixed points, which we present only up to one loop due to space constraints.
\begin{equation}
    \Delta_{\left(g,y\right)_{1}} = \frac{1}{10} \left(5 \mp \sqrt{113}-\sqrt{2 \left(69 \pm 5 \sqrt{113}\right)}\right) \varepsilon
\end{equation}
\begin{equation}
    \Delta_{\left(g,y\right)_{2}} = \frac{1}{10} \left(5 \mp \sqrt{113}+\sqrt{2 \left(69 \pm 5 \sqrt{113}\right)}\right) \varepsilon
\end{equation}
\section{Pad\'e approximates of scaling dimensions}
We also computed Pad\'e approximates of scaling dimensions for all of the fixed points of the three models in this paper. In this section, we present results in $d=3$ for the stable $[susy]$ fixed point at $N_f = \frac{1}{2}$ in the large $N$ limit, and for $N=3$ in each of the three models. Additional results, including Pad\'e approximates for unstable non-supersymmetric fixed points, and extrapolations to $d=2$, are presented in Appendix \ref{app:pade}.

Our Pad\'e approximations can be compared to bootstrap results in \cite{Rong:2019qer} for the $SU(N)$ adjoint supersymmetric fixed points, and the $SO(3)$ symmetric traceless supersymmetric fixed point. It should be possible to study the other fixed points identified in this paper using bootstrap using techniques similar to those in \cite{Rong:2018okz, Rong:2019qer, Erramilli:2022kgp}. 

\subsection{Comments on Gross-Neveu Yukawa CFTs with a gap}
One of our primary motivations for studying these theories is the possibility of an interesting holographic dual, with little or no supersymmetry.
Suppose one of the fixed points identified in this paper (or from some other large-$N$ GNY model) is dual to Einstein gravity or $\mathcal N=1$ supergravity plus a finite number of other fields with finite or zero mass. This would imply that the anomalous dimension of a generic unprotected operator must diverge in $d=3$. How would we see this in the epsilon expansion? 

Using the epsilon-expansion, one can obtain Pad\'e approximations for the scaling dimensions, say of a higher-spin operator:
\begin{equation}
    \Delta_s(d) = \frac{p(d)}{q(d)}
\end{equation}
where $p(d) = \sum p_k d^k$ and $q(d) = \sum q_k d^k$ are polynomials in $d$ whose coefficients depend on $N$. More precisely, we expect coefficients that can be expanded as a power series in $1/N^2$, as follows: $p_k=p_k^0  + p_k^1/N^2 + p_k^2/N^4 + \ldots$.

$\Delta_s(d)$ may have a simple pole at some $d=d^*(N)=d^*_0-d^*_1/N^2 + O(1/N^4)$. This means, near $d=d^*$, $$ \Delta_s(d) \sim \frac{A}{d-d^*},$$ where $A$ is the residue of the pole at $d^*$. If $d^*_0=3$, then $$ \Delta_s(3) = N^2\frac{A}{d^*_1}(1+O(1/N^2)),$$ and the anomalous dimension is proportional to $N^2$ in the large-$N$ limit. We would require that $A/d_1^*$ be positive. 

There is a  problem with the above approach. While Pad\'e approximations yield remarkably precise estimates for scaling dimensions in theories such as the critical Ising model or the $N=1$ GNY model \cite{Fei:2016sgs}, one generally requires that the Pad\'e approximation should not contain a pole in the region of interest ($2<d<4$ for the GNY model). If the Pad\'e approximation did contain a pole in this region, one might argue that it should not be trusted.  Therefore, it seems natural to also demand that the Pad\'e approximation must not contain any real poles between $3<d<4$ for the theory at finite $N$. In this case, if the large-$N$ Pad\'e approximation contains a pole at $d=3$, it must have a small imaginary part $d^*=3\pm i\frac{a}{N}$ when $1/N$ corrections are taken into account. Such a pole would be absent at finite $N$, and only emerge as $N\to \infty$. Therefore, for any finite $N$, one has a Pad\'e approximation free from poles that is (presumably) as reliable as those used to estimate scaling dimensions in other finite $N$ theories. 

The above discussion means that we must look for a Pad\'e approximation that takes the form:
$$p(d)/q(d) \sim \frac{A}{(d-3)^2+\frac{a^2}{N^2}}$$ near $d=3$. This manifests itself as a second-order pole in the large-$N$ scaling dimensions, which can only be present in $(m,n)$-Pad\'e approximations with $n\geq 2$. 

The no-crossing theorem \cite{Sourlas:2017wyf}, would indicate that, if a generic operator $\mathcal O$ near $d=4$ diverges when $d \to 3$, we would expect the same for all operators with the same quantum numbers as $\mathcal O$, whose scaling dimensions exceed $\Delta_{\mathcal O}$ slightly below $d=4$, in the absence of crossing. If one could find an example of a large-$N$ CFT one can study in the epsilon-expansion that contains even a single operator whose scaling dimension diverges as $d \to 3$, in the large-$N$ limit, as described above, it would be relatively surprising.

In particular, to help identify the holographic dual, we are interested the twists (recall that the twist $\tau$ of a field is conventionally defined in terms of its spin $s$ and scaling dimension $\Delta$ via $\tau = \Delta-s$) of higher spin currents. If the twists of all higher spin currents with spin greater than $2$ were to diverge, it would imply that all higher-spin gauge fields in the bulk dual are infinitely massive, and the dual is an Einsteinian theory of gravity (rather than a higher-spin gauge theory or finite tension string theory). While we have not computed the higher spin spectrum in this paper, we expect that the lowest-twist spin-$s$ operator will be a mixture of fermion bilinears and scalar bilinears. Therefore we expect that $\lim_{s \to \infty} (\Delta_s -s) = \text{min}( 2\Delta_\phi , 2 \Delta_\psi-1 )$, and we can study the large $s$ limit of the higher spin spectrum via $\Delta_\phi$ and $\Delta_\psi$. Pad\'e approximates suggest these anomalous dimensions remain finite in $d=3$, though perhaps these are not conclusive. We, therefore, suspect that the bulk dual $4$-dimensional string-theoretic description of our fixed point is an $\mathcal N=1$ string theory at finite tension, rather than a theory of Einstein gravity. 

It appears somewhat unlikely to us that $\Delta_\psi$ or $\Delta_\phi$ would diverge in any multi-scalar or GNY fixed point, although we do not have rigorous proof of this. In a gauge theory, such as a Chern-Simons theory with matter in $d=3$, $\psi$, and $\phi$ are not gauge invariant, so their scaling dimension is not meaningful. Moreover $\tau_s \sim \log s$ in the limit $s \to \infty$, rather than $2\Delta_\psi -1$ or $2\Delta_\phi$. Because $\Delta_s > d$ for $s>2$ near $d=4$, the divergence of the higher-spin spectrum does not require any low-lying operators, whose scaling dimension satisfies $\Delta<d$, for $d \approx 4$, to cross the line $\Delta=d$ separating relevant from irrelevant operators. Therefore it would be interesting to study infrared fixed points of gauge theories in $d=3$, using the epsilon expansion, to see if such a divergence can be found.

\subsection{Large \texorpdfstring{$N$}{N} scaling dimensions}

\begin{figure}[H]
    \centering
    {\includegraphics[width=12 cm]{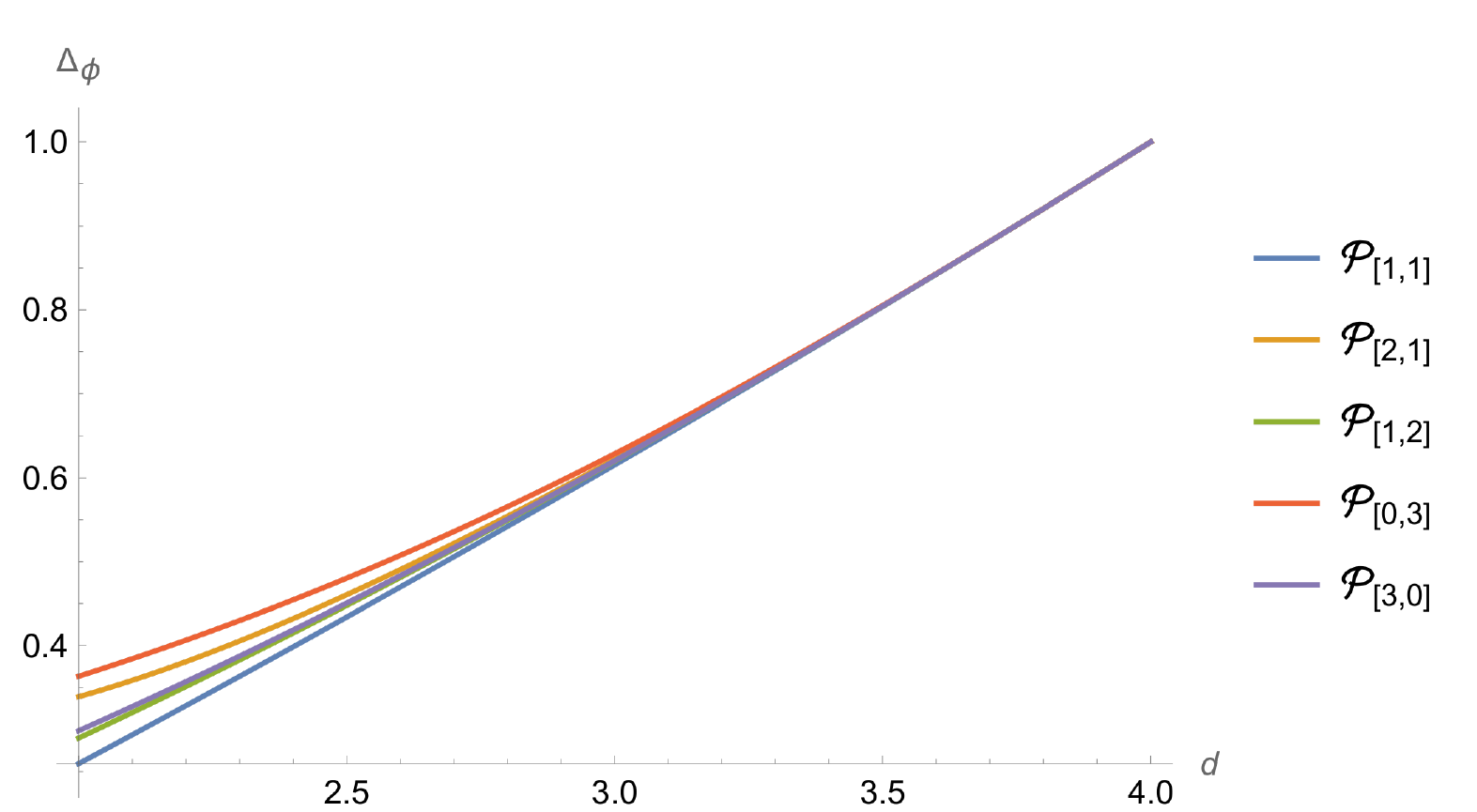} }
    \caption{Pad\'e approximates of $\Delta_\phi$ for $2<d<4$ in the large $N$ supersymmetric fixed point.}
    \label{figure: pade-scaling-dimension-phi-psi}
\end{figure}

Using the three-loop results computed in this paper, we can obtain Pad\'e approximates up to order three, e.g., Pad\'e-[1,2] or [2,1] approximates. We present plots of the Pad\'e-[1,1], [1,2] and [2,1] approximates for the scaling dimensions $\Delta_\phi$ as a function of $d$ at the fixed point $[susy]$, in Figure \ref{figure: pade-scaling-dimension-phi-psi}.  In all cases, there is no sign of a pole at any value of $d$ between $2$ and $4$. The Pad\'e-[1,2] approximates for scaling dimensions of composite operators in the supersymmetric fixed point of the common large-$N$ limit of the $SU(N)$ and $SO(N)$ models studied in the paper are also plotted as function of $d$ in Figure \ref{figure: crossing scaling-dimension-largeN-[1,2]}. (The Pad\'e-[2,1] approximate is  similar in the range $2<d<4$, and is not shown.) We see that all scaling dimensions appear to be well behaved in this region, with no sign of a divergence, indicating the putative holographic dual is a string theory of finite string tension.

It is interesting to try to assess the validity of the epsilon expansion for this theory.\footnote{We thank Z. Komargodski for discussions on this point.} When the epsilon expansion results imply that the scaling dimensions of two operators (with the same quantum numbers) cross, it may suggest a failure of the epsilon expansion. Essentially, scaling dimensions are eigenvalues of the dilatation operator, which we consider to be a continuous function of a single parameter -- the dimension $d$ -- and eignevalues of this Hermitian operator are not expected to cross due to the no-crossing theorem \cite{Sourlas:2017wyf}.  A violation of the no-crossing theorem would suggest that the theory is not generic and the epsilon expansion might not be trustworthy near or below the value of $d$ where crossing occurs. See \cite{Giombi:2015haa, DiPietro:2015taa, DiPietro:2017kcd, DiPietro:2017vsp} for some related discussion in conformal QED${}_3$. From Figure \ref{figure: crossing scaling-dimension-largeN-[1,2]}, we find no crossing of scaling dimensions before $d=3$, although some relevant operators become irrelevant below $d=3$, and cross before $d=2$. Some operators of different parity do cross -- however this does not indicate a violation of the no-crossing theorem, as we expect the dilation operator to commute with parity and therefore be block diagonal, and there is no reason eigenvalues of different blocks cannot cross. Near $d\approx 2.3$, we find the scaling dimensions of the two mixtures of $\phi^3$ and $\bar{\psi}\psi$ cross, suggesting possible difficulties in extrapolating to $d=2$.

Numerical values of all possible Pad\'e approximates of the scaling dimensions of various operators in $d=3$ in this fixed point are presented in Table \ref{tab:largeNpade}. The numerical values for this case appear quite consistent across different Pad\'e approximates for this case. The same is not true for some of the fixed points, as can be seen in Appendix \ref{app:pade}. 

\begin{figure}[H]
    \centering
    \includegraphics[width= 15cm]{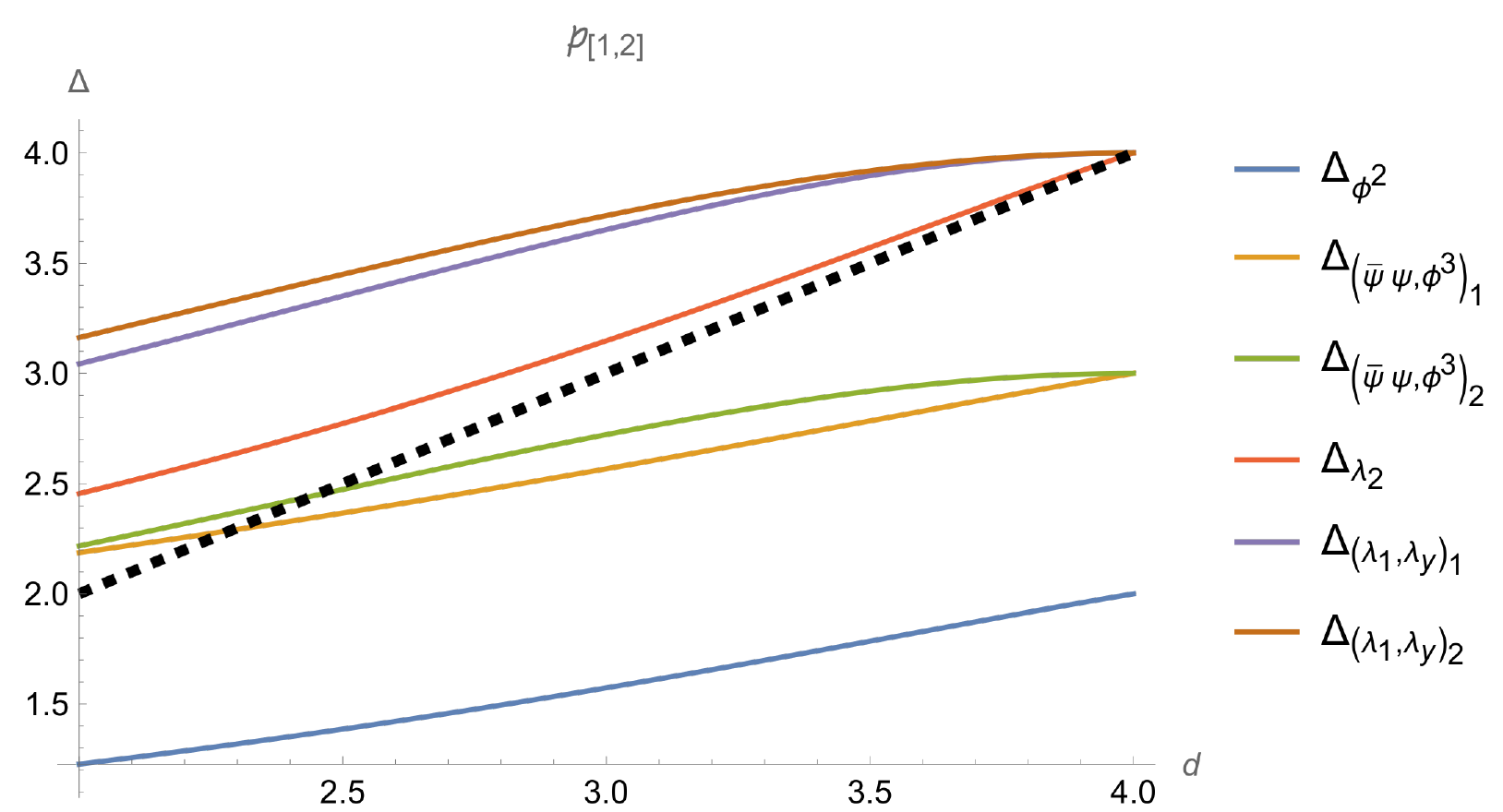}
    \caption{$\text{Pad\'e}_{[1,2]}$ approximates of the scaling dimensions of composite operators as a function of $d$ in the supersymmetric fixed point of the common large-$N$ limit. The dashed line is $\Delta=d$, and separates relevant from irrelevant operators.}
    \label{figure: crossing scaling-dimension-largeN-[1,2]}
\end{figure}

\begin{table}[H]\centering
        \resizebox{\columnwidth}{!}{
	\begin{tabular}{|m{2 cm}|m{1.5cm}|m{1.5cm}|m{1.5cm}|m{1.6cm}|m{1.6cm}|m{1.5cm}|m{1.8cm}|m{1.8cm}|}\hline
	\vspace{0.3 cm} & $\Delta_{\phi}$ & $\Delta_{\psi}$ & $\Delta_{\phi^{2}}$ & $\Delta_{\left(\bar\psi\psi,\phi^{3}\right)_{1}}$ & $\Delta_{\left(\bar\psi\psi,\phi^3\right)_{2}}$ & $\Delta_{\left(\phi^2\right)^2}$ & $\Delta_{\left(\phi^4,\phi\bar\psi\psi\right)_{1}}$ & $\Delta_{\left(\phi^4,\phi\bar\psi\psi\right)_{2}}$\\
		\hline
		\hline
   $\text{Pad\'e}_{[1,1]}$ & 0.615 & 1.115 & 1.459 & 2.459 & 3. & 2.919 &
   4. & 4. \\
 $\text{Pad\'e}_{[1,2]}$ & 0.62 & 1.12 & 1.574 & 2.568 & 2.723 & 3.148 &
   3.653 & 3.717 \\
 $\text{Pad\'e}_{[2,1]}$ & 0.622 & 1.122 & 1.55 & 2.55 & 2.695 & 3.1 &
   3.62 & 3.695 \\
 $\text{Pad\'e}_{[0,3]}$ & 0.629 & 1.121 & 1.582 & 2.587 & 2.731 & 3.164
   & 3.675 & 3.725 \\
 $\text{Pad\'e}_{[3,0]}$ & 0.62 & 1.12 & 1.609 & 2.609 & 2.705 & 3.219 &
   3.646 & 3.705 \\
    \hline
    \end{tabular}}
    \caption{$\text{Pad\'e}$ approximates of the scaling dimensions in $d=3$ for $[susy]$ fixed points in the common large-$N$ limit of the $SU(N)$ adjoint, $SO(N)$ $S_2$ and $S_2-A_2$ models. \label{tab:largeNpade}}
\end{table}

\subsection{\texorpdfstring{$SU(3)$}{SU(3)} adjoint scaling dimensions}
Here, we present Pad\'e approximates for the $SU(3)$ ajoint GNY model. In Figure \ref{figure: pade [1,2] SU(3) scaling-dimension-N=3}, we plot the Pad\'e-[1,2] approximates of the composite operators in this theory, as a function of $d$, for $2<d<4$. We find that the scaling dimensions of one combination of the classically marginal couplings and the scaling dimension of $\phi^2$ cross near $d=2.8$. These scaling dimensions also intersect the line $\Delta=d$, near the same value of $d$. It is therefore possible that the epsilon expansion is not reliable near this $d=2.8 \approx 3$. However, the reasonably good agreement with bootstrap suggests that the epsilon expansion is valid at least until $d=3$.%

Numerical values of various possible Pad\'e-approximates of scaling dimensions  for different operators in $d=3$ are given in Tables \ref{table: SU(3) pade-scaling-[susy]-d=3}.
\begin{table}[H]\centering
    \begin{tabular}{|m{2 cm}|m{2cm}|m{1.7cm}|m{1.5cm}|m{1.6cm}|m{1.6cm}|m{1.5cm}|m{1.5cm}|}\hline
    \vspace{0.2cm} & $\Delta_{\phi}$ & $\Delta_{\psi}$ & $\Delta_{\phi^{2}}$ & $\Delta_{\left(\bar\psi\psi,\phi^{3}\right)_{1}}$ & $\Delta_{\left(\bar\psi\psi,\phi^{3}\right)_{2}}$ & $\Delta_{\left(g,y\right)_{1}}$ & $\Delta_{\left(g,y\right)_{2}}$ \\
    \hline
    \hline
    $\text{Pad\'e}_{[1,1]}$ & 0.813 & 1.313 & 2.456 & 3. & 3.456 & 4. & 5.19 \\
    \hline
    $\text{Pad\'e}_{[1,2]}$ & 0.818 & 1.306 & 2.793 & 2.332 & 3.732 & 3.293 & 6.701 \\
    \hline
    $\text{Pad\'e}_{[2,1]}$ & 0.762 & 1.262 & 2.626 & 2.141 & 3.626 & 3.141 & 6.029 \\
    \hline
    $\text{Pad\'e}_{[0,3]}$ & 0.928 & 1.442 & -8.69 & 2.349 & 15.64 & 3.312 & -1.228 \\
    \hline
    $\text{Pad\'e}_{[3,0]}$ & 0.965 & 1.465 & 4.403 & 2.169 & 5.403 & 3.169 & 17.21 \\
    \hline
    \end{tabular}\caption{Pad\'e approximates of scaling dimensions of various operators in $d=3$ for the supersymmetric fixed point of $SU(3)$ adjoint theory.\label{table: SU(3) pade-scaling-[susy]-d=3}}
\end{table}
We included all possible Pad\'e approximates in Table \ref{table: SU(3) pade-scaling-[susy]-d=3} to facilitate comparison to \cite{Rong:2019qer}. We see that the Pad\'e-[1,2] approximation seems to be in good agreement with the bootstrap result for the $SU(3)$ theory: the result for $\Delta_\phi \approx .82$ is in good agreement with the prediction in Figure 5 of \cite{Rong:2019qer}, and $\Delta_{\phi^2} \approx 2.79$, which is in good agreement with the bootstrap result of $2.77$ in Table II of \cite{Rong:2019qer}.\footnote{The result for $\Delta_{(\psi\psi, \phi^3)_1, 2,1} \approx 2.33$ is not in as good agreement with the result of 2.50 in Table II of \cite{Rong:2019qer}. At large-$N$, Pad\'e-[2,1] approximation of the scaling dimension $\Delta_\phi \approx .62$, which is in good but not perfect agreement with Figure 5 of \cite{Rong:2019qer}.}

This indicates that the kink discovered in \cite{Rong:2019qer} does indeed correspond to the $SU(3)$ adjoint GNY fixed point theory. This is in apparent conflict with the duality conjectured in \cite{Gaiotto:2018yjh}, which conjectures that GNY fixed point is dual to the IR limit of $\mathcal N=2$ sQED. As discussed in \cite{Rong:2019qer}, one possible resolution is that $\mathcal N=2$ sQED contains a relevant operator, and when deformed by that operator flows to the $[susy]$ fixed point of the  $SU(3)$ adjoint model. However, it is interesting to note that, without resummation, $(\Delta_\phi)_{[3,0]} \approx 0.97$, which is in agreement with predictions from duality conjecture of \cite{Gaiotto:2018yjh}, so, even at three loops results are not conclusive.

\begin{figure}[H]
    \centering
    \includegraphics[width= 15 cm]{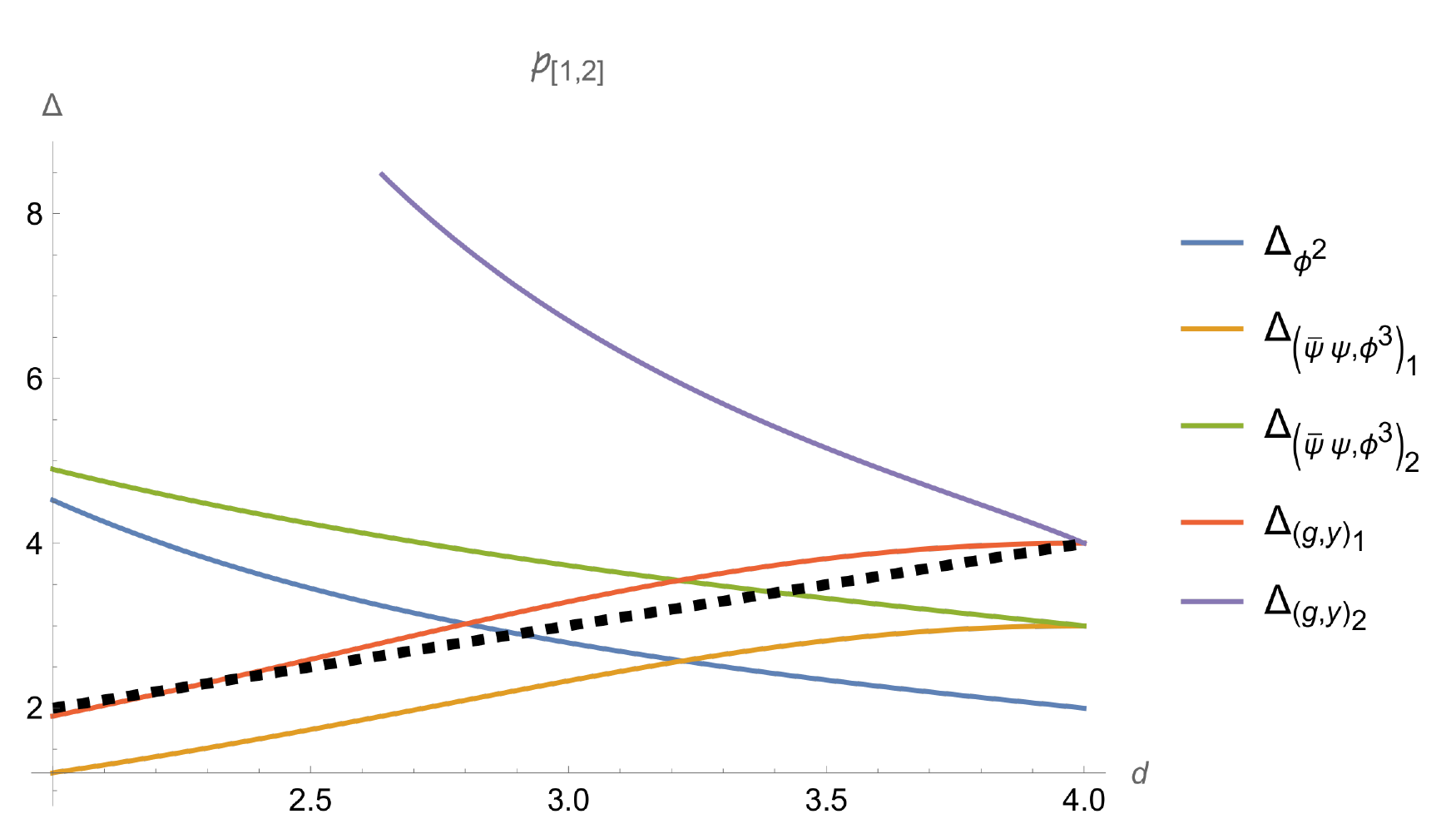}
    \caption{$\text{Pad\'e}_{[1,2]}$ approximates of the scaling dimensions of composite operators, as a function of  $d$, for the supersymmetric fixed point of the $SU(3)$ adjoint theory. The dashed line is $\Delta=d$, and separates relevant from irrelevant operators. We see a crossing of operators with the same parity near $d=2.8$.}
    \label{figure: pade [1,2] SU(3) scaling-dimension-N=3}
\end{figure}

\subsection{\texorpdfstring{$SO(3)~S_2$}{SO(3)-S2} scaling dimensions}

We plot the Pad\'e approximate of the scaling dimensions of composite operators as a function of $d$ in Figure \ref{figure: pade [1,2] SO(3) s2 scaling-dimension-N=3}. Near $d=2.5$, a relevant operator becomes irrelevant.

At the $[susy]$ fixed point of $SO(3) - S_2$ model, various possible Pad\'e approximates of scaling dimensions for different operators are given in Table \ref{table: SO(3) pade-scaling-[susy]-d=3}.

\begin{table}[H]\centering
    \begin{tabular}{|m{2 cm}|m{2cm}|m{1.7cm}|m{1.5cm}|m{1.6cm}|m{1.6cm}|m{1.5cm}|m{1.5cm}|}\hline
    \vspace{0.2cm} & $\Delta_{\phi}$ & $\Delta_{\psi}$ & $\Delta_{\phi^{2}}$ & $\Delta_{\left(\bar\psi\psi,\phi^{3}\right)_{1}}$ & $\Delta_{\left(\bar\psi\psi,\phi^{3}\right)_{2}}$ & $\Delta_{\left(g,y\right)_{1}}$ & $\Delta_{\left(g,y\right)_{2}}$ \\
    \hline
    \hline
    $\text{Pad\'e}_{[1,1]}$ & 0.739 & 1.239 & 2.193 & 3. & 3.193 & 4. & 4.89 \\
    $\text{Pad\'e}_{[1,2]}$ & 0.768 & 1.262 & 2.334 & 2.721 & 3.325 & 3.714 & 5.881 \\
    $\text{Pad\'e}_{[2,1]}$ & 0.733 & 1.233 & 2.308 & 2.692 & 3.308 & 3.692 & 5.568 \\
    $\text{Pad\'e}_{[0,3]}$ & 0.784 & 1.291 & 11.7 & 2.817 & 6.515 & 3.814 & -1.106 \\
    $\text{Pad\'e}_{[3,0]}$ & 0.811 & 1.311 & 3.55 & 2.805 & 4.55 & 3.805 & 18.15 \\
    \hline
    \end{tabular}\caption{Pad\'e Approximates of scaling dimensions of various operators in $d=3$ for the supersymmetric fixed point of $SO(3)$ symmetric traceless theory.\label{table: SO(3) pade-scaling-[susy]-d=3}}
\end{table}

\begin{figure}[H]
    \centering
    \includegraphics[width= 15 cm]{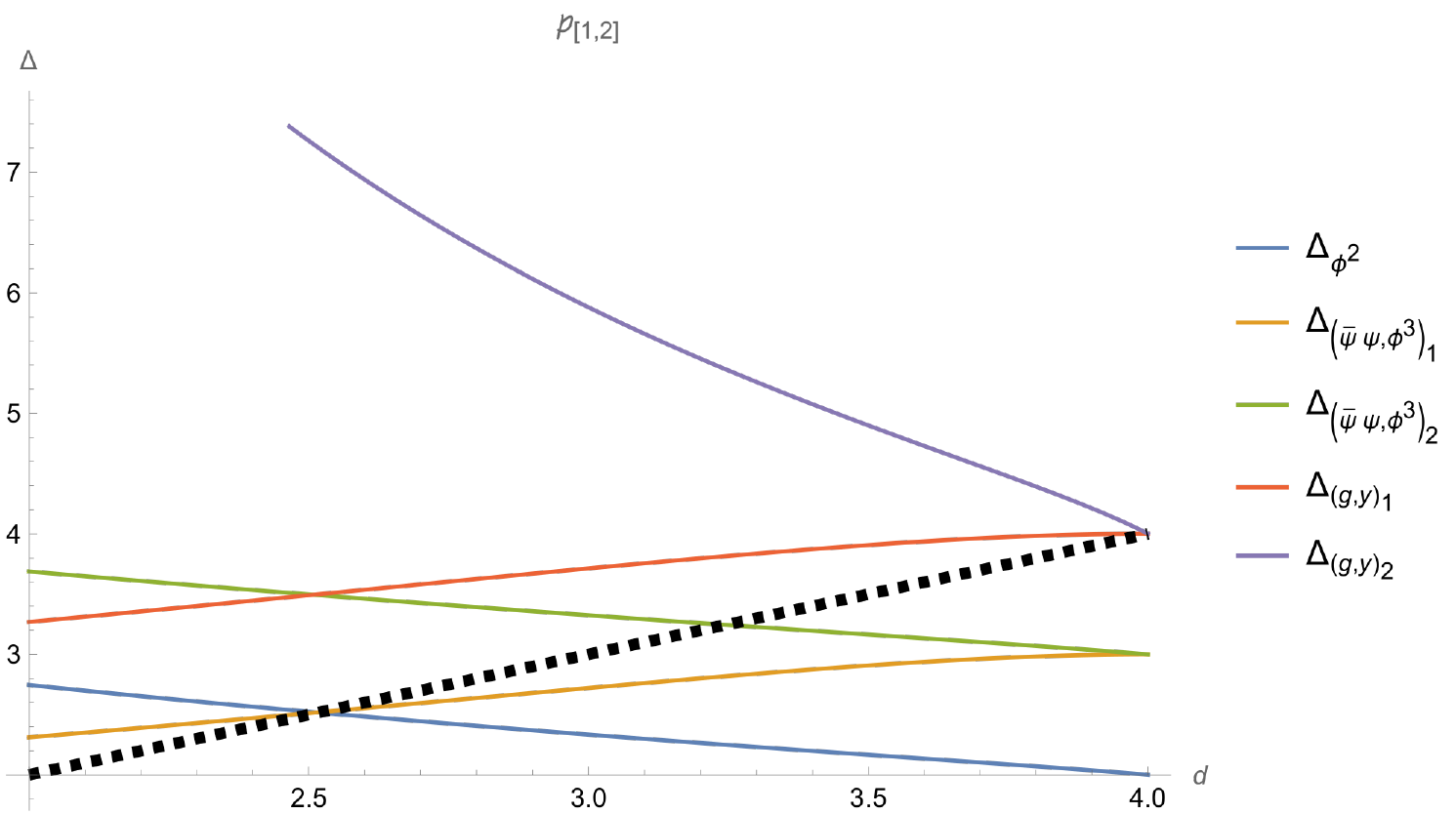}
    \caption{$\text{Pad\'e}_{[1,2]}$ approximates of the scaling dimension of composite operators, as a function of $d$, for the supersymmetric fixed point of the $SO(3)$ $S_2$ theory. The dashed line is $\Delta=d$.}
    \label{figure: pade [1,2] SO(3) s2 scaling-dimension-N=3}
\end{figure}
\subsection{\texorpdfstring{$SO(3)$ $S_2-A_2$}{SO(3) S2 - A2} scaling dimensions}
We plot the Pad\'e approximate of the scaling dimensions of composite operators in the stable $[ns_+]$ fixed point of the $SO(3)$ $S_2$-$A_2$ model as a function of $d$ in Figure \ref{figure: pade [1,2] SO(3) s2-a2 scaling-dimension-N=3}. There, we see that a relevant operator becomes irrelevant near $d=3$, but there is no other sign of operator crossing. Various possible Pad\`e approximates of scaling dimensions for different operators in $d=3$ at this fixed point are given in Table \ref{table: S2-A2 pade-scaling-[ns+]-d=3}.

\begin{table}[H]\centering
    \begin{tabular}{|m{2 cm}|m{2cm}|m{1.7cm}|m{1.5cm}|m{1.6cm}|m{1.6cm}|m{1.5cm}|m{1.5cm}|}\hline
    \vspace{0.2cm} & $\Delta_{\phi}$ & $\Delta_{\psi}$ & $\Delta_{\phi^{2}}$ & $\Delta_{\left(\bar\psi\psi,\phi^{3}\right)_{1}}$ & $\Delta_{\left(\bar\psi\psi,\phi^{3}\right)_{2}}$ & $\Delta_{\left(g,y\right)_{1}}$ & $\Delta_{\left(g,y\right)_{2}}$ \\
    \hline
    \hline
    $\text{Pad\'e}_{[1,1]}$ & 0.603 & 1.189 & 2.095 & 4.665 & 3. & 4.727 & 5.054 \\
    $\text{Pad\'e}_{[1,2]}$ & 0.547 & 1.933 & 1.903 & 2.24 & 3.02 & 5.009 & 5.75 \\
    $\text{Pad\'e}_{[2,1]}$ & 0.6 & 1.174 & 1.899 & 2.123 & 3.02 & 4.907 & 5.534 \\
    $\text{Pad\'e}_{[0,3]}$ & 0.604 & 1.161 & 2.065 & 2.497 & 2.457 & 7.153 & -4.611 \\
    $\text{Pad\'e}_{[3,0]}$ & 0.566 & 1.142 & 2.073 & 2.816 & 2.337 & 5.763 & 9.701 \\
    \hline
    \end{tabular}\caption{Pad\'e approximates of scaling dimensions of various operators in $d=3$ for the $[ns_+]$ fixed point of $S_2-A_2$ theory.\label{table: S2-A2 pade-scaling-[ns+]-d=3}}
\end{table}

\begin{figure}[H]
    \centering
    \includegraphics[width= 15 cm]{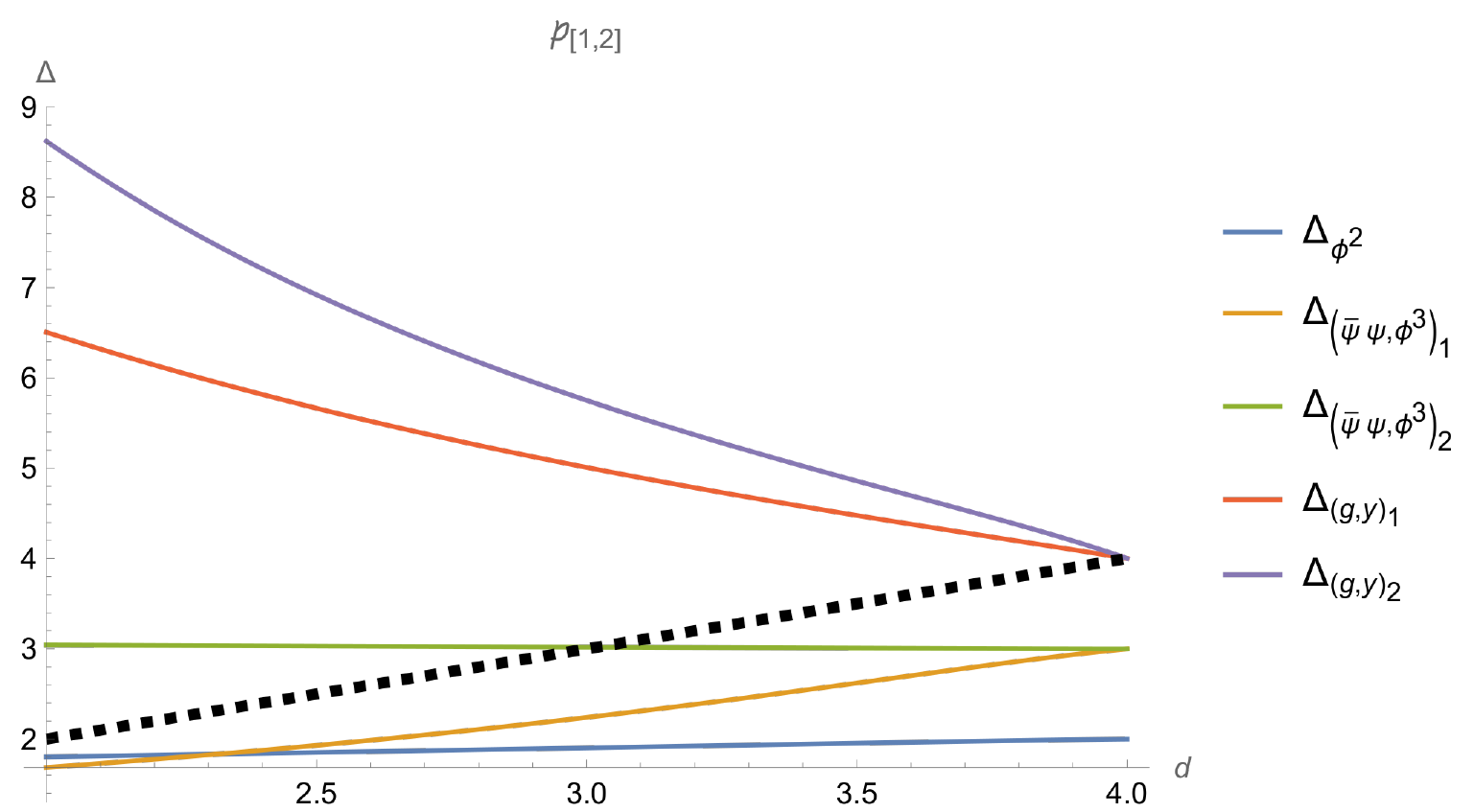}
    \caption{$\text{Pad\'e}_{[1,2]}$ approximates of the scaling dimension of composite operators, as a function of $d$, for the $[ns_+]$ fixed point of the $SO(3)$ $S_2-A_2$ theory. The dashed line is $\Delta=d$. One of the operators becomes irrelevant near $d=3$. }
    \label{figure: pade [1,2] SO(3) s2-a2 scaling-dimension-N=3}
\end{figure}
\section{\label{sec:discussion}Discussion}
In this paper, we studied interacting  infrared fixed points of Gross-Neveu Yukawa models in $d=4-\varepsilon$ that possess a large $N$ limit dominated by planar diagrams. In particular, we determined a stable interacting large-$N$ fixed point, which can be obtained from three different theories whose observables are equivalent in common sectors in the large-$N$ limit -- the $SU(N)$ adjoint theory, the  $SO(N)$ $S_2$ theory, and the $SO(N)$ $S_2$-$A_2$ theory. The $SU(N)$ adjoint and $SO(N)$ $S_2$ fixed points possess emergent $\mathcal N=1$ supersymmetry in $d=3$. The fixed point of the $SO(N)$ $S_2$-$A_2$ theory is not supersymmetric, but all correlation functions of single-trace operators (involving a finite number of fields) in the theory obey the constraints of supersymmetry in the large $N$ limit. We identified three other large-$N$ non-supersymmetric fixed points, that may define conformal field theories in $d=3$, although they are unphysical in $d=4$, due to the classical instability of the classical scalar potential. The supersymmetric fixed point arises when one considers a GNY model defined using a single matrix-valued two-component Majorona fermion; modifying the theory to contain $2N_f$ flavours of matrix-valued two-component Majorana fermions, with an additional $O(2N_f)$ flavour symmetry, leads to a stable, interacting, non-supersymmetric fixed point, with a positive definite classical scalar potential. 

We also studied fixed points at finite $N$. When $N$ is sufficiently small but greater than $3$, the supersymmetric fixed points of the adjoint and $S_2$ models acquire one unstable direction, and one of the non-supersymmetric fixed points becomes stable. For finite $N$, some of the non-supersymmetric fixed points give rise to scalar potentials which are positive definite. It is interesting that the $SO(N)$ $S_2$-$A_2$ model is manifestly non-supersymmetric at finite $N$, as it possesses unequal bosonic and fermionic degrees of freedom. Yet, for all $N>7.1$, it possesses a stable fixed point, which, although non-supersymmetric, exhibits many features of a supersymmetric fixed point in the large-$N$ limit, in the sense that observables such as scaling dimensions of certain operators (including all those we computed in this paper) obey the constraints implied by $\mathcal N=1$ supersymmetry up to $1/N$ corrections. However, it is important to note that, even in the large $N$ limit, the fixed point is not supersymmetric -- as is clear from the fact that, certain single-trace operators such as $\Tr (\psi \phi)$ present in the supersymmetric fixed points, vanish in the stable fixed point of the $S_2$-$A_2$ theory.

It is interesting to interpret the large $N$ equivalence in a holographic language. Fixed-points of the common large-$N$ limit of the two supersymmetric theories may be interpreted as classical string theories in $AdS_4$ (with $\mathcal N=1$ supersymmetry for the $N_f=1/2$ $[susy]$ fixed point), which exist for particular finite values of string tension. This string theory can be quantized in two inequivalent ways, corresponding to the finite-$N$ $SU(N)$ adjoint and $SO(N)$ $S_2$ GNY models. The holographic dual of fixed points of the large-$N$ $SO(N)$ $S_2-A_2$ model, correspond to non-supersymmetric string theories, with an unequal number of bosonic and fermionic states at any given mass, which, for many, but not all, states, obey the constraints of $\mathcal N=1$ supersymmetry.


At large-$N$, the IR fixed point defined by the $O(N)$ GNY models in $d=3$, can also be studied in $d=2+\varepsilon$ via the Gross-Neveu model \cite{Gross:1974jv}, which allows for two-sided Pad\'e approximations \cite{Fei:2016sgs}. However, the $N=1$ supersymmetric fixed point of the GNY model is instead believed to extend to the tricritical Ising model when $d=2$, which also exhibits emergent supersymmetry \cite{Fei:2016sgs}; and candidates for the non-supersymmetric $N=1$ fixed point in $d=2$ have also been identified \cite{Nakayama:2022svf}. Do the large-$N$ fixed points studied here extend to known CFTs in $d=2$, which also exhibit emergent supersymmetry? If one can identify such CFTs, it would provide additional information to constrain the Pad\'e approximations used to study the IR fixed points in $d=3$. 

While constructions based on multi-scalar theories do not appear to give rise to interacting large-$N$ fixed points dominated by planar diagrams, our findings suggest multi-component GNY models allow for a landscape of such fixed points, with and without supersymmetry. Pad\'e approximations of scaling dimensions based on three-loop computations suggest that any dual description of these interacting large-$N$ CFTs in $d=3$ should contain higher spin fields of finite mass, and thus correspond to string theories of finite tension, similar to, e.g., the SYK model \cite{Sachdev:1992fk, KitaevTalk} and theories \cite{Chang:2021fmd, Chang:2021wbx, Prakash:2022gvb} which attempt to generalize it to $d=3$. However, this expectation does not hold when these theories are deformed by a Chern-Simons gauge field, leaving open the possibility of a landscape of theories with potentially interesting holographic duals, such as $\mathcal N=1$ supergravity, when the Chern-Simons gauge field is taken to strong coupling -- although, like the conjectures in \cite{ Banerjee:2013nca, Gurucharan:2014cva, GuruCharan:2017ftx}, studying these theories directly appears difficult. It would be interesting to ask if one can engineer more sophisticated fixed points from multi-component Gross-Neveu Yukawa theories with some of the properties expected for the putative CFT's dual to various constructions of $AdS_4$ vacuua in string theory\footnote{We thank Anshuman Maharana for discussions on this point.}, as identified in, e.g., \cite{Aharony:2008wz, deAlwis:2014wia, Conlon:2018vov, Conlon:2020wmc, Conlon:2021cjk, Apers:2022tfm, Ning:2022zqx, Apers:2022vfp}.

\section*{Acknowlegements}

The authors thank D. Gaiotto, Z. Komargodski, H. Osborn, and  A. Stergiou for discussions and comments on  earlier drafts of this manuscript. SP also acknowledges the Harish Chandra-Research Institute for hospitality when part of this work was completed, and thanks A. Maharana, D. Jatkar, and others present for useful discussions and comments when preliminary results were presented in a seminar in December, 2022. SS thanks K.P. Yogendran for the discussions. The work of SP was supported in part by a DST-SERB grant (CRG/2021/009137).

\appendix

\section{Three-loop \texorpdfstring{$\beta$}{beta}-functions}
In this, appendix we present three-loop expressions for the $\beta$-functions in the large-$N$ limit. We present the large-$N$ $\beta$-functions for the adjoint model; the large-$N$ $\beta$-functions for the other two models can be obtained by rescaling the coupling constants as described in the main text. 

Three loop results here were obtained using the general $\beta$-functions in \cite{Jack:2023zjt}. To obtain the $\beta$-functions at finite $N$ from \cite{Jack:2023zjt}, one must make the following replacements for $y^a$ and $\lambda^{abcd}$. Let us denote the entries of the matrix $y^a$ of \cite{Jack:2023zjt} as $(y^{a})^{ij}$. 
\begin{align}
    (y^{a})^{ij} &=  \frac{y}{2} \bigg[\tr({\mathcal T}^a {\mathcal T'}^i {\mathcal T'}^j)+ \tr({\mathcal T}^a {\mathcal T'}^j {\mathcal T'}^i)\bigg], 
\end{align}
and 
\begin{align}
    \lambda^{abcd} &= 4 g_1 \bigg[\tr({\mathcal T}^a{\mathcal T}^b{\mathcal T}^c{\mathcal T}^d) + \tr({\mathcal T}^a{\mathcal T}^b{\mathcal T}^d{\mathcal T}^c) + \tr({\mathcal T}^a{\mathcal T}^c{\mathcal T}^b{\mathcal T}^d) + \tr({\mathcal T}^a{\mathcal T}^c{\mathcal T}^d{\mathcal T}^b)\nn\\
    & \qquad  + \tr(\mathcal T^a{\mathcal T}^d{\mathcal T}^b{\mathcal T}^c) + \tr({\mathcal T}^a{\mathcal T}^d{\mathcal T}^c{\mathcal T}^b)\bigg] + 8 g_2 \bigg[\tr({\mathcal T}^a{\mathcal T}^b)\tr({\mathcal T}^c{\mathcal T}^d) \nn\\
& \quad +\tr({\mathcal T}^a{\mathcal T}^d)\tr({\mathcal T}^b{\mathcal T}^c)+\tr({\mathcal T}^a{\mathcal T}^c)\tr({\mathcal T}^b{\mathcal T}^d)\bigg].
\end{align}
Note that, with these replacements, the constraint $N_f=1/2$ is built-in to the results of \cite{Jack:2023zjt}. We also computed two-loop results computed using RGBeta \cite{Thomsen:2021ncy}, for the model $SU(N)\times O(N_f)$, which agree with results computed using \cite{Jack:2023zjt} when we substitute $N_f=1/2$. 

Three-loop expressions for anomalous dimensions of composite operators were obtained from results in \cite{Jack:2023zjt} using the dummy field method  \cite{Martin:1993zk,Luo:2002ti,Schienbein:2018fsw}. The three-loop $\beta$-functions at finite $N$ for the three theories we study in this paper span tens of pages, so we do not reproduce them here. They are available from the authors in electronic form on request.

We write the $\beta$-function for any coupling constant $\lambda$ in the loop expansion \eqref{eq: beta-loop-exp}, where $l$ denotes the number of loops, as follows,
\be\label{eq: beta-loop-exp}
\beta_{\lambda} = \sum_{l=0}^{\infty} \left(4\,\pi\right)^{-2l}\beta^{(l)}_{\lambda},
\ee
We also use a similar notation to present the scaling dimension at the fixed point in terms of $\varepsilon$ -- expansion up to three orders. 

The relationship between scaling and anomalous dimensions for various operators in terms of space-time dimension $d$ is given by
\[
\Delta_{\phi} = \frac{d- 2}{2} + \gamma_{\phi},\quad \Delta_{\psi} = \frac{d- 1}{2} + \gamma_{\psi}, \quad  \Delta_{\phi^2} = 2 \left(\frac{d - 2}{2}\right) + \gamma_{\phi^2} 
\]
\[
\Delta_{\left(\Bar{\psi}\psi,\phi^3\right)} = 3 + \gamma_{\left(\Bar{\psi}\psi,\phi^3\right)},\quad \Delta_{g} = d + \gamma_{g},
\]
where the anomalous dimension of any operator $\mathcal{O}$ can also be written in loop expansion, using the same conventions as the loop expansion of the $\beta$-function,
\be
\gamma_{\mathcal{O}} = \sum_{l=0}^{\infty} \left(4\pi\right)^{-2l}\gamma^{(l)}_{\mathcal{O}}.
\ee

For the anomalous dimension for field $\phi$ and $\psi$, and for the scalar mass operator , $\gamma^{(0)}_{\mathcal{O}} = 0$.

The expression for the $\beta$-functions for the classically marginal 't Hooft couplings at two loops is,
\begin{align}
    \beta^{(2)}_{\lambda_{1}} &=  \frac{1}{2048} (-12288 \lambda_{1}^3 - 1024 \lambda_{1}^2 \lambda_{y}^2 N_f
     - 64 \lambda_{1} \lambda_{y}^4 N_f
    + 5 \lambda_{y}^6 N_f)\\
    \beta^{(2)}_{\lambda_{2}} &= \frac{1}{2048} (-49152 \lambda_{1}^3 - 20480 \lambda_{1}^2 \lambda_{2} - 1536 \lambda_{1}^2 \lambda_{y}^2 N_f - 2048 \lambda_{1} \lambda_{2} \lambda_{y}^2 N_f\nonumber \\
    &\quad - 512 \lambda_{2}^2 \lambda_{y}^2 N_f + 96 \lambda_{1} \lambda_{y}^4 N_f
    - 32 \lambda_{2} \lambda_{y}^4 N_f
    + 15 \lambda_{y}^6 N_f).\\
    \beta^{(2)}_{\lambda_{y}} &=  \frac{1}{1024} \lambda_{y} \bigl(512 \lambda_{1}^2
    - 64 \lambda_{1} \lambda_{y}^2
    + \lambda_{y}^4 (-7
    - 18 N_f)\bigr).
\end{align}
At three loops, (with $N_f=1/2$),
\begin{align}
    \beta^{(3)}_{\lambda_1} &= \frac{1}{262144} (3407872 \lambda_{1}^4 + 69632 \lambda_{1}^3 \lambda_{y}^2 + 27136 \lambda_{1}^2 \lambda_{y}^4 - 976 \lambda_{1} \lambda_{y}^6 - 33 \lambda_{y}^8)\\
    \beta^{(3)}_{\lambda_2} &= \frac{1}{262144} \Bigl(131072 \lambda_{1}^2 \bigl((193 + 96 \zeta_{3}) \lambda_{1}^2 + 120 \lambda_{1} \lambda_{2} + 3 \lambda_{2}^2\bigr) + 12288 \lambda_{1}^2 (32 \lambda_{1}\nonumber \\
    & \quad + 11 \lambda_{2}) \lambda_{y}^2 + 1536 \bigl(8 (5 + 3 \zeta_{3}) \lambda_{1}^2 + (23 + 16 \zeta_{3}) \lambda_{1} \lambda_{2} + 2 (3 + \zeta_{3}) \lambda_{2}^2\bigr) \lambda_{y}^4\nonumber \\
    & \quad - 16 \bigl(8 (55 + 18 \zeta_{3}) \lambda_{1} + (157 - 24 \zeta_{3}) \lambda_{2}\bigr) \lambda_{y}^6 -  (151 + 84 \zeta_{3}) \lambda_{y}^8\Bigr)\\
    \beta^{(3)}_{\lambda_y} &= \frac{1}{65536} \lambda_{y} (-32768 \lambda_{1}^3 - 9984 \lambda_{1}^2 \lambda_{y}^2 + 2016 \lambda_{1} \lambda_{y}^4 + 83 \lambda_{y}^6).
\end{align}
The $\beta$-functions for relevant couplings at one loop are,
\begin{align}
    \beta^{(1)}_{\chi} &= \frac{1}{2} \left(36 \lambda_{\alpha}
   ^2-\lambda_{y}^2 N_{f}  \left(3 \rho
   ^2-\chi \right)+16 (2 \lambda_{1}+\lambda_{2}) \chi \right)\\
   \beta^{(1)}_{\rho} &=  \frac{3 \lambda_{y}^2 \rho }{16}, \quad 
   \beta^{(1)}_{\lambda_\alpha} = \frac{1}{16} \left(96 \lambda_{1} \lambda_{\alpha} +\lambda_{y}^2 N_{f}  (3 \lambda_{\alpha} -\lambda_{y} \rho )\right).
\end{align}
At two loops,
\begin{align}
    \beta^{(2)}_{\chi} &= \frac{1}{128} \Bigl(-16 \lambda_{2} \bigl(36 \lambda_{\alpha}^2 -  \lambda_{y}^2 N_{f} (\rho^2 - 2 \chi)\bigr) + 16 \lambda_{1} \bigl(-216 \lambda_{\alpha}^2\nonumber \\
    & \quad  + 2 \lambda_{y}^2 N_{f} (\rho^2 - 2 \chi)\bigr)+ \lambda_{y}^2 N_{f} \bigl(-72 \lambda_{\alpha}^2 + 12 \lambda_{y} \lambda_{\alpha} \rho + \lambda_{y}^2 (19 \rho^2\nn\\
    &\quad -  \chi)\bigr) - 640 \lambda_{1}^2 \chi \Bigr).\\
    \beta^{(2)}_{\rho} &= \frac{1}{1024} (-48 \lambda_{y}^3 \lambda_{\alpha} - 13 \lambda_{y}^4 \rho - 14 \lambda_{y}^4 N_{f} \rho)\\
    \beta^{(2)}_{\lambda_\alpha} &= \frac{1}{512} (-8448 \lambda_{1}^2 \lambda_{\alpha} - 384 \lambda_{1} \lambda_{y}^2 \lambda_{\alpha} N_{f} - 12 \lambda_{y}^4 \lambda_{\alpha} N_{f} + 32 \lambda_{1} \lambda_{y}^3 N_{f} \rho \nn\\
    &\quad+ 11 \lambda_{y}^5 N_{f} \rho).
\end{align}
At three loops, (with $N_f=1/2$),
\begin{align}
    \beta^{(3)}_{\chi} &= \frac{1}{32768}\bigg[983040 \lambda_1^3 \chi
     + 768 \lambda_1^2 \bigl(48 (103
     + 48 \zeta_{3}) \lambda_{\alpha}^2
     + 64 \lambda_2 \chi
     + \lambda_{y}^2 (32 \rho^2
     + 11 \chi)\bigr)\nonumber \\
    & \quad -  \lambda_{y}^4 \Bigl(-4608 \lambda_{\alpha}^2
     + 144 (19
     + 6 \zeta_{3}) \lambda_{y} \lambda_{\alpha} \rho
     + \lambda_{y}^2 \bigl((338
     + 240 \zeta_{3}) \rho^2
     + (157\nn \\
    & \quad - 24 \zeta_{3}) \chi \bigr)\Bigr)
     + 384 \lambda_2 \lambda_{y}^2 \Bigl(24 \lambda_{\alpha}^2
     + 4 (-1
     + 3 \zeta_{3}) \lambda_{y} \lambda_{\alpha} \rho
     + \lambda_{y}^2 \bigl(2 (-4
     + \zeta_{3}) \rho^2
     \nn \\
    & \quad + (3+ \zeta_{3}) \chi \bigr)\Bigr)
     + 96 \lambda_1 \biggl(10752 \lambda_2 \lambda_{\alpha}^2
     + \lambda_{y}^2 \Bigl(576 \lambda_{\alpha}^2
     + 64 (1
     + 3 \zeta_{3}) \lambda_{y} \lambda_{\alpha} \rho \nn \\
    & \quad + \lambda_{y}^2 \bigl(-2 (37
     + 8 \zeta_{3}) \rho^2
     + (23
     + 16 \zeta_{3}) \chi \bigr)\Bigr)\biggr)\bigg]\\
    \beta^{(3)}_{\rho} &= \frac{\lambda_{y}^2}{65536}  \Bigl(96 \lambda_{y} (80 \lambda_{1} + 11 \lambda_{y}^2) \lambda_{\alpha} - 2 \bigl(7424 \lambda_{1}^2 - 640 \lambda_{1} \lambda_{y}^2  + (33 - 96 \zeta_{3}) \lambda_{y}^4\bigr) \rho \Bigr)\\
    \beta^{(3)}_{\alpha} &= \frac{1}{131072} \Bigl(6 \bigl(32768 (38 + 8 \zeta_{3}) \lambda_{1}^3 + 20736 \lambda_{1}^2 \lambda_{y}^2 + 3232 \lambda_{1} \lambda_{y}^4\nonumber \\
    & \quad + (-81- 64 \zeta_{3}) \lambda_{y}^6\bigr) \lambda_{\alpha} + 16 \lambda_{y}^3 \bigl(512 (1 + 3 \zeta_{3}) \lambda_{1}^2 + 4 (-99\nonumber \\
    &\quad  + 12 \zeta_{3}) \lambda_{1} \lambda_{y}^2 - 3 (3 + \zeta_{3}) \lambda_{y}^4\bigr) \rho \Bigr).
\end{align}
Anomalous dimension for the fields $\phi$ and $\psi$ at one and two-loop are,
\be 
    \gamma^{(1)}_\phi = \frac{\lambda_y^2 N_f}{16},\quad\gamma^{(1)}_{\psi} = \frac{\lambda_y^2}{32},\quad 
    \gamma^{(2)}_{\phi} = \frac{64\lambda_1^2 - \lambda_y^4 N_f}{128},\quad \gamma^{(2)}_{\psi} =  - \frac{\lambda_y^4 (1 + 6 N_f)}{2048}. 
\ee
At three loops, with $N_f=1/2$,
\begin{align}
    \gamma^{(3)}_{\phi} &= \frac{1}{65536} (-32768 \lambda_{1}^3 - 3840 \lambda_{1}^2 \lambda_{y}^2 + 160 \lambda_{1} \lambda_{y}^4 + 31 \lambda_{y}^6)
    \\
    \gamma^{(3)}_{\psi} &= \frac{\lambda_{y}^2}{65536}(-2816 \lambda_{1}^2 + 128 \lambda_{1} \lambda_{y}^2 + 21 \lambda_{y}^4).
\end{align}
The anomalous dimension for the scalar mass operator is,
\begin{align}
    \gamma^{(1)}_{\phi^2} &= 4 \lambda_1 + 2 \lambda_2 + \frac{1}{8} \lambda_{y}^2 N_{f}
    \\
    \gamma^{(2)}_{\phi^2} &= -5 \lambda_1^2
     -  \frac{1}{2} \lambda_1 \lambda_{y}^2 N_{f}
     -  \frac{1}{128} \lambda_{y}^2 (32 \lambda_2
     + \lambda_{y}^2) N_{f}
    \\
    \gamma^{(3)}_{\phi^2} &= \frac{1}{32768} \Bigl(983040 \lambda_1^3
     + 96 (23
     + 16 \zeta_{3}) \lambda_1 \lambda_{y}^4
     + 768 \lambda_1^2 (64 \lambda_2
     + 11 \lambda_{y}^2)
    \nonumber\\
    & \quad  + \lambda_{y}^4 \bigl(384 (3+ \zeta_{3}) \lambda_2
     + (-157
     + 24 \zeta_{3}) \lambda_{y}^2\bigr)\Bigr),
\end{align}
where $N_f=1/2$ for the three-loop result.
The anomalous dimension matrix for mixtures of the fermion mass and scalar tri-linear operators, are, up to one-loop,
\begin{equation}
\gamma^{(0)}_{\left(\Bar{\psi}\psi,\phi^3\right)} = \left(\begin{array}{cc} - \varepsilon &  0  \\ 0 & -\frac{3 \varepsilon}{2} \\\end{array}\right) , \quad
\gamma^{(1)}_{\left(\Bar{\psi}\psi,\phi^3\right)} = \left(\begin{array}{cc} \frac{3 \lambda_{y}^2}{16} &   -\frac{\lambda_{y}^3 N_{f}}{16}    \\ 0 & 6 \lambda_{1}+\frac{3   \lambda_{y}^2 N_{f} }{16} \\\end{array}\right).
\end{equation}
At two loops,
\begin{equation}
\gamma^{(2)}_{\left(\Bar{\psi}\psi,\phi^3\right)} = \left(
\begin{array}{cc}
 -\frac{\lambda_{y}^4 (13+14
   N_{f} )}{1024} & \frac{\lambda_{y}^3 \left(32
   \lambda_{1}+11
   \lambda_{y}^2\right) N_{f}}{512} 
   \\
 -\frac{3
   \lambda_{y}^3}{64} &
   \frac{-3 \left(704
   \lambda_{1}^2+32
   \lambda_{1} \lambda_{y}^2 N_{f} +\lambda_{y}^4
   N_{f} \right)}{128}  \\
\end{array}
\right).
\end{equation}
At three loops, with $N_f=1/2$,
\begin{equation}
\resizebox{\linewidth}{!}{%
    $\gamma^{(3)}_{\left(\Bar{\psi}\psi,\phi^3\right)} = \left(
\begin{array}{cc}
 -\frac{\lambda_{y}^2
   \left(7424 \lambda_{1}^2-640 \lambda_{1}
   \lambda_{y}^2+(33-96
   \zeta_3) \lambda_{y}^4\right)}{32768} &
   \frac{\lambda_{y}^3
   \left(512 (1+3 \zeta_3)
   \lambda_{1}^2+12 (-33+4
   \zeta_3) \lambda_{1} \lambda_{y}^2-3
   (3+\zeta_3)
   \lambda_{y}^4\right)}{8192} \\
 \frac{3 \lambda_{y}^3
   \left(80 \lambda_{1}+11
   \lambda_{y}^2\right)}{2048} & \frac{3
   \left(65536 (19+4 \zeta_3) \lambda_{1}^3+20736
   \lambda_{1}^2
   \lambda_{y}^2+3232
   \lambda_{1} \lambda_{y}^4-(81+64 \zeta_3)
   \lambda_{y}^6\right)}{65536} \\
\end{array}
\right)$}.
\end{equation}
The anomalous dimension of $\lambda_2$ is,
\begin{align}
   \gamma^{(0)}_{\lambda_{2}} &= -\varepsilon,\quad \gamma^{(1)}_{\lambda_{2}} =  8 \lambda_{1}+4 \lambda_{2}+\frac{\lambda_{y}^2
   N_{f} }{4}\\
   \gamma^{(2)}_{\lambda_{2}} &= -10 \lambda_{1}^2-\lambda_{1} \lambda_{y}^2 N_{f} -\frac{1}{64} \lambda_{y}^2 \left(32 \lambda_{2}+\lambda_{y}^2\right) N_{f} \\
   \gamma^{(3)}_{\lambda_{2}} &= \frac{1}{16384} \bigg[983040 \lambda_{1}^3+96 (23+16 \zeta_{3}) \lambda_{1} \lambda_{y}^4+384 (3+\zeta_{3}) \lambda_{2} \lambda_{y}^4 \nn\\
   & \quad +(-157+24 \zeta_{3}) \lambda_{y}^6+768 \lambda_{1}^2 \left(64 \lambda_{2}+11 \lambda_{y}^2\right)\bigg],
\end{align}
with $N_f=1/2$ at three loops.

The anomalous dimension matrix for the mixture of marginal operators corresponding to $\lambda_1$ and $\lambda_y$, is, up to one loop,
\be
\gamma^{(0)}_{\left(\lambda_1, \lambda_y\right)} = \left(
\begin{array}{cc}
 - \varepsilon
   & 0 \\
 0 & \frac{- \varepsilon}{2} \\
\end{array}
\right) , \quad \gamma^{(1)}_{\left(\lambda_{1},\lambda_{y}\right)} = \left(
\begin{array}{cc}
 8 \lambda_{1}+\frac{\lambda_{y}^2 N_{f} }{4} & 0 \\
 -\frac{\lambda_{y} \left(-16 \lambda_{1}+\lambda_{y}^2\right) N_{f}}{32}   & \quad \frac{3 \lambda_{y}^2 (2+N_{f} )}{16}  \\
\end{array}
\right).
\ee
At two loops,
\be
\gamma^{(2)}_{\left(\lambda_{1},\lambda_{y}\right)} = \left(
\begin{array}{cc}
 -18 \lambda_{1}^2-\lambda_{1} \lambda_{y}^2 N_{f}
   -\frac{\lambda_{y}^4 N_{f} }{32} & \lambda_{1}
   \lambda_{y}-\frac{\lambda_{y}^3}{16} \\
 -\lambda_{1}^2 \lambda_{y} N_{f} -\frac{1}{8}
   \lambda_{1} \lambda_{y}^3 N_{f} +\frac{15
   \lambda_{y}^5 N_{f} }{1024} & \frac{\lambda_{1}^2}{2}-\frac{3 \lambda_{1} \lambda_{y}^2}{16}-\frac{5 \lambda_{y}^4 (7+18 N_{f} )}{1024} \\
\end{array}
\right).
\ee
At three loops,
\be
\gamma^{(3)}_{\left(\lambda_{1},\lambda_{y}\right)} = \left(
\begin{array}{cc}
 52 \lambda_{1}^3+\frac{51 \lambda_{1}^2 \lambda_{y}^2}{64}+\frac{53 \lambda_{1} \lambda_{y}^4}{256}-\frac{61 \lambda_{y}^6}{16384} & \frac{3
   \lambda_{y} \left(-1024 \lambda_{1}^2-208
   \lambda_{1} \lambda_{y}^2+21 \lambda_{y}^4\right)}{2048} \\
 \frac{17408 \lambda_{1}^3 \lambda_{y}+13568
   \lambda_{1}^2 \lambda_{y}^3-732 \lambda_{1}
   \lambda_{y}^5-33 \lambda_{y}^7}{32768} & \frac{-32768
   \lambda_{1}^3-29952 \lambda_{1}^2 \lambda_{y}^2+10080 \lambda_{1} \lambda_{y}^4+581
   \lambda_{y}^6}{65536} \\
\end{array}
\right),
\ee
with $N_f=1/2$.

The anomalous dimension matrix of the quartic and Yukawa coupling, up to one loop
\[
\gamma^{(0)}_{\left(g,y\right)} = \left(
\begin{array}{cc}
 - \varepsilon
   & 0 \\
 0 & \frac{- \varepsilon}{2} \\
\end{array}
\right) , \quad \gamma^{(1)}_{\left(g,y\right)} = \left(
\begin{array}{cc}
 \frac{147456 g+960 y^2 N_{f} }{2304}
   & 0 \\
 \frac{1920 g y N_{f} -68 y^3 N_{f}
   }{2304} & \frac{1}{16} y^2 (2+5
   N_{f} ) \\
\end{array}
\right),
\]
and at two loops,
\[
\gamma^{(2)}_{\left(g,y\right)} = \left(
\begin{array}{cc}
 -1368 g^2-\frac{40}{3} g y^2 N_{f}
   -\frac{7 y^4 N_{f} }{144} & 20 g
   y-\frac{17 y^3}{24} \\
 \frac{1}{3} (-40) g^2 y N_{f}
   -\frac{7}{36} g y^3 N_{f}
   +\frac{193 y^5 N_{f} }{18432} & \quad 10
   g^2-\frac{17 g y^2}{8}-\frac{35
   y^4 (-7+30 N_{f} )}{9216} \\
\end{array}
\right).
\]
The three-loop expressions is rather unwieldy so we do not include it here.

\section{Classical moduli space}
\label{moduli-space}
Here we explicitly work out the moduli space of classical vacuua at the supersymmetric fixed point for $N>3$ for the $SU(N)$ theory with adjoint matter. Similar results also apply to the $SO(N)$ $S_2$ model.

For supersymmetric fixed points, the potential is 
\begin{equation}
    V(\phi) = g_1 \left(  \tr \phi^4 - \frac{1}{N} (\tr \phi^2)^2\right).
\end{equation}
When $N$ is odd, $V(\phi)=0$ implies $\phi=0$.
Let us determine the solutions to $V(\phi)=0$, when $N$ is even.

The field $\phi_{ij}$ forms an $N\times N$ traceless Hermitian matrix. The potential is invariant under any similarity transformation of $\phi$. We choose a unitary similarity transformation to make $\phi$ diagonal. This is not unique. Denote the diagonalized field as $\phi = \text{diag}(x_1,~x_2,~\ldots,~x_N)$.  The problem is thus transformed to finding the solutions to
\begin{equation}
    v(x_i) = \sum_{i=1}^{N} x_i^4 - \frac{1}{N} (\sum_{i=1}^N x_i^2)^2,
 \end{equation}
 subject to the constraint $ \sum_i x_i =0,$.
 
We use a Lagrange multiplier $\lambda$, and minimize
\begin{eqnarray}
    \tilde{V} & = & V(x_i) - 4 \lambda \sum_i x_i \\
    \partial_j \tilde{V} & = & 4 x_j^3 -  4 \langle x^2 \rangle x_j - 4 \lambda = 0. \label{lmeq}
\end{eqnarray}
From $\sum_j \partial_j \tilde{V}=0$, we have: 
\begin{equation}
    \lambda = \frac{1}{N}\sum x_j^3 \equiv \langle x^3 \rangle.
\end{equation}

If $(x_j,\lambda)$ is a solution to equation \eqref{lmeq}, then so is the rescaled solution $(\tilde{x}_i=R x_i, ~\tilde{\lambda}=R^3 \lambda)$. We pick only one representative solution from this family, by choosing $R$ such that $\langle \tilde{x}^2 \rangle=1$, and $\tilde{\lambda}>0$. Then, we must solve
\begin{equation}
 \tilde{x}_j^3 -  \tilde{x}_j - \tilde{\lambda} = 0. \label{moduli-eq} 
\end{equation}

\begin{figure}
    \centering
    \includegraphics[width=9cm]{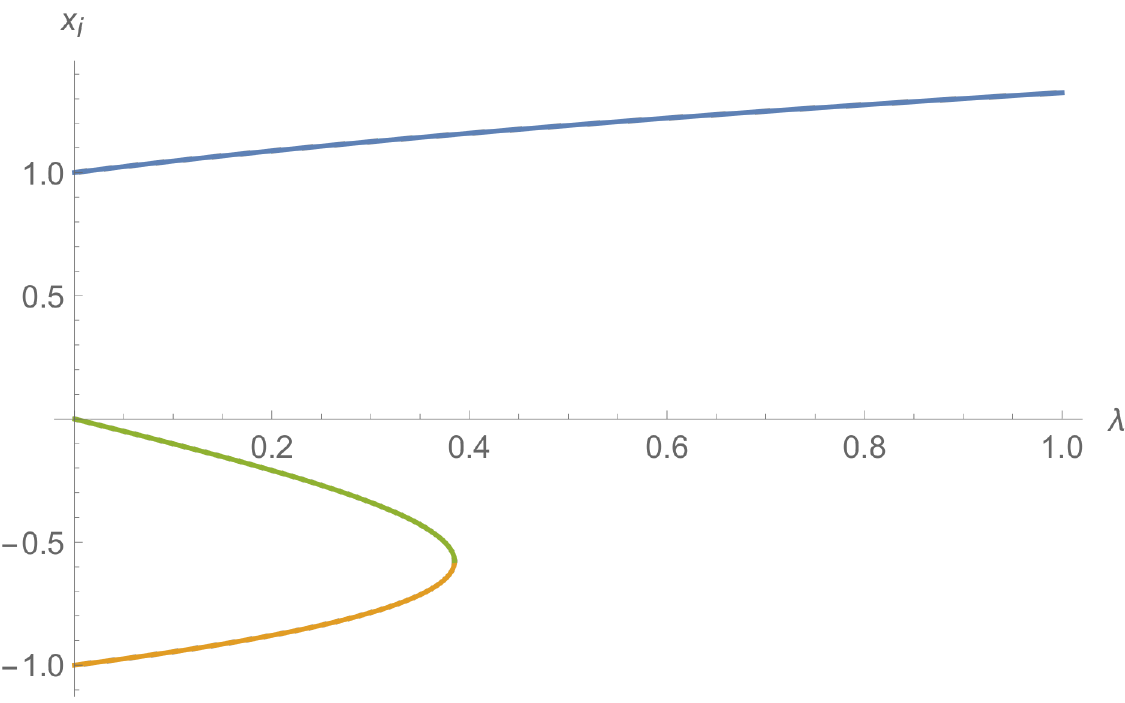}
    \caption{This plot shows the three solutions of equation \eqref{moduli-eq} as a function of $\tilde{\lambda}$. 
    \label{moduli-fig}}
\end{figure}

Equation \eqref{moduli-eq} is a cubic equation with three solutions for $\tilde{x}_j$. The solutions are plotted as a function of $\tilde{\lambda}$ in Figure \ref{moduli-fig}. For $\tilde{\lambda}<\frac{2}{3\sqrt{3}}$, all three solutions are real. We denote these as $y_0(\tilde\lambda)$, $y_+(\tilde{\lambda})$ and $y_-(\tilde\lambda)$. Here, $y_-<y_+<y_0$, with $y_-$ merging with $y_+$ at $\tilde{\lambda}=\frac{2}{3\sqrt{3}}$. 

Let $n_0$ of the $N$ different $\tilde{x}_j$'s be equal to $y_0$, $n_+$ of the $N$ different $\tilde{x}_j$'s be equal to $y_+$, and
$n_-$ of the $N$ different $\tilde{x}_j$'s be equal to $y_-$. $n_0$, $n_\pm$ are three non-negative integers satisfying 
\begin{eqnarray}
    n_0+n_+ + n_-  & = &  N \\
    \sum_i x_i=n_0 y_0+n_+ y_+ + n_- y_- &= & 0.
\end{eqnarray}
Because $y_0>0$, we require $n_+$ or $n_-$ to be non-zero. We also require $\tilde{x}_j$ to be real. This means we need $y_\pm$ to be real. We therefore need to choose a $0 \leq \tilde{\lambda} \leq 2/(3\sqrt{3})$. 

We have, 
\begin{eqnarray}
    n_0 + n_+ + n_-  & = &  N \\
   (n_+ - n_0) y_+ + (n_- - n_0) y_- & = & 0.
\end{eqnarray}
Using, $y_0+y_++y_- = 0$,
\begin{eqnarray}
    n_0 + n_+ + n_-  & = &  N \\
   \frac{(n_0-n_+)} {(n_0- n_-)} =  -y_-/y_+.
\end{eqnarray}
We also have 
\begin{equation}
    n_0 y_0^2+n_+ y_+^2 + n_- y_-^2=N,
\end{equation} because $\langle \tilde{x}^2 \rangle =1$. Finally, for a minimum, we also require $n_0 y_0^4+n_+y_+^4+n_-y_-^4 - 1=0$.  Putting these constraints, and requiring non-negativity, the unique solution is $\tilde{\lambda}=0$, and $n_0=n_+=\frac{N}{2}$. This corresponds to $\phi$ given by:
\begin{equation}
    \phi_0 = R \begin{pmatrix} \mathbf 1_{N/2 \times N/2} & 0 \\ 0 & - \mathbf 1_{N/2 \times N/2}\end{pmatrix}. \label{generic-point}
\end{equation} 

\subsection{Expanding the action around a generic point in moduli space}

When we expand the action around a generic point in moduli space, the symmetry group is broken from $SU(N)$ to $U(N/2)\times U(N/2)/U(1)$. Let us see this explicitly.

The subgroup of $SU(N)$ that commute with $\phi_0$ in equation \eqref{generic-point} include the matrices which are of the form
\begin{equation}
    \begin{pmatrix} T_a & 0 \\ 0 & 0 \end{pmatrix}, \quad     \begin{pmatrix} 0 & 0 \\ 0 & T_b \end{pmatrix},
\end{equation}
where $T_a$ is a generator of $SU(N/2)$. Any matrix of the form 
\begin{equation}
    \begin{pmatrix} O & T \\ T^\dagger & 0 \end{pmatrix}
\end{equation}
does not preserve $\phi_0$. Taking into account the diagonal generators of $SU(N)$, we find the symmetry group is broken to $U(N/2)\times U(N/2)/U(1)$.

Let us expand the action around an arbitrary point in moduli space. Let $\tilde{\phi}= \phi-\phi_0$. The matrix $\tilde{\phi}$ can be written as:
\begin{equation}
    \tilde{\phi} = \begin{pmatrix} A & B \\ B^\dagger & D \end{pmatrix}
\end{equation}
where $A$ and $D$ are Hermitian and satisfy $\tr A = -\tr D$. Let $A_+=A - \tr A$, and $D_+ = D- \tr D$, and $E=\tr A$. 

In the same way, we can expand the Yukawa term in the action around a point in a moduli space. We write the field $\psi$ in the matrix representation
\be
\psi  = \begin{pmatrix}
        e  & f\\
        f^\dagger & h
\end{pmatrix},\qquad 
\Bar{\psi} = \begin{pmatrix}
        \bar e  & \bar f\\
        \bar f^\dagger & \bar h
\end{pmatrix}
\end{equation}
where $e$ and $h$ are hermitian matrices, and $\tr e = -\tr h$. Let $e_+=e - \tr e$, and $h_+ = h- \tr h$, and $\xi=\tr e$.

Let us list the transformation properties of these fields under the $SU(N/2)\times SU(N/2)$ subgroup of $U(N/2)\times U(N/2)/U(1)$, $B$ transforms in the bifundamental representation of $SU(N/2)\times SU(N/2)$. $A_+$ transforms in the adjoint representation of the first $SU(N/2)$, $A_-$ transforms in the adjoint representation of the second $SU(N/2)$ and $E$ is a singlet. Similar results apply to the fermionic fields.

Written in terms of these fields, the scalar potential becomes,
\begin{align}
    V(\phi) &= 4\,R ^2 \big(   \tr\left(A_{+}A_{+}\right)    +  \tr\left(D_{+}D_{+}\right)   \big) + 4\, R  \big(   \tr\left(A_{+}A_{+}A_{+}\right)    - \tr\left(D_{+}D_{+}D_{+}\right)\nn \\
    &\quad +2   \tr\left(A_{+}A_{+}\right) \tr(A)     +2   \tr\left(D_{+}D_{+}\right) \tr(A)     + \tr\left(B\,B^{\dagger}A_{+}\right)  -   \tr\left(B^{\dagger}B\,D_{+}\right)  \big)\nn\\
    &\quad + \frac{1}{N }\bigg[-\tr\left(A_{+}A_{+}\right)^2-2 \tr\left(A_{+}A_{+}\right)   \tr\left(D_{+}D_{+}\right)-\tr\left(D_{+}D_{+}\right)^2\nn\\
    &\quad -4   \tr\left(A_{+}A_{+}\right) \tr\left(B\,B^{\dagger}\right) -4   \tr\left(D_{+}D_{+}\right) \tr\left(B\,B^{\dagger}\right) -4   \tr\left(B\,B^{\dagger}\right) ^2\nn\\
    &\quad +4 \tr\left(A_{+}A_{+}B\,B^{\dagger}\right)   N +\tr\left(A_{+}A_{+}A_{+}A_{+}\right) N +2   \tr\left(B\,B^{\dagger}B\,B^{\dagger}\right)   N\nn\\
    &\quad + 4   \tr\left(D_{+}D_{+}B^{\dagger}B\right)   N +\tr\left(D_{+}D_{+}D_{+}D_{+}\right) N +4   \tr\left(A_{+}A_{+}A_{+}\right) \tr(A)    N\nn\\
    &\quad - 4 \tr\left(D_{+}D_{+}D_{+}\right) \tr(A)  N +4   \tr\left(A_{+}A_{+}\right) \tr(A) ^2   N +4   \tr\left(D_{+}D_{+}\right) \tr(A) ^2   N \nn\\
    &\quad + 4 \tr\left(A_{+}B\,D_{+}B^{\dagger}\right)    N +4 \tr(A)    \tr\left(B\,B^{\dagger}A_{+}\right) N   -4 \tr(A)  \tr\left(B^{\dagger}B\,D_{+}\right)   N \bigg]
\end{align}
The Yukawa coupling becomes:
\begin{align}
    \tr\left(\phi\bar\psi \psi\right) &= \tr\left(A_{+}\bar{e}_{+}e_{+}\right)+2   \tr\left(A_{+}\bar{e}_{+}\xi\right)+\tr\left(B\bar{h}_{+}f^{\dagger}\right)+\tr\left(A_{+}\bar{f}f^{\dagger}\right)+\tr\left(B\bar{f}^{\dagger}e_{+}\right)\nn\\
    &\quad +\tr\left(D_{+}\bar{h}_{+}h_{+}\right)-2   \tr\left(D_{+}\bar{h}_{+}N_{f}\right)+\tr\left(D_{+}\bar{f}^{\dagger}f\right)+\tr\left(B^{\dagger}\bar{e}_{+}f\right) +\tr\left(B^{\dagger}\bar{f}h_{+}\right)\nn\\
    &\quad +\tr\left(A\right)    \tr\left(\bar{e}_{+}e_{+}\right)+R    \tr\left(\bar{e}_{+}e_{+}\right)-\tr\left(A\right)    \tr\left(\bar{h}_{+}h_{+}\right)-R    \tr\left(\bar{h}_{+}h_{+}\right)
\end{align}

We find that the fields $B$ and $b$ are massless, as are $E$ and $e$. The adjoint fields, $A_+$, $a+$, $D_+$, and $d_+$ acquire masses proportioanl to $R^2$. There are also cubic couplings involving the adjoint fields proportional to $R$. When one integrates out the massive fields, in the limit $R\to \infty$, one finds a free action in terms of the bifundamental and scalar fields $B$, $b$, $E$ and $e$.

\section{Unstable fixed points at large \texorpdfstring{$N$}{N}}\label{App: non-supersymmetric fp large N}
Here we present the scaling dimensions at the various unstable, non-supersymmetric large-$N$ fixed points, for $N_f=1/2$.

%

\subsection{The fixed point \texorpdfstring{$[ns_2]$}{[ns2]}}
The scaling dimension of $\phi$ , $\psi$ and the two mixtures of $\tr \phi^3$ and $\tr \Bar{\psi}\psi$ -- at the $[ns_2]$ fixed point are the same as at the $[susy]$ fixed point presented in the main text. The scaling dimension of the relevant operator $\tr \phi^2$ is
\be
\Delta_{\phi^2} = 2 - \frac{3 \varepsilon}{5} + \frac{13 \varepsilon^2}{125}+\frac{2 \varepsilon}{5}+\frac{(13-600 \zeta_{3}) \varepsilon ^3}{6250}.
\ee
The scaling dimension of the double trace quartic coupling is
\begin{equation}
    \Delta_{\lambda_2} = 4 -\frac{6 \varepsilon }{5}+\frac{26\varepsilon ^2}{125}+\frac{\varepsilon ^3(13-600 \zeta_3)}{3125}.
\end{equation}
The scaling dimensions of the two mixtures of the Yukawa coupling and single-trace quartic coupling are
\begin{equation}
    \Delta_{\left(\lambda_1,\lambda_y\right)_1} = 4 -\frac{12 \varepsilon^2}{25}+\frac{79 \varepsilon^3}{625}, \quad \Delta_{\left(\lambda_1,\lambda_y\right)_1} =  4 -\frac{9 \varepsilon^2}{25}+\frac{81 \varepsilon^3}{1250}.
\end{equation}

\subsection{The fixed points \texorpdfstring{[$ns_\pm$]}{ns pm}}
The scaling dimensions of $\phi$ and $\psi$ at the $[ns_\pm]$ fixed point are,
\begin{equation}
    \Delta_\phi =  1 - \frac{2 \varepsilon}{5} - \frac{9 \varepsilon^2}{1000}+\frac{357 \varepsilon^3}{100000}, \quad \Delta_\psi = \frac{3}{2} - \frac{2 \varepsilon}{5} + \frac{39 \varepsilon^2}{4000}+\frac{4653 \varepsilon^3}{400000}.
\end{equation}
The scaling dimensions of the relevant operators -- $\tr \Phi^2$ and two mixtures of $\tr \phi^3$ and $\tr \Bar{\psi}\psi$ --  are given by
\be
\Delta_{\phi^2} =2 -  \frac{5 \pm \sqrt{46}}{10} \varepsilon \pm \frac{1049 ~ \varepsilon ^2}{1000 \sqrt{46}}  \mp \frac{3 \sqrt{46} \left(662400  \zeta_{3}+896053 \right) \varepsilon ^3}{423200000},
\ee
\begin{align}
    \Delta_{(\bar{\psi}\psi,\phi^3)_1} &= 3 -\frac{3 \varepsilon}{2} + \frac{51 \varepsilon
   ^2}{1100}-\frac{3 \varepsilon ^3 (1478703+3460600
   \zeta_{3})}{106480000},
   \\
   \Delta_{(\bar \psi \psi , \phi^3)_2} &= 3 - \frac{2 \varepsilon}{5} + \frac{487 \varepsilon
   ^2}{22000}+\frac{\varepsilon ^3 (-33083621+25555200
   \zeta_{3})}{266200000}.
\end{align}
The scaling dimension of the classically marginal operators are, for the double-trace coupling,
\begin{equation}
    \Delta_{\lambda_2} = 4 - \frac{5 \pm \sqrt{46} }{5} \varepsilon \pm \frac{1049 \varepsilon ^2}{500 \sqrt{46}} \mp \frac{3 \sqrt{46} \varepsilon ^3 \left(896053 + 662400 \zeta_{3}\right)}{211600000},
\end{equation}
and, for the two mixtures of single-trace couplings,
\begin{equation}
    \Delta_{\left(\lambda_1,\lambda_y\right)_1} = 4-2 \varepsilon +\frac{21 \varepsilon^2}{200}-\frac{6749\varepsilon ^3}{160000}, \quad \Delta_{\left(\lambda_1,\lambda_y\right)_2} = 4  -\frac{119 \varepsilon^2}{400}-\frac{16407\varepsilon ^3}{160000}.
\end{equation}
\section{Stable planar large \texorpdfstring{$N$}{N} fixed point without supersymmetry}

\label{App: stable-large-N-nonsusy}
By generalizing the computations to $N_f>1/2$, we obtain a real, stable  fixed point dominated by planar diagrams in the large-$N$ limit without supersymmetry. 

\begin{figure}[H]
    \centering
    \includegraphics[width= 10 cm]{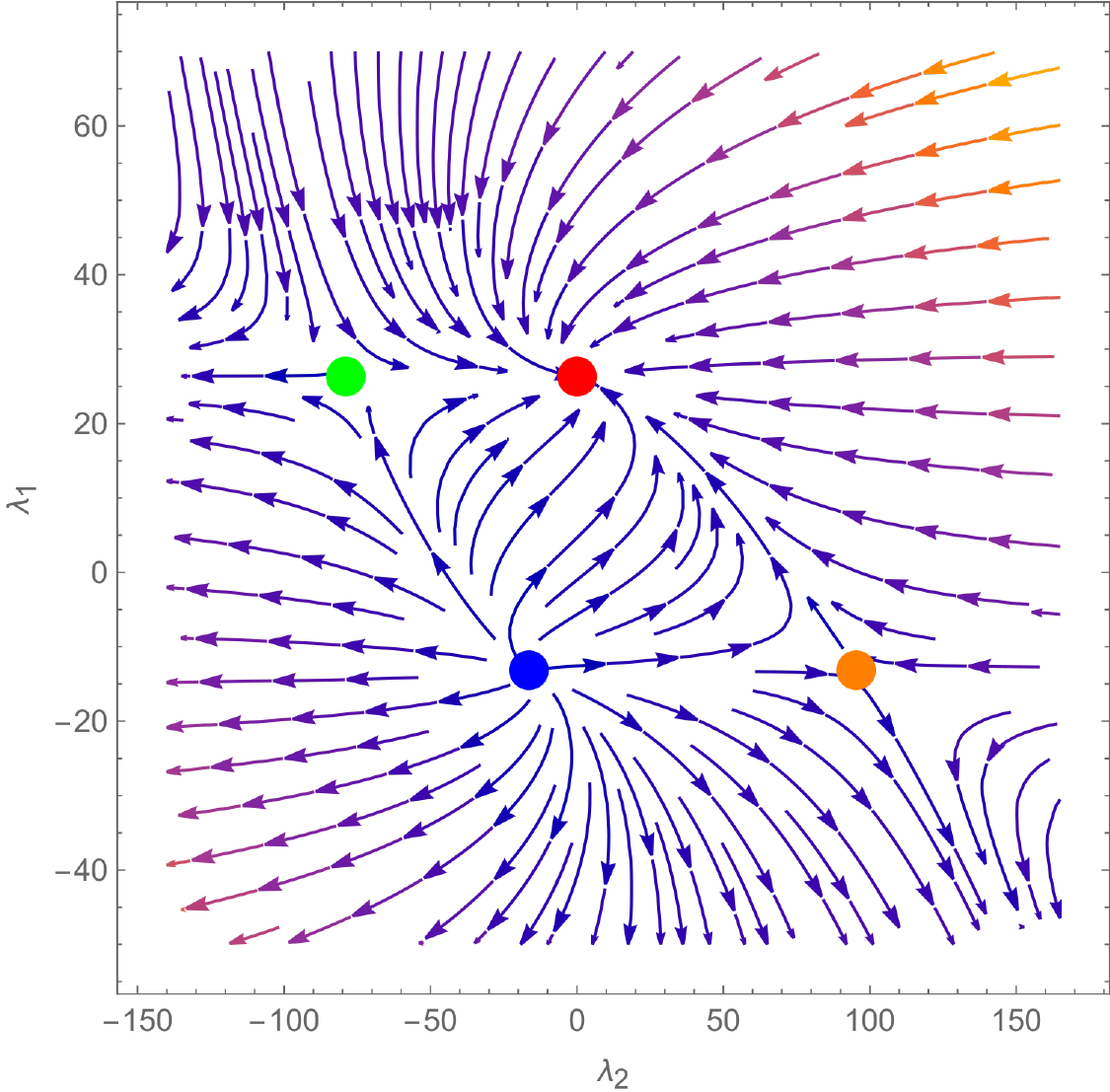}
    \caption{Flows and fixed points of the one-loop $\beta$-functions in the $\lambda_1$-$\lambda_2$ plane, with $\lambda_y$ tuned to criticality, for the theory when $N_f=1$. The red dot is the stable fixed point. \label{fig:flowsNf=1}}
\end{figure}

Let us describe the fixed points of the large-$N$ $\beta-$ functions for an arbitrary $N_f$. There continue to be four fixed points with non-zero Yukawa coupling, which we refer to as $[ns_\pm]_{N_f}$, $[ns_2]_{N_f}$, and $[susy]_{N_f}$. The generalization of $[susy]$ to an arbitrary $N_f$ is stable, but not supersymmetric when $N_f \neq \frac12$, but we still refer to it as $[susy]_{N_f}$ for convenience. For $N_f=1$, which corresponds to two flavours of Majorana fermions in $d=3$, a plot of the fixed points and flows is shown in Figure \ref{fig:flowsNf=1}. We see that the behaviour is qualitatively similar to the case of $N_f=1/2$.

The couplings at the $[susy]_{N_f}$  and $[ns_2]_{N_f}$ fixed points up to two loops are
\begin{align}
    \frac{\lambda_1^*}{\left(4 \pi\right)^2} &= \frac{1  }{2(2+N_f) } \varepsilon + \frac{3  (1+N_f )}{4(2+N_f )^3} \varepsilon^2, \quad \frac{\left(\lambda_y^*\right)^2}{\left(4 \pi\right)^2} = \frac{8 }{2+N_f } \varepsilon  + \frac{9  (1+2 N_f )}{(2+N_f )^3} \varepsilon ^2,
\end{align}
and,
\begin{align}
   \frac{\lambda_2^*}{\left(4 \pi\right)^2} &= \frac{\varepsilon}{4}  \left(-1 \pm \frac{\sqrt{(2+N_f )^2 (-8+N_f 
   (16+N_f ))}}{(2+N_f )^2}\right) + \frac{ \varepsilon ^2
  }{16(2+N_f )^4 \left(-8+16 N_f + N_f ^2\right)} \times \nn\\
  & \quad  \bigg[-86 N_f ^5-4 N_f ^6 + 32 \left(2 \pm \sqrt{(2+N_f )^2
   \left(-8+16 N_f +N_f ^2\right)}\right) + N_f ^4
   \big(-347 \nn\\
   & \quad \pm 4 \sqrt{(2+N_f )^2 \left(-8 + 16 N_f +N_f
   ^2\right)}\big) + 11 N_f ^2 \left(16 \pm 9 \sqrt{(2+N_f )^2
   \left(-8+16 N_f +N_f ^2\right)}\right) \nn\\
   & \quad - 8 N_f  \left(14 \pm 17
   \sqrt{(2+N_f )^2 \left(-8+16 N_f +N_f ^2\right)}\right) + 2 N_f ^3 \big(-129 \nn\\
   & \quad \pm 23 \sqrt{(2+N_f )^2 \left(-8+16 N_f
   +N_f ^2\right)}\big)\bigg].
\end{align}
The couplings at $[ns_{\mp}]_{N_f}$ fixed points up to two loops are,
\begin{align}
   \frac{\lambda_1^*}{\left(4 \pi\right)^2} &= -\frac{N_f }{4(2+N_f)} \varepsilon + \frac{ N_f  (13-2 N_f  (4+N_f ))}{16(2+N_f )^3} \varepsilon ^2 , \quad \frac{\left(\lambda_y^*\right)^2}{\left(4 \pi\right)^2} = \frac{8}{2+N_f } \varepsilon - \frac{ (-14+(-32+N_f ) N_f )}{2(2+N_f )^3} \varepsilon ^2,
\end{align}
and
\begin{align}
    \frac{\lambda^*_2}{\left(4 \pi\right)^2} &=  \left(1 \pm \frac{\sqrt{2} \sqrt{-(2+N_f )^2
   (-2+(-8+N_f ) N_f )}}{(2+N_f )^2}\right) \frac{\varepsilon}{4} + \frac{3 
    N_f \varepsilon ^2}{32 (2+N_f )^4 \left(-2-8 N_f +N_f ^2\right)} \times \nn\\
   & \quad \bigg[ 24 + 84 N_f -6 N_f ^4+2 N_f ^5 \pm 32
   \sqrt{2} \sqrt{(2+N_f )^2 \left(2+8 N_f - N_f
   ^2\right)} \nn\\
   & \quad \mp 52 N_f \sqrt{2} \sqrt{(2+N_f )^2 \left(2+8
   N_f - N_f ^2\right)} + 3 N_f ^3 \left(-26 \pm \sqrt{2}
   \sqrt{(2+N_f )^2 \left(2 + 8 N_f - N_f
   ^2\right)}\right) \nn\\
   & \quad -4 N_f ^2 \left(20 \pm \sqrt{2}
   \sqrt{-(2+N_f )^2 \left(-2-8 N_f +N_f
   ^2\right)}\right)\bigg].
\end{align}
The $[susy]_{N_f}$ fixed point is stable for $N_f > - 8 + 6 \sqrt{2} \approx 0.485$. The classical potential satisfies the condition $\lambda_1 > - \lambda_2$ for any $N_f > \frac12$, and is therefore positive definite. We, therefore, believe that this fixed point defines a large-$N$ non-supersymmetric interacting 
CFT in $d=3$ that is dominated by planar diagrams for any value of $N_f>1/2$, e.g., $N_f=1$. Of course, as described in the footnote at the end of section \ref{sec:theories}, there are actually two possibilities for the global symmetry group in the 3d CFT, which contains either a factor of $O(2N_f)$ or $O(N_f)^2/\mathbb Z_2$. These two global symmetry groups define two distinct CFTs \cite{Erramilli:2022kgp}, that are indistinguishable at three-loop order for the observables we compute.

Scaling dimensions for operators at the $[susy]_{
N_f}$ fixed point up to two loops are,
\begin{equation}
    \Delta_{\phi} = 1-\frac{\varepsilon }{2+N_f }+\frac{\varepsilon ^2 \left(4-5 N_f
   +10 N_f ^2\right)}{16 (2+N_f )^3},
\end{equation}
\begin{equation}
    \Delta_{\psi} = \frac{3}{2} +  
   \left(-\frac{1}{2}+\frac{1}{4 (2+N_f )}\right) \varepsilon + \frac{\varepsilon ^2 \left(7+5 N_f -6 N_f
   ^2\right)}{32 (2+N_f )^3} ,
\end{equation}
\begin{align}
    \Delta_{\phi^2} &= 2 + \left(-1+\frac{2+N_f +\sqrt{-8+16 N_f +N_f
   ^2}}{2 (2+N_f )}\right) \varepsilon + \frac{\varepsilon ^2 \left(32-72 N_f +3 N_f ^2-26 N_f
   ^3\right)}{8 (2+N_f )^3 \sqrt{-8+16 N_f +N_f
   ^2}} ,
\end{align}
\begin{align}
    \Delta_{\left(\Bar{\psi}\psi,\phi^3\right)_1} &= 3 - 
 \frac{\varepsilon  (1+2 N_f )}{2 (2+N_f )}+\frac{\varepsilon ^2
   \left(1-81 N_f -36 N_f ^2-28 N_f ^3\right)}{16 (2+N_f )^3
   (1+2 N_f )},
\end{align}
\begin{equation}
     \Delta_{\left(\Bar{\psi}\psi,\phi^3\right)_2} = 3-\frac{3  \left(20+45 N_f +64 N_f ^2+12 N_f
   ^3\right)}{16 (2+N_f )^3 (1+2 N_f )} \varepsilon ^2,
\end{equation}
\begin{equation}
    \Delta_{\lambda_2} = 4 +  \left(-1+\frac{\sqrt{-8+16 N_f +N_f
   ^2}}{2+N_f }\right) \varepsilon  + \frac{\varepsilon ^2 \left(32-72 N_f +3 N_f ^2-26 N_f
   ^3\right)}{4 (2+N_f )^3 \sqrt{-8+16 N_f +N_f
   ^2}}, 
\end{equation}
and
\begin{equation}
    \Delta_{\left(\lambda_1,\lambda_y\right)_1} = 4-\frac{3 \varepsilon ^2 \left(14+37 N_f +10 N_f ^2+\left| -2-7
   N_f +2 N_f ^2\right| \right)}{16 (2+N_f )^3},
\end{equation}
\begin{equation}
    \Delta_{\left(\lambda_1,\lambda_y\right)_2} = 4-\frac{3 \varepsilon ^2 \left(14+37 N_f +10 N_f ^2 - \left| -2-7
   N_f +2 N_f ^2\right| \right)}{16 (2+N_f )^3}.
\end{equation}

\section{More Pad\'e approximates}
\label{app:pade}
In this appendix, we present some Pad\'e approximates to $d=3$ for the various unstable fixed points in the $N=3$ theories and the common large-$N$ limit. We also include Pad\'e approximates to $d=2$, in case that helps in identifying candidate $2d$ CFTs that these theories may reduce to in $d=2$.

\subsection{Scaling dimension at large-\texorpdfstring{$N$}{N}}
Here we present results in the common large-$N$ limit of all the models we study. Table \ref{table:d=2-susy-largeN} lists all Pad\'e approximates for scaling dimensions in $d=2$ of the large-$N$ limit of the stable fixed point. 

\begin{table}[H]\centering
        \resizebox{\columnwidth}{!}{
	\begin{tabular}{|m{2 cm}|m{1.5cm}|m{1.5cm}|m{1.5cm}|m{1.6cm}|m{1.6cm}|m{1.5cm}|m{1.8cm}|m{1.8cm}|}\hline
	\vspace{0.3 cm} & $\Delta_{\phi}$ & $\Delta_{\psi}$ & $\Delta_{\phi^{2}}$ & $\Delta_{\left(\bar\psi\psi,\phi^{3}\right)_{1}}$ & $\Delta_{\left(\bar\psi\psi,\phi^3\right)_{2}}$ & $\Delta_{\left(\phi^2\right)^2}$ & $\Delta_{\left(\phi^4,\phi\bar\psi\psi\right)_{1}}$ & $\Delta_{\left(\phi^4,\phi\bar\psi\psi\right)_{2}}$\\
		\hline
		\hline
   $\text{Pad\'e}_{[1,1]}$ & 0.259 & 0.759 & 0.333 & 1.333 & 3. & 0.667 & 4. & 4. \\
$\text{Pad\'e}_{[1,2]}$ &  0.29 & 0.789 & 1.228 & 2.187 & 2.217 & 2.455
   & 3.043 & 3.163 \\
 $\text{Pad\'e}_{[2,1]}$ & 0.339 & 0.839 & 1.069 & 2.069 & 1.941 & 2.138
   & 2.742 & 2.941 \\
 $\text{Pad\'e}_{[0,3]}$ & 0.363 & 0.807 & 1.294 & 2.367 & 2.295 & 2.589
   & 3.259 & 3.251 \\
 $\text{Pad\'e}_{[3,0]}$ & 0.299 & 0.799 & 1.691 & 2.691 & 2.078 & 3.381
   & 3.091 & 3.078 \\
    \hline
    \end{tabular}}
    \caption{$\text{Pad\'e}$ estimates of the scaling dimensions in $d=2$ for $[susy]$ fixed points in the common large-$N$ limit of the $SU(N)$ adjoint, $SO(N)$ $S_2$ and $S_2-A_2$ models. \label{table:d=2-susy-largeN}}
\end{table}

Tables \ref{table:d=2-[ns1]-largeN} and \ref{table:d=3-[ns1]-largeN} list all Pad\'e approximates for scaling dimensions in $d=2$ and $d=3$ of the large-$N$ limit of the $[ns_1]$ fixed point. 

\begin{table}[H]\centering
        \resizebox{\columnwidth}{!}{
	\begin{tabular}{|m{2 cm}|m{1.5cm}|m{1.5cm}|m{1.5cm}|m{1.6cm}|m{1.6cm}|m{1.5cm}|m{1.8cm}|m{1.8cm}|}\hline
	\vspace{0.3 cm} & $\Delta_{\phi}$ & $\Delta_{\psi}$ & $\Delta_{\phi^{2}}$ & $\Delta_{\left(\bar\psi\psi,\phi^{3}\right)_{1}}$ & $\Delta_{\left(\bar\psi\psi,\phi^3\right)_{2}}$ & $\Delta_{\left(\phi^2\right)^2}$ & $\Delta_{\left(\phi^4,\phi\bar\psi\psi\right)_{1}}$ & $\Delta_{\left(\phi^4,\phi\bar\psi\psi\right)_{2}}$\\
		\hline
		\hline
   $\text{Pad\'e}_{[1,1]}$ & 0.162 & 0.737 & 0.134 & 0.175 & 2.28 & 0.267
   & 0.38 & 4. \\
 $\text{Pad\'e}_{[1,2]}$ & 0.193 & 0.801 & -0.369 & -6.933 & 2.167 &
   -0.739 & 0.098 & 2.043 \\
 $\text{Pad\'e}_{[2,1]}$ & 0.18 & 0.672 & -0.055 & 0.024 & 2.249 & -0.11
   & 0.233 & 0.169 \\
$\text{Pad\'e}_{[0,3]}$ &  0.331 & 0.809 & 0.446 & 0.708 & 2.247 & 0.891
   & 1.061 & 2.662 \\
 $\text{Pad\'e}_{[3,0]}$ & 0.193 & 0.832 & -0.389 & -1.085 & 2.217 &
   -0.777 & 0.083 & 1.99 \\
    \hline
    \end{tabular}}
    \caption{$\text{Pad\'e}$ estimates of the scaling dimensions in $d=2$ for $[ns_1]$ fixed points in the common large-$N$ limit of the $SU(N)$ adjoint, $SO(N)$ $S_2$ and $S_2-A_2$ models. \label{table:d=2-[ns1]-largeN}}
\end{table}

\begin{table}[H]\centering
        \resizebox{\columnwidth}{!}{
	\begin{tabular}{|m{2 cm}|m{1.5cm}|m{1.5cm}|m{1.5cm}|m{1.6cm}|m{1.6cm}|m{1.5cm}|m{1.8cm}|m{1.8cm}|}\hline
	\vspace{0.3 cm} & $\Delta_{\phi}$ & $\Delta_{\psi}$ & $\Delta_{\phi^{2}}$ & $\Delta_{\left(\bar\psi\psi,\phi^{3}\right)_{1}}$ & $\Delta_{\left(\bar\psi\psi,\phi^3\right)_{2}}$ & $\Delta_{\left(\phi^2\right)^2}$ & $\Delta_{\left(\phi^4,\phi\bar\psi\psi\right)_{1}}$ & $\Delta_{\left(\phi^4,\phi\bar\psi\psi\right)_{2}}$\\
		\hline
		\hline
   $\text{Pad\'e}_{[1,1]}$ &  0.591 & 1.11 & 0.958 & 1.545 & 2.621 & 1.917
   & 2.1 & 4. \\
 $\text{Pad\'e}_{[1,2]}$ & 0.595 & 1.119 & 0.902 & 1.325 & 2.612 & 1.804
   & 2.065 & 3.592 \\
 $\text{Pad\'e}_{[2,1]}$ & 0.594 & 1.05 & 0.923 & 1.51 & 2.616 & 1.846 &
   2.075 & 3.546 \\
 $\text{Pad\'e}_{[0,3]}$ & 0.611 & 1.12 & 0.994 & 1.581 & 2.615 & 1.987
   & 2.182 & 3.636 \\
 $\text{Pad\'e}_{[3,0]}$ & 0.595 & 1.121 & 0.895 & 1.388 & 2.613 & 1.79
   & 2.063 & 3.6 \\
    \hline
    \end{tabular}}
    \caption{$\text{Pad\'e}$ estimates of the scaling dimensions in $d=3$ for $[ns_1]$ fixed points in the common large-$N$ limit of the $SU(N)$ adjoint, $SO(N)$ $S_2$ and $S_2-A_2$ models. \label{table:d=3-[ns1]-largeN}}
\end{table}

Tables \ref{table:d=2-[ns2]-largeN} and \ref{table:d=3-[ns2]-largeN} list all Pad\'e approximates for scaling dimensions in $d=2$ and $d=3$ of the large-$N$ limit of the $[ns_2]$ fixed point. 

\begin{table}[H]\centering
        \resizebox{\columnwidth}{!}{
	\begin{tabular}{|m{2 cm}|m{1.5cm}|m{1.5cm}|m{1.5cm}|m{1.6cm}|m{1.6cm}|m{1.5cm}|m{1.8cm}|m{1.8cm}|}\hline
	\vspace{0.3 cm} & $\Delta_{\phi}$ & $\Delta_{\psi}$ & $\Delta_{\phi^{2}}$ & $\Delta_{\left(\bar\psi\psi,\phi^{3}\right)_{1}}$ & $\Delta_{\left(\bar\psi\psi,\phi^3\right)_{2}}$ & $\Delta_{\left(\phi^2\right)^2}$ & $\Delta_{\left(\phi^4,\phi\bar\psi\psi\right)_{1}}$ & $\Delta_{\left(\phi^4,\phi\bar\psi\psi\right)_{2}}$\\
		\hline
		\hline
   $\text{Pad\'e}_{[1,1]}$ &  0.259 & 0.759 & 1.109 & 1.333 & 3. & 2.218 &
   4. & 4. \\
 $\text{Pad\'e}_{[1,2]}$ & 0.29 & 0.789 & 1.321 & 2.187 & 2.217 & 2.643
   & 3.043 & 3.163 \\
 $\text{Pad\'e}_{[2,1]}$ & 0.339 & 0.839 & 0.931 & 2.069 & 1.941 & 1.862
   & 2.742 & 2.941 \\
 $\text{Pad\'e}_{[0,3]}$ & 0.363 & 0.807 & 0.921 & 2.367 & 2.295 & 1.842
   & 3.259 & 3.251 \\
 $\text{Pad\'e}_{[3,0]}$ & 0.299 & 0.799 & 0.309 & 2.691 & 2.078 & 0.619
   & 3.091 & 3.078 \\
    \hline
    \end{tabular}}
    \caption{$\text{Pad\'e}$ estimates of the scaling dimensions in $d=2$ for $[ns_2]$ fixed points in the common large-$N$ limit of the $SU(N)$ adjoint, $SO(N)$ $S_2$ and $S_2-A_2$ models. \label{table:d=2-[ns2]-largeN}}
\end{table}

\begin{table}[H]\centering
        \resizebox{\columnwidth}{!}{
	\begin{tabular}{|m{2 cm}|m{1.5cm}|m{1.5cm}|m{1.5cm}|m{1.6cm}|m{1.6cm}|m{1.5cm}|m{1.8cm}|m{1.8cm}|}\hline
	\vspace{0.3 cm} & $\Delta_{\phi}$ & $\Delta_{\psi}$ & $\Delta_{\phi^{2}}$ & $\Delta_{\left(\bar\psi\psi,\phi^{3}\right)_{1}}$ & $\Delta_{\left(\bar\psi\psi,\phi^3\right)_{2}}$ & $\Delta_{\left(\phi^2\right)^2}$ & $\Delta_{\left(\phi^4,\phi\bar\psi\psi\right)_{1}}$ & $\Delta_{\left(\phi^4,\phi\bar\psi\psi\right)_{2}}$\\
		\hline
		\hline
   $\text{Pad\'e}_{[1,1]}$ &  0.615 & 1.115 & 1.489 & 2.459 & 3. & 2.977 &
   4. & 4. \\
   $\text{Pad\'e}_{[1,2]}$ & 0.62 & 1.12 & 1.666 & 2.568 & 2.723 & 3.333 &
   3.653 & 3.717 \\
   $\text{Pad\'e}_{[2,1]}$ & 0.622 & 1.122 & 1.45 & 2.55 & 2.695 & 2.9 &
   3.62 & 3.695 \\
   $\text{Pad\'e}_{[0,3]}$ & 0.629 & 1.121 & 1.438 & 2.587 & 2.731 & 2.877
   & 3.675 & 3.725 \\
  $\text{Pad\'e}_{[3,0]}$ &  0.62 & 1.12 & 1.391 & 2.609 & 2.705 & 2.781 &
   3.646 & 3.705 \\
    \hline
    \end{tabular}}
    \caption{$\text{Pad\'e}$ estimates of the scaling dimensions in $d=3$ for $[ns_2]$ fixed points in the common large-$N$ limit of the $SU(N)$ adjoint, $SO(N)$ $S_2$ and $S_2-A_2$ models. \label{table:d=3-[ns2]-largeN}}
\end{table}

Tables \ref{table:d=2-[ns3]-largeN} and \ref{table:d=3-[ns3]-largeN} list all Pad\'e approximates for scaling dimensions in $d=2$ and $d=3$ of the large-$N$ limit of the $[ns_3]$ fixed point. 

\begin{table}[H]\centering
        \resizebox{\columnwidth}{!}{
	\begin{tabular}{|m{2 cm}|m{1.5cm}|m{1.5cm}|m{1.5cm}|m{1.6cm}|m{1.6cm}|m{1.5cm}|m{1.8cm}|m{1.8cm}|}\hline
	\vspace{0.3 cm} & $\Delta_{\phi}$ & $\Delta_{\psi}$ & $\Delta_{\phi^{2}}$ & $\Delta_{\left(\bar\psi\psi,\phi^{3}\right)_{1}}$ & $\Delta_{\left(\bar\psi\psi,\phi^3\right)_{2}}$ & $\Delta_{\left(\phi^2\right)^2}$ & $\Delta_{\left(\phi^4,\phi\bar\psi\psi\right)_{1}}$ & $\Delta_{\left(\phi^4,\phi\bar\psi\psi\right)_{2}}$\\
		\hline
		\hline
   $\text{Pad\'e}_{[1,1]}$ & 0.162 & 0.737 & 2.13 & 0.175 & 2.28 & 4.261 &
   0.38 & 4. \\
 $\text{Pad\'e}_{[1,2]}$ & 0.193 & 0.801 & 2.061 & -6.933 & 2.167 &
   4.122 & 0.098 & 2.043 \\
 $\text{Pad\'e}_{[2,1]}$ & 0.18 & 0.672 & 2.055 & 0.024 & 2.249 & 4.11 &
   0.233 & 0.169 \\
$\text{Pad\'e}_{[0,3]}$ &  0.331 & 0.809 & 2.772 & 0.708 & 2.247 & 5.544
   & 1.061 & 2.662 \\
$\text{Pad\'e}_{[3,0]}$ &  0.193 & 0.832 & 2.389 & -1.085 & 2.217 &
   4.777 & 0.083 & 1.99 \\
    \hline
    \end{tabular}}
    \caption{$\text{Pad\'e}$ estimates of the scaling dimensions in $d=2$ for $[ns_3]$ fixed points in the common large-$N$ limit of the $SU(N)$ adjoint, $SO(N)$ $S_2$ and $S_2-A_2$ models. \label{table:d=2-[ns3]-largeN}}
\end{table}

\begin{table}[H]\centering
        \resizebox{\columnwidth}{!}{
	\begin{tabular}{|m{2 cm}|m{1.5cm}|m{1.5cm}|m{1.5cm}|m{1.6cm}|m{1.6cm}|m{1.5cm}|m{1.8cm}|m{1.8cm}|}\hline
	\vspace{0.3 cm} & $\Delta_{\phi}$ & $\Delta_{\psi}$ & $\Delta_{\phi^{2}}$ & $\Delta_{\left(\bar\psi\psi,\phi^{3}\right)_{1}}$ & $\Delta_{\left(\bar\psi\psi,\phi^3\right)_{2}}$ & $\Delta_{\left(\phi^2\right)^2}$ & $\Delta_{\left(\phi^4,\phi\bar\psi\psi\right)_{1}}$ & $\Delta_{\left(\phi^4,\phi\bar\psi\psi\right)_{2}}$\\
		\hline
		\hline
   $\text{Pad\'e}_{[1,1]}$ &  0.591 & 1.11 & 2.095 & 1.545 & 2.621 & 4.191 & 2.1 & 4. \\
   $\text{Pad\'e}_{[1,2]}$ & 0.595 & 1.119 & 2.078 & 1.325 & 2.612 & 4.155 & 2.065 & 3.592 \\
   $\text{Pad\'e}_{[2,1]}$ & 0.594 & 1.05 & 2.077 & 1.51 & 2.616 & 4.154 & 2.075 & 3.546 \\
   $\text{Pad\'e}_{[0,3]}$ & 0.611 & 1.12 & 2.125 & 1.581 & 2.615 & 4.251 & 2.182 & 3.636 \\
   $\text{Pad\'e}_{[3,0]}$ & 0.595 & 1.121 & 2.105 & 1.388 & 2.613 & 4.21 & 2.063 & 3.6 \\
    \hline
    \end{tabular}}
    \caption{$\text{Pad\'e}$ estimates of the scaling dimensions in $d=3$ for $[ns_3]$ fixed points in the common large-$N$ limit of the $SU(N)$ adjoint, $SO(N)$ $S_2$ and $S_2-A_2$ models. \label{table:d=3-[ns3]-largeN}}
\end{table}


%
\subsection{\texorpdfstring{$SU(3)$}{SU(3)} adjoint model}
Here we present results in the $SU(3)$ adjoint model.  Table \ref{table: SU(3) pade-scaling-[susy]-d=2} lists various Pad\'e approximates for scaling dimensions in $d=2$ of the stable supersymmetric fixed point. At the unstable $[ns]$ fixed point of $SO(3) - S_2$ model, various possible Pad\`e approximates of scaling dimension for different operators are given in Tables \ref{table: SU(3) pade-scaling-[ns]-d=3} and \ref{table: SU(3) pade-scaling-[ns]-d=2}. For the $[ns]$ fixed point all Pad\'e approximates except $[1,2]$, we find negative scaling dimensions, suggesting that a unitary CFT may not exist, or that the only acceptable resummation is the Pad\'e $[1,2]$ approximate. For the $[susy]$ fixed point in $d=2$, only the $[1,1]$ and $[1,2]$ approximates give rise to a spectrum with only positive scaling dimensions. 

 \begin{table}[H]\centering
    \begin{tabular}{|m{2 cm}|m{2cm}|m{1.7cm}|m{1.5cm}|m{1.6cm}|m{1.6cm}|m{1.5cm}|m{1.5cm}|}\hline
    \vspace{0.2cm} & $\Delta_{\phi}$ & $\Delta_{\psi}$ & $\Delta_{\phi^{2}}$ & $\Delta_{\left(\bar\psi\psi,\phi^{3}\right)_{1}}$ & $\Delta_{\left(\bar\psi\psi,\phi^{3}\right)_{2}}$ & $\Delta_{\left(g,y\right)_{1}}$ & $\Delta_{\left(g,y\right)_{2}}$ \\
    \hline
    \hline
    $\text{Pad\'e}_{[1,1]}$ & 0.676 & 1.176 & 2.692 & 3. & 3.692 & 4. & 5.532 \\
    \hline
    $\text{Pad\'e}_{[1,2]}$ & 0.692 & 1.157 & 4.523 & 1.215 & 4.898 & 1.904 & 15.7 \\
    \hline
    $\text{Pad\'e}_{[2,1]}$ & 0.529 & 1.029 & 3.247 & -1.405 & 4.247 & -0.405 & 7.922 \\
    \hline
    $\text{Pad\'e}_{[0,3]}$ & 3.997 & 3.288 & -0.253 & 1.318 & -0.633 & 2.043 & -0.117 \\
    \hline
    $\text{Pad\'e}_{[3,0]}$ & 1.889 & 2.389 & 18.46 & -0.831 & 19.46 & 0.169 & 106.9 \\
		\hline
	\end{tabular}\caption{Pad\'e approximates of scaling dimensions of various operators in $d=2$ for the supersymmetric fixed point of the  $SU(3)$ adjoint theory. Only the Pad\'e [1,1] and [1,2] approximates contain no operators with negative scaling dimension. \label{table: SU(3) pade-scaling-[susy]-d=2}}
\end{table}

\begin{table}[H]\centering
    \begin{tabular}{|m{2 cm}|m{2cm}|m{1.7cm}|m{1.5cm}|m{1.6cm}|m{1.6cm}|m{1.5cm}|m{1.5cm}|}\hline
    \vspace{0.2cm} & $\Delta_{\phi}$ & $\Delta_{\psi}$ & $\Delta_{\phi^{2}}$ & $\Delta_{\left(\bar\psi\psi,\phi^{3}\right)_{1}}$ & $\Delta_{\left(\bar\psi\psi,\phi^{3}\right)_{2}}$ & $\Delta_{\left(g,y\right)_{1}}$ & $\Delta_{\left(g,y\right)_{2}}$ \\
    \hline
    \hline
    $\text{Pad\'e}_{[1,1]}$ & -1.003 & -0.888 & 1.006 & 1.612 & 3.158 & -0.418 & 4. \\
    \hline
    $\text{Pad\'e}_{[1,2]}$ & 0.768 & 1.259 & 1.665 & 2.183 & 3.158 & 2.186 & 4.007 \\
    \hline
    $\text{Pad\'e}_{[2,1]}$ & 0.729 & 1.232 & 0.909 & 1.625 & 3.158 & -0.663 & 4.007 \\
    \hline
    $\text{Pad\'e}_{[0,3]}$ & 1.043 & 1.678 & 0.936 & 1.233 & -2.801 & 0.411 & 2.117 \\
    \hline
    $\text{Pad\'e}_{[3,0]}$ & 1.189 & 1.757 & 0.446 & -0.368 & 8.378 & -19.76 & 0.441 \\
    \hline
    \end{tabular}\caption{Pad\'e Approximates of scaling dimensions of various operators in $d=3$ for the $[ns]$ fixed point of $SU(3)$ adjoint theory. Only the Pad\'e [1,2] approximate contains no operators with negative scaling dimension. \label{table: SU(3) pade-scaling-[ns]-d=3}}
\end{table}
\begin{table}[H]\centering
    \begin{tabular}{|m{2 cm}|m{2cm}|m{1.7cm}|m{1.5cm}|m{1.6cm}|m{1.6cm}|m{1.5cm}|m{1.5cm}|}\hline
    \vspace{0.2cm} & $\Delta_{\phi}$ & $\Delta_{\psi}$ & $\Delta_{\phi^{2}}$ & $\Delta_{\left(\bar\psi\psi,\phi^{3}\right)_{1}}$ & $\Delta_{\left(\bar\psi\psi,\phi^{3}\right)_{2}}$ & $\Delta_{\left(g,y\right)_{1}}$ & $\Delta_{\left(g,y\right)_{2}}$ \\
    \hline
    \hline
    $\text{Pad\'e}_{[1,1]}$ & 1.571 & 2.046 & 0.568 & 0.198 & 3.18 & -4.389 & 4. \\
    \hline
    $\text{Pad\'e}_{[1,2]}$ & 0.611 & 1.074 & 0.896 & 1.689 & 3.178 & 1.462 & 4.015 \\
    \hline
    $\text{Pad\'e}_{[2,1]}$ & 0.445 & 0.952 & 0.191 & 0.25 & 3.178 & -5.326 & 4.015 \\
    \hline
    $\text{Pad\'e}_{[0,3]}$ & -0.607 & -1.1 & 0.432 & 0.339 & -0.174 & 0.068 & 0.483 \\
    \hline
    $\text{Pad\'e}_{[3,0]}$ & 4.636 & 5.693 & -4.808 & -15.65 & 50.58 & -159.1 & -25.15 \\
    \hline
    \end{tabular}\caption{Pad\'e approximates of scaling dimensions of various operators in $d=2$ for the $[ns]$-fixed point in the  $SU(3)$ adjoint theory. Only the Pad\'e [1,2] approximate contains no operators with negative scaling dimension. \label{table: SU(3) pade-scaling-[ns]-d=2}}
\end{table}

\subsection{\texorpdfstring{$SO(3)~S_2$}{SO(3) s2} model}
Here we present results in the $SO(3)$-$S_2$ model.  Table \ref{table: SO(3) pade-scaling-[susy]-d=2} lists various Pad\'e approximates for scaling dimensions in $d=2$ of the stable supersymmetric fixed point. At the unstable $[ns]$ fixed point of $SO(3) - S_2$ model, various possible Pad\`e approximates of scaling dimension for different operators are given in Tables \ref{table: SO(3) pade-scaling-[ns]-d=3} and \ref{table: SO(3) pade-scaling-dimension-[ns]-d=2}. Again, only the Pad\'e $[1,2]$ approximate seems to give results that are free from operators with negative scaling dimensions.

\begin{table}[H]\centering
    \begin{tabular}{|m{2 cm}|m{2cm}|m{1.7cm}|m{1.5cm}|m{1.6cm}|m{1.6cm}|m{1.5cm}|m{1.5cm}|}\hline
    \vspace{0.2cm} & $\Delta_{\phi}$ & $\Delta_{\psi}$ & $\Delta_{\phi^{2}}$ & $\Delta_{\left(\bar\psi\psi,\phi^{3}\right)_{1}}$ & $\Delta_{\left(\bar\psi\psi,\phi^{3}\right)_{2}}$ & $\Delta_{\left(g,y\right)_{1}}$ & $\Delta_{\left(g,y\right)_{2}}$ \\
    \hline
    \hline
    $\text{Pad\'e}_{[1,1]}$ & 0.49 & 0.99 & 2.254 & 3. & 3.254 & 4. & 5.093 \\
    \hline
    $\text{Pad\'e}_{[1,2]}$ & 0.614 & 1.086 & 2.746 & 2.311 & 3.689 & 3.27 & 9.436 \\
    \hline
    $\text{Pad\'e}_{[2,1]}$ & 0.466 & 0.966 & 2.595 & 2.106 & 3.595 & 3.106 & 6.945 \\
    \hline
    $\text{Pad\'e}_{[0,3]}$ & 0.745 & 1.345 & -0.367 & 3.476 & -0.949 & 4.458 & -0.106 \\
    \hline
    $\text{Pad\'e}_{[3,0]}$ & 1.067 & 1.567 & 13.72 & 3.411 & 14.72 & 4.411 & 119.1 \\
    \hline
    \end{tabular}\caption{Pad\'e approximates of scaling dimensions of various operators in $d=2$ for the supersymmetric fixed point of the $SO(3)$ $S_2$ model. Only the Pad\'e [1,2] and [2,1] approximates contains no operators with negative scaling dimension. \label{table: SO(3) pade-scaling-[susy]-d=2}}
\end{table}
\begin{table}[H]\centering
    \begin{tabular}{|m{2 cm}|m{2cm}|m{1.7cm}|m{1.5cm}|m{1.6cm}|m{1.6cm}|m{1.5cm}|m{1.5cm}|}\hline
    \vspace{0.2cm} & $\Delta_{\phi}$ & $\Delta_{\psi}$ & $\Delta_{\phi^{2}}$ & $\Delta_{\left(\bar\psi\psi,\phi^{3}\right)_{1}}$ & $\Delta_{\left(\bar\psi\psi,\phi^{3}\right)_{2}}$ & $\Delta_{\left(g,y\right)_{1}}$ & $\Delta_{\left(g,y\right)_{2}}$ \\
    \hline
    \hline
    $\text{Pad\'e}_{[1,1]}$ & -0.042 & 0.492 & 1.108 & 1.871 & 3.071 & 0.296 & 4. \\
    \hline
    $\text{Pad\'e}_{[1,2]}$ & 0.734 & 1.224 & 1.217 & 1.977 & 2.974 & 2.134 & 4.031 \\
    \hline
    $\text{Pad\'e}_{[2,1]}$ & 0.685 & 1.189 & 1.007 & 1.53 & 2.97 & -0.364 & 4.031 \\
    \hline
    $\text{Pad\'e}_{[0,3]}$ & 0.951 & 1.584 & 1.019 & 1.164 & -4.067 & 0.466 & 2.064 \\
    \hline
    $\text{Pad\'e}_{[3,0]}$ & 1.144 & 1.705 & 0.593 & -1.119 & 7.761 & -17.95 & 0.249 \\
    \hline
    \end{tabular}\caption{Pad\'e approximates of scaling dimensions of various operators in $d=3$ for the $[ns]$ fixed point of $SO(3)$ symmetric traceless theory. Only the Pad\'e [1,2] approximate contains no operators with negative scaling dimension.\label{table: SO(3) pade-scaling-[ns]-d=3}}
\end{table}
\begin{table}[H]\centering
    \begin{tabular}{|m{2 cm}|m{2cm}|m{1.7cm}|m{1.5cm}|m{1.6cm}|m{1.6cm}|m{1.5cm}|m{1.5cm}|}\hline
    \vspace{0.2cm} & $\Delta_{\phi}$ & $\Delta_{\psi}$ & $\Delta_{\phi^{2}}$ & $\Delta_{\left(\bar\psi\psi,\phi^{3}\right)_{1}}$ & $\Delta_{\left(\bar\psi\psi,\phi^{3}\right)_{2}}$ & $\Delta_{\left(g,y\right)_{1}}$ & $\Delta_{\left(g,y\right)_{2}}$ \\
    \hline
    \hline
    $\text{Pad\'e}_{[1,1]}$ & 2.092 & 2.633 & 0.704 & 1.236 & 3.077 & -2.396 & 4. \\
    \hline
    $\text{Pad\'e}_{[1,2]}$ & 0.566 & 1.023 & 0.858 & 1.474 & 2.85 & 1.431 & 4.066 \\
    \hline
    $\text{Pad\'e}_{[2,1]}$ & 0.355 & 0.866 & 0.311 & 0.077 & 2.827 & -4.727 & 4.065 \\
    \hline
    $\text{Pad\'e}_{[0,3]}$ & -0.791 & -1.318 & 0.488 & 0.294 & -0.205 & 0.078 & 0.451 \\
    \hline
    $\text{Pad\'e}_{[3,0]}$ & 4.547 & 5.522 & -4.125 & -22.98 & 46.16 & -148.5 & -27.5 \\
    \hline
    \end{tabular}\caption{Pad\'e approximates of scaling dimension of various operators in $d=2$ for the $[ns]$-fixed point of the $SO(3)$ symmetric traceless theory. Only the Pad\'e [1,2] approximate contains no operators with negative scaling dimension.\label{table: SO(3) pade-scaling-dimension-[ns]-d=2}}
\end{table}

\subsection{\texorpdfstring{$SO(3)~S_2-A_2$}{SO(3) S2-A2} model}
Here we present results in the $SO(3)$ $S_2-A_2$ model.  Table \ref{table: S2-A2 pade-scaling-[ns+]-d=2} lists all Pad\'e approximates for scaling dimensions in $d=2$ of the stable fixed point. At the unstable $[ns_-]$ fixed point of $SO(3)$ $S_2-A_2$ model, all Pad\'e approximates of scaling dimension for different operators are given in Tables \ref{table: S2-A2 pade-scaling-[ns-]-d=3} and \ref{table: S2-A2 pade-scaling-[ns-]-d=2}. For the stable $[ns_+]$ fixed point in $d=2$, only the [2,1] and [1,1] Pad\'e approximates contain no operators with negative scaling dimension. The $d=3$ Pad\'e approximates for the unstable $[ns_-]$ fixed point are mostly well behaved. However, all of the $d=2$ Pad\'e approximates for $[ns_-]$ contain operators with negative scaling dimensions, suggesting that the 2d CFT might not exist.

\begin{table}[H]\centering
    \begin{tabular}{|m{2 cm}|m{2cm}|m{1.7cm}|m{1.5cm}|m{1.6cm}|m{1.6cm}|m{1.5cm}|m{1.5cm}|}\hline
    \vspace{0.2cm} & $\Delta_{\phi}$ & $\Delta_{\psi}$ & $\Delta_{\phi^{2}}$ & $\Delta_{\left(\bar\psi\psi,\phi^{3}\right)_{1}}$ & $\Delta_{\left(\bar\psi\psi,\phi^{3}\right)_{2}}$ & $\Delta_{\left(g,y\right)_{1}}$ & $\Delta_{\left(g,y\right)_{2}}$ \\
    \hline
    \hline
    $\text{Pad\'e}_{[1,1]}$ & 0.213 & 0.916 & 2.055 & 3.677 & 3. & 5.143 & 5.402 \\
    $\text{Pad\'e}_{[1,2]}$ & -1.825 & 1.13 & 1.803 & 1.685 & 3.043 & 6.505 & 8.621 \\
    $\text{Pad\'e}_{[2,1]}$ & 0.201 & 0.852 & 1.786 & 1.093 & 3.042 & 5.788 & 6.881 \\
    $\text{Pad\'e}_{[0,3]}$ & 0.311 & 0.803 & 6.249 & 9.876 & 1.017 & -2.422 & -0.285 \\
    $\text{Pad\'e}_{[3,0]}$ & -0.09 & 0.542 & 3.428 & 8.023 & -2.848 & 13.61 & 45.5 \\
    \hline
    \end{tabular}\caption{Pad\'e approximates of scaling dimensions of various operators in $d=2$ for the $[ns_+]$ fixed point of $S_2-A_2$ theory. Only the Pad\'e [1,1] and [2,1] approximates contain no operators with negative scaling dimension.\label{table: S2-A2 pade-scaling-[ns+]-d=2}}
\end{table}
\begin{table}[H]\centering
    \begin{tabular}{|m{2 cm}|m{2cm}|m{1.7cm}|m{1.5cm}|m{1.6cm}|m{1.6cm}|m{1.5cm}|m{1.5cm}|}\hline
    \vspace{0.2cm} & $\Delta_{\phi}$ & $\Delta_{\psi}$ & $\Delta_{\phi^{2}}$ & $\Delta_{\left(\bar\psi\psi,\phi^{3}\right)_{1}}$ & $\Delta_{\left(\bar\psi\psi,\phi^{3}\right)_{2}}$ & $\Delta_{\left(g,y\right)_{1}}$ & $\Delta_{\left(g,y\right)_{2}}$ \\
    \hline
    \hline
    $\text{Pad\'e}_{[1,1]}$ & 0.539 & 1.11 & 1.041 & 1.386 & 3. & 2.377 & 5.06 \\
    $\text{Pad\'e}_{[1,2]}$ & 0.587 & 1.169 & 0.937 & 1.285 & 2.81 & 2.647 & 5.514 \\
    $\text{Pad\'e}_{[2,1]}$ & 0.571 & 1.148 & 0.987 & 1.608 & 2.798 & 1.984 & 5.005 \\
    $\text{Pad\'e}_{[0,3]}$ & 0.597 & 1.176 & 1.032 & 1.506 & 3.265 & 1.637 & 4.246 \\
    $\text{Pad\'e}_{[3,0]}$ & 0.591 & 1.198 & 0.914 & 1.306 & 3.244 & -0.745 & 4.448 \\
    \hline
    \end{tabular}\caption{Pad\'e approximates of scaling dimensions of various operators in $d=3$ for the $[ns_-]$ fixed point of $S_2-A_2$ theory. Only the Pad\'e [3,0] approximate, i.e. the result without resummation, contains an operator with negative scaling dimension.\label{table: S2-A2 pade-scaling-[ns-]-d=3}}
\end{table}
\begin{table}[H]\centering
    \begin{tabular}{|m{2 cm}|m{2cm}|m{1.7cm}|m{1.5cm}|m{1.6cm}|m{1.6cm}|m{1.5cm}|m{1.5cm}|}\hline
    \vspace{0.2cm} & $\Delta_{\phi}$ & $\Delta_{\psi}$ & $\Delta_{\phi^{2}}$ & $\Delta_{\left(\bar\psi\psi,\phi^{3}\right)_{1}}$ & $\Delta_{\left(\bar\psi\psi,\phi^{3}\right)_{2}}$ & $\Delta_{\left(g,y\right)_{1}}$ & $\Delta_{\left(g,y\right)_{2}}$ \\
    \hline
    \hline
    $\text{Pad\'e}_{[1,1]}$ & -0.088 & 0.561 & 0.413 & -0.363 & 3. & 1.375 & 6.254 \\
    $\text{Pad\'e}_{[1,2]}$ & 0.258 & 0.916 & -1.233 & -1.624 & 2.567 & 1.975 & 8.681 \\
    $\text{Pad\'e}_{[2,1]}$ & 0.12 & 0.788 & 0.153 & 0.033 & 2.493 & -0.011 & 6.01 \\
    $\text{Pad\'e}_{[0,3]}$ & 0.317 & 0.969 & 0.494 & 0.642 & -9.358 & 0.45 & 2.171 \\
    $\text{Pad\'e}_{[3,0]}$ & 0.336 & 1.279 & -0.699 & -0.999 & 6.962 & -23.84 & 1.356 \\
    \hline
    \end{tabular}\caption{Pad\'e approximates of scaling dimensions of various operators in $d=2$ for the $[ns_-]$ fixed point of $S_2-A_2$ theory. All Pad\'e  approximates contain operators with negative scaling dimension.\label{table: S2-A2 pade-scaling-[ns-]-d=2}}
\end{table}

\addcontentsline{toc}{section}{References}
\bibliographystyle{JHEP}
\bibliography{cft}

\end{document}